\documentclass[12pt]{article}
\usepackage{amsmath}
\usepackage{graphicx}
\usepackage{enumerate}
\usepackage{bbm}
\usepackage{natbib}
\usepackage{dsfont}
\usepackage{url} 
\usepackage{relsize,color}
\usepackage[dvipsnames]{xcolor}
\usepackage{setspace}
\usepackage{etoolbox}
\AtBeginEnvironment{tabular}{\onehalfspacing}
\usepackage{multirow}

\usepackage{xr}

\usepackage{algpseudocode}
\usepackage{algorithm}
\usepackage{epsfig}
\usepackage{ulem} 
\usepackage{xspace} 
\usepackage{graphicx}
\usepackage{sidecap}
\usepackage{times,epsfig,makeidx,amsfonts,amsmath}
\usepackage{lscape} 

\newcommand{\Norm}{\text{N}} 
\newcommand{\Gam}{\text{Gamma}} 

\newcommand{\di}{{\rm d}}

\newcommand{\bfs}{\textbf{s}}
\newcommand{\bfY}{\textbf{Y}}
\newcommand{\bfg}{\textbf{g}}
\newcommand{\bfn}{\textbf{n}}
\newcommand{\bfa}{\textbf{a}}
\newcommand{\bft}{\textbf{t}}
\newcommand{\bfV}{\textbf{V}}
\newcommand{\bfE}{\textbf{E}}
\newcommand{\bfT}{\textbf{T}}

\usepackage{ulem}

\newcommand{\blind}{1}

\begin{document}

\def\spacingset#1{\renewcommand{\baselinestretch}%
{#1}\small\normalsize} \spacingset{1}

	\vspace{-5in}
\if1\blind
{

  \title{\bf The Tajima heterochronous $n$-coalescent: inference from heterochronously sampled 
   molecular data}
  \author{Lorenzo Cappello$^1$\thanks{The authors gratefully acknowledge partial funding from the France-Stanford Center for Interdisciplinary Studies. JAP acknowledges support from National Institutes of Health grant
  		R01-GM-131404 and the Alfred P. Sloan Foundation. AV acknowledges partial funding from the chaire program Mathematical Modeling and Biodiversity (Ecole polytechnique, Museum National d'Histoire Naturelle, Veolia Environment, Foundation X).},  Amandine V\'eber$^{2*}$, Julia A. Palacios$^{1,3*}$\\
    $^1$Department of Statistics, Stanford University\\
   $^2$MAP5, CNRS, Universit\'e de Paris\\
$^3$Department of Biomedical Data Science, Stanford Medicine}
  \maketitle
} \fi

\if0\blind
{
  \bigskip
  \bigskip
  \bigskip
  \begin{center}
    {\LARGE\bf The Tajima heterochronous $n$-coalescent: inference from heterochronously sampled 
    	molecular data}
\end{center}
  \medskip
} \fi
\bigskip

\begin{abstract}

The observed sequence variation at a locus informs about the evolutionary history of the sample and past population size dynamics. The Kingman coalescent is used in a generative model of molecular sequence variation to infer evolutionary parameters. However, it is well understood that inference under this model does not scale well with sample size. Here, we build on recent work based on a lower resolution coalescent process, the Tajima coalescent, to model longitudinal samples. While the Kingman coalescent models the ancestry of labeled individuals, the heterochronous Tajima coalescent models the ancestry of individuals labeled by their sampling time. We propose a new inference scheme for the reconstruction of effective population size trajectories based on this model with the potential to improve computational efficiency. 
Modeling of longitudinal samples is necessary for applications (\textit{e.g.} ancient DNA and RNA from rapidly evolving pathogens like viruses) and statistically desirable (variance reduction and parameter identifiability). 
We propose an efficient algorithm to calculate the likelihood and employ a Bayesian nonparametric procedure to infer the population size trajectory. We provide a new MCMC sampler to explore the space of heterochronous Tajima's genealogies and model parameters. We compare our procedure with state-of-the-art methodologies in simulations and applications.

\end{abstract}

\noindent%
\vfill

\newpage
\spacingset{1.5} 
\section{Introduction}
\label{sec:intro}

Statistical inference of evolutionary parameters from a sample 
of $n$ DNA sequences \bfY{} accounts for the dependence among samples and models observed variation
through two stochastic processes: an ancestral process of the sample represented by a genealogy \bfg{}, and a mutation process with a given set of parameters $\mu$ that, conditionally on \bfg{}, models the phenomena that have given rise to the sequences. A standard choice for modeling \bfg{} is the Kingman $n$-coalescent, \citep{kingn82,king82}, a model that depends on a parameter called \textit{effective population size} $(N_e(t))_{t\geq 0}$ (henceforth $N_e=(N_e(t))_{t\geq 0}$). The function $N_e$ is a measure of genetic diversity that, in the absence of natural selection, can be used to approximate census population size when direct estimates are difficult to obtain due to high costs, challenging sampling designs, or simply because past estimates are not available. Hence, inference of $N_e$ has important applications in many fields, such as genetics, anthropology, and public health.

Standard approaches to do Bayesian inference of $N_{e}$ stochastically approximates the posterior distribution $\pi(N_e| \bfY{}, \mu)$ through Markov chain Monte Carlo (MCMC). This approximation requires the definition of Markov chains (MCs) on genealogies, whose state space is the product space $\mathcal{G}_n\times \mathbb{R}_+^{n-1}$ of tree topologies ($g \in \mathcal{G}_n$) and coalescent times $\textbf{t} \in \mathbb{R}^{n-1}_{+}$ (times between consecutive coalescence events in the topology). The convergence of MCs in these type of spaces is notoriously challenging: the posterior is highly multi-modal \citep{whi15}, with most of the posterior density concentrating on few separated tree topologies; in addition, theoretical results on mixing times of MCs on tree topologies $\mathcal{G}_n$ in simpler settings (i.e. with uniform stationary distribution on $\mathcal{G}_n$) show polynomial mixing times in the number of leaves ($n$) \citep{ald83}. The issue is exacerbated as the sample size increases because the cardinality of $\mathcal{G}_n$ grows superexponentially with $n$ for the standard coalescent ($|\mathcal{G}_n|=n!(n-1)!/2^{n-1}$). The result is that state-of-the-art methodologies are not scalable to the amount of data available.

To resolve this computational bottleneck, much research has focused on algorithms that are known to scale to large datasets and less prone to ``get stuck" into local modes, such as sequential Monte Carlo \citep{bou12,wan15,fou17,din17}, Hamiltonian Monte Carlo \citep{din17hmc}, and variational Bayes \citep{zha18}. We argue that an alternative (or perhaps complementary) solution to this problem consists of considering a lower resolution ancestral (genealogical) process. The Tajima $n$-coalescent \citep{taj83,sai15,pal19} is a lumping of the Kingman $n$-coalescent whose realizations are in bijection with the set of timed and \textit{unlabeled} binary trees with $n$ leaves, a space of trees with a drastically smaller cardinality than that of the space of Kingman trees \citep{di13}. Mathematically speaking, this amounts to taking equivalence classes of Kingman trees, in which only the ranking of the coalescence events is retained, but leaf labels are removed so that the external tree branches are all considered equivalent. Intuitively, the likelihood values conditional on this type of tree should be ``more concentrated",  in the sense that, for a fixed dataset,  the range of possible likelihood values is drastically smaller. We conjecture that this property, along with the cardinality reduction, contributes to a more efficient exploration of the tree space. We elaborate on this argument through the following example.

\begin{SCfigure}[2][t]
	\hspace{-0.2cm}\includegraphics[width=.5\textwidth]{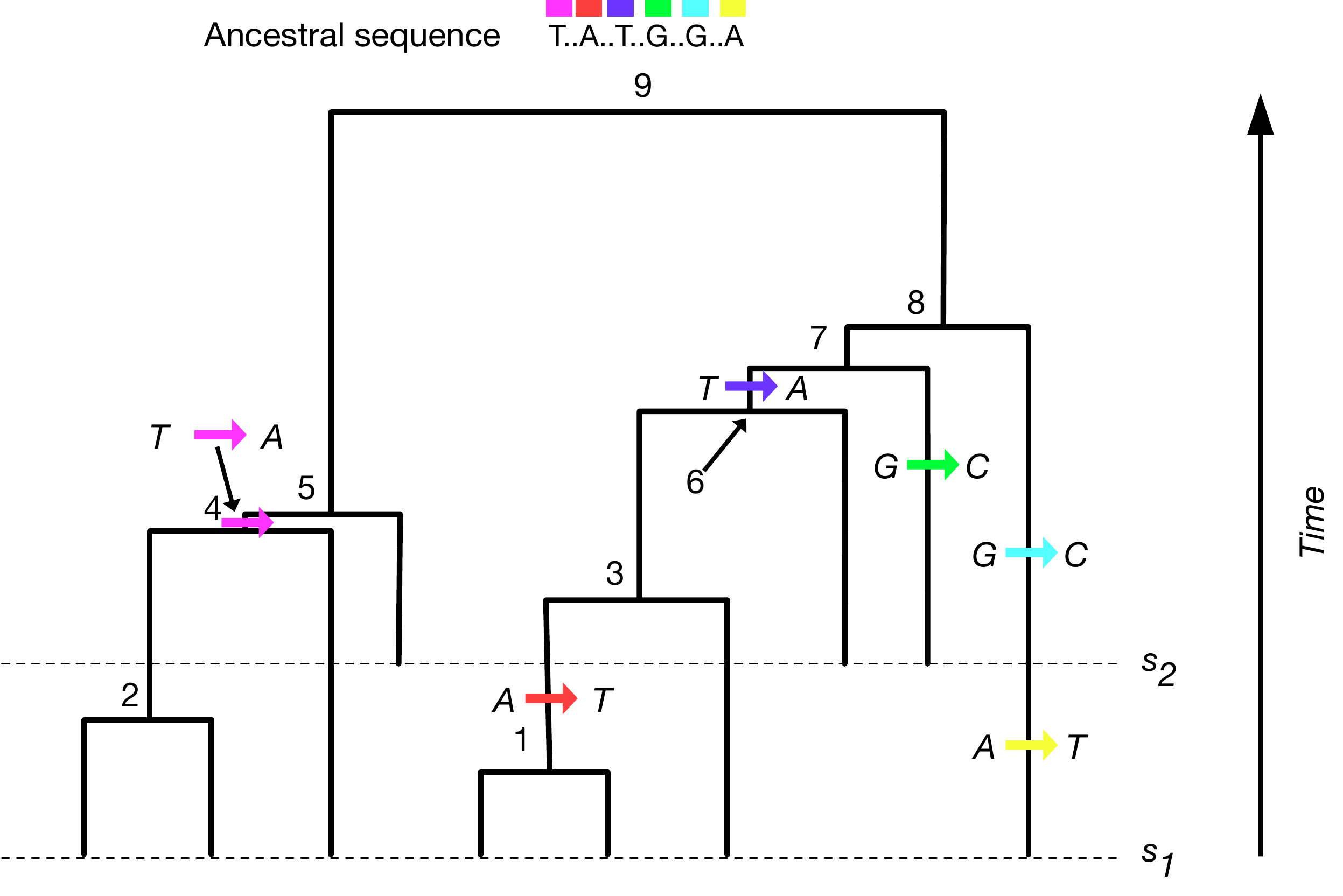}
	\caption{\small{\textbf{Coalescence and mutation.} A genealogy of 10 individuals at a locus of 100 sites is depicted as a bifurcating tree. 
			Six mutations (at different sites) along the branches of the tree give rise to the 10 sequences. Black dots represent the 94 sites that do not mutate in the ancestral sequence. The nucleotides at the polymorphic sites are shown, and the colored arrows depict how ancestral sites are modified by mutation}.}
	\label{fig:coalescent}
\end{SCfigure}

\begin{figure}[ht]
	\centering
	\includegraphics[scale=0.4]{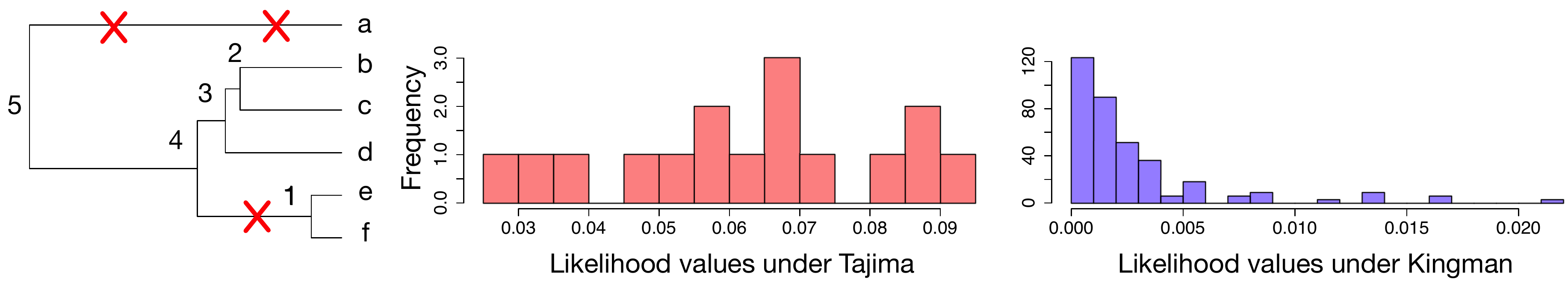}
	\caption{\small{\textbf{Distributions of the likelihood values under the Kingman and the Tajima $n$-coalescent for a given dataset.} The first plot shows a realization of a genealogy of $n=6$ samples (tips) with three mutations superimposed (marked as X).  The second plot shows the histogram of the likelihood values conditionally on all possible Tajima tree topologies, and the third plot shows the histogram of the likelihood values conditionally on all possible Kingman trees with $6$ leaves. We assumed the same coalescent times across all trees.}}
	\label{fig:likTajKing}
\end{figure}

We generated a ranked binary tree with $n=6$ tips and superimposed $3$ mutations along the branches of the tree at $3$ different sites (Figure \ref{fig:likTajKing}), yielding an (unlabeled) sequence alignment made of one sequence carrying two mutations and two sequences carrying a third mutation.	We compute the likelihood conditionally on all Kingman and Tajima tree topologies with $6$ leaves, assuming all trees have the same ``true" coalescent times and a mutation model called the infinite sites model (ISM) (\cite{kim69}, details described below in the paper). There are respectively $360$ Kingman topologies and $16$ Tajima topologies with positive likelihood. Figure \ref{fig:likTajKing} plots the distribution of the normalized likelihood values along with their frequencies. Under Kingman's coalescent, the maximum likelihood value is about $823.2$ times larger than the minimum likelihood value. Under Tajima's coalescent, this ratio between the maximum and minimum likelihood value is about $3.3$. Besides, the profiles are remarkably different: under Kingman's coalescent, there are many trees with a negligible likelihood and a few with higher values; under Tajima's coalescent, the more frequent likelihood values are closer to the center. We conjecture that Tajima's likelihood profile should make MCs exploration of the whole state space more manageable, with higher acceptance probabilities and allowing moves between modes.

The difference observed in Figure \ref{fig:likTajKing} follows from the type of topologies used. The Tajima $n$-coalescent partitions the space of Kingman's trees into equivalence classes, where each Tajima's topology corresponds to a set of Kingman's topologies. When we compute the likelihood under Tajima, we generally account for a ``large number" of Kingman's topologies. In the example discussed, we are effectively summing over many topologies having a small likelihood. We stress that there is no loss of information when lumping states of tree topologies; the two marginal likelihood functions only differ by a constant. Although this example relies on the ISM assumption, intuitively, the likelihood profiles would have similar differences under more general mutation models: many Kingman trees with zero likelihood under ISM will have a very small likelihood under alternative mutation models. More details on this example are given in the supplementary material.

\cite{pal19} proposed to use the Tajima coalescent and introduced a new algorithm based on this encoding of the hidden genealogies for the likelihood calculation and inference of $N_e$. Despite the advances in that paper, there are still many challenges to be addressed for the Tajima $n$-coalescent to be a viable alternative to the Kingman $n$-coalescent. First, the algorithm for the likelihood calculation of \cite{pal19} can be prohibitively expensive; a loose upper bound of the current algorithm's complexity is $\mathcal{O} (n!)$. Second, the definition of the likelihood relies on several restrictive modeling assumptions such as the ISM mutation model, no recombination, no population structure, and the fact that all samples are obtained at a single point in time. 
To make Tajima-based inference attractive, further research is needed given the large body of literature and software programs developed for the standard Kingman coalescent. 

This paper includes the following contributions: we introduce a new algorithm for likelihood calculation whose upper bound complexity is $\mathcal{O} (n^2)$, and we extend the Tajima modeling framework to sequences observed at different time points like those at the tips of the genealogy in Figure~\ref{fig:coalescent}, \textit{i.e.}, heterochronous data. We also extend the methodology to allow for joint estimation of the mutation rate $\mu$, $N_e$, and other parameters, from data collected at multiple independent loci. These extensions will enable us to investigate further the use of the Tajima $n$-coalescent for inference of effective population size trajectories while allowing practitioners to use it in more readily applicable settings immediately.

Out of the many possible directions that may have been pursued from \cite{pal19}, the extension to heterochronous data was prioritized for several reasons: $(i)$ data are collected longitudinally in many applications (\textit{e.g.}, ancient DNA and viral DNA), $(ii)$ employing longitudinal data reduces the variance of the estimators of $N_e$ \citep{rod99} and $(iii)$ the model becomes identifiable for joint estimation of mutation rates and effective population sizes \citep{dru02, par19}. While our current implementation is limited to a single mutation model (ISM), we note that research employing the ISM is still very active, both in terms of method development \citep{spe19}, and emerging research areas in evolutionary biology, such as cancer dynamics \citep{rub20,Quinneabc1944} and single-cell lineage tracing studies \citep{jon20}.

To give an example of the applications that can be handled with the current model, we include two real data applications: we analyze ancient samples of bison in North America \citep{fro17}, revisiting the question of why the Beringian bison went extinct \citep{sha04}, and in a second study, we analyze viral samples of SARS-CoV-2, the virus responsible for the current COVID-19 pandemic.

Dealing with longitudinal data requires the definition of a continuous time Markov chain, which is a lumping of the Kingman heterochronous $n$-coalescent. \cite{rod99} introduced the Kingman \textit{heterochronous} $n$-coalescent as a model for ranked labeled heterochronous genealogical trees. We refer to the lower resolution of this process (the lumped process) as the Tajima \textit{heterochronous} $n$-coalescent. This process differs from the Tajima $n$-coalescent \citep{sai15, pal19}  in that sequences sampled at different time points are not exchangeable. The Tajima $n$-coalescent distinguishes between singletons and \textit{vintaged} lineages, where a singleton lineage refers to a lineage that subtends a leaf in \bfg{}, and a vintaged lineage refers to a lineage that subtends an internal node in \bfg{}. Singletons are indistinguishable, while vintages are labeled by the ranking of the coalescence event at which they were created. When dealing with heterochronous samples, singletons are instead implicitly labeled by their underlying sampling times so that only singletons sampled simultaneously are indistinguishable.

Fast likelihood calculation is essential for the usability of the methodology. The algorithm to compute the likelihood relies on a graphical representation of the data as a tree structure. We note that this tree structure extends the tree structure representation of isochronously sampled data (also called the gene tree \citep{grif94}, the perfect phylogeny \citep{gus14}, and the directed acyclic graph (DAG) \citep{pal19}) to heterochronous data. We also stress that although all these graphs are tree structures, they are graphical representations of the data \bfY{} under the ISM and not representations of the underlying genealogical tree.

The rest of the paper proceeds as follows. In Section~\ref{sec:hettaj}, we define the Tajima heterochronous $n$-coalescent. In Section~\ref{lik}, we introduce the mutation model we shall assume, describe the data, define the likelihood and the new algorithm to compute it. Section~\ref{mcmc} describes the MCMC algorithm for posterior inference, and in Section~\ref{sim}, we present a comprehensive simulation study outlining how the model works and comparing our method to state-of-the-art alternatives. In Section~\ref{app}, we analyze modern and ancient bison sequences described in \cite{fro17}. In Section \ref{covid}, we apply our method to SARS-CoV-2 viral sequences collected in France and Germany. Section~\ref{concl} concludes. An open-source implementation is available. 

\section{The Tajima heterochronous $n$-coalescent}
\label{sec:hettaj}

\begin{figure}[ht]
	\centering
\includegraphics[scale=0.4]{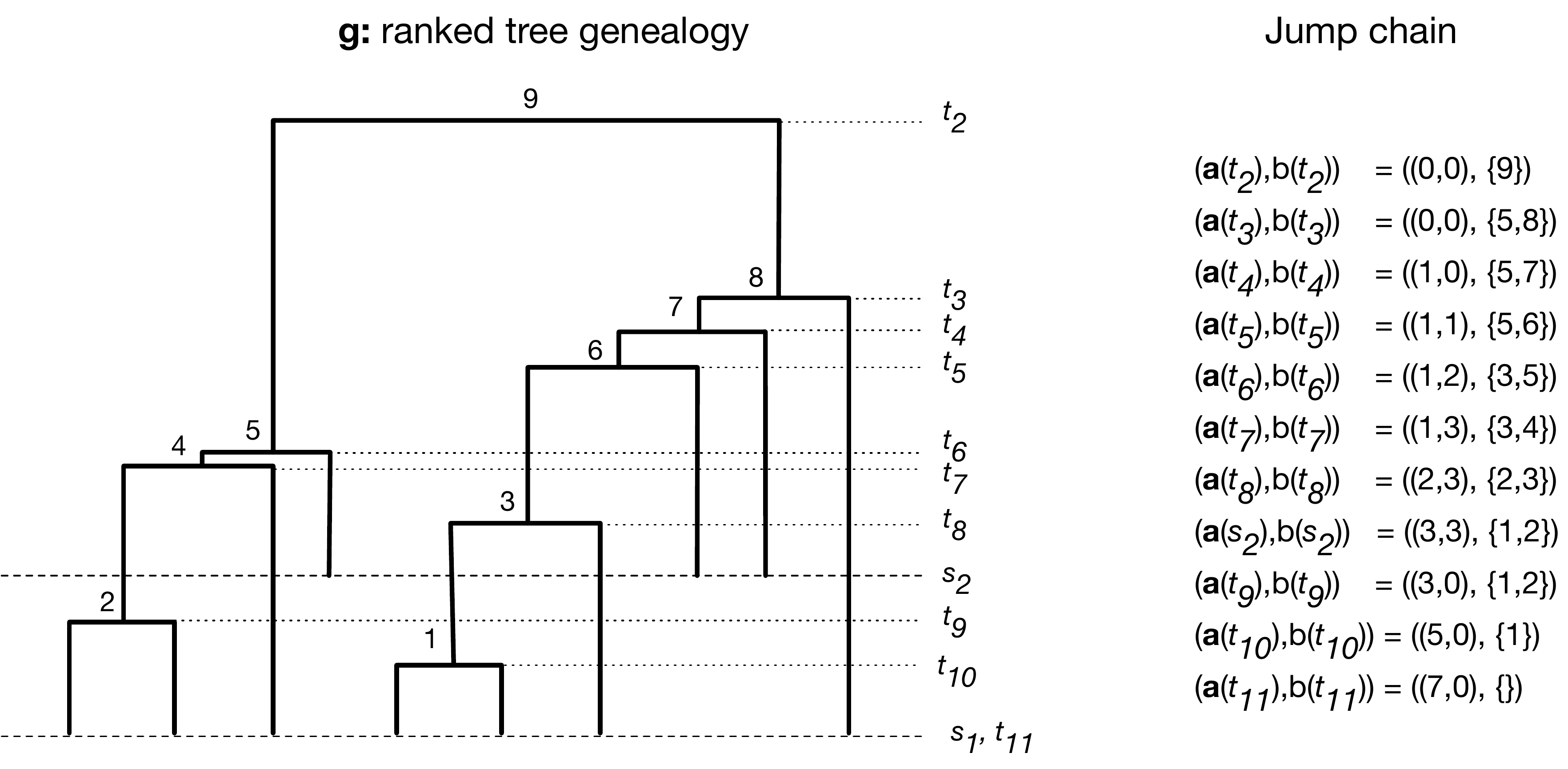}
	\caption{\small{\textbf{Example of a Tajima heterochronous genealogy and its jump chain.} A realization of a Tajima heterochronous $n$-coalescent with $\bfn{}=(7,3)$ and $\bfs{}=(s_1,s_2)$, represented as a ranked tree shape with coalescence and sampling times, denoted \bfg{}. The column to the right displays the corresponding jump chain (see the text for notation).}}
	\label{fig:hetTaj}
\end{figure}

The Tajima heterochronous $n$-coalescent is an inhomogeneous continuous-time Markov chain that describes the ancestral relationships of a set of $n$ individuals sampled, possibly at different times, from a large population. The set of ancestral relationships of the sample is represented by a ranked genealogy, for example the one depicted in Figure \ref{fig:hetTaj}. Every organism is dated and labeled according to the time in which the organism lived (if ancient, by radiocarbon date) or in which the living organism was sequenced. 
In this generalization of the Tajima coalescent, each pair of extant ancestral lineages merges into a single lineage at an instantaneous rate depending on the current effective population size $N_e(t)$, and new lineages are added when one of the prescribed sampling times is reached. This is the same mechanism of the heterochronous $n$-coalescent of \cite{rod99}. We do not model the stochasticity of sampling times, but we condition on them as being fixed.


Let us introduce some notation. Let $m$ be the number of sampling time points and $n$ be the total number of samples. Let $\bfn{}=(n_1,\ldots,n_{m})$ denote the number of sequences  collected at times $\bfs{}=(s_{1},\ldots,s_{m})$, with $s_{1}=0$ denoting the present time, and $s_{j}>s_{j-1}$ for $j=2,\ldots,m$ (time goes from present towards the past). We refer to the sequences counted in $n_i$ as ``belonging to sampling group $s_i$''. Let $\bft{}=(t_{n+1}, \ldots, t_{2})$ be the vector of coalescent times with $t_{n+1}=0<t_{n}<...<t_{2}$; these are the times when two lineages have a common ancestor. 
Note that the subscript in $t_{k}$ does not indicate the current number of lineages, as it is often done in the coalescent literature, but it indicates the number of lineages that have yet to coalesce (some sequences may not have been sampled yet). We use the rank order of the coalescent events (bottom-up) to label the internal nodes of the genealogy. 
That is, the node corresponding to the coalescent event occurring at time $t_n$ is labeled $1$ (see $t_{10}$ in Figure~\ref{fig:hetTaj}), the node corresponding to the coalescence event occurring at time $t_{n-1}$ is labeled $2$, \textit{etc}. We refer to the internal node labels as \textit{vintages} (\textit{i.e.},  rankings).

The Tajima heterochronous $n$-coalescent is the process $(\bfa{}(t), b(t))_{t \geq 0}$ that keeps track of $\bfa{} (t)$, a vector of length $m$ whose $j$-th position indicates the number of singletons (\textit{i.e.}, lineages that have not been involved in a coalescence event) from sampling group $s_j$ at time $t$, and $b (t)$ is the set of vintaged lineages at time $t$.  The process starts at $t=0$ in state $(\bfa{}(0)=(n_1,0,\dots,0), b(0)=\emptyset)$, jumps deterministically at every sampling time and jumps stochastically at every random coalescent time until it reaches the unique absorbing state $(\bfa{(t_2)}=(0,\ldots, 0), b(t_2)=\{n-1\})$ at time $t_2$, when all $n$ samples have a single most recent common ancestor at the root (Figure \ref{fig:hetTaj}). At each sampling time $s_i$, the state of the Tajima coalescent jumps deterministically as follows:
$$(\bfa{}(s_i), b(s_i))=(\bfa{}(s_i-)+ n_i\textbf{e}_i, b(s_i-)),$$
where $f(s_i-)$ denotes the left-limit of the function $f$ at $s_i$ and $\textbf{e}_i$ is the $i$-th unit vector.

Let us now turn to the embedded jump chain at coalescent times. 
At time $t_i$, two extant lineages coalesce to create a new lineage with vintage $n+1-i$. Four types of coalescence transitions are possible depending on which and how many sampling groups are involved: (1) two singletons of the same sampling group coalesce (up to $m$ possible moves for the chain), (2) two singletons of different sampling groups coalesce (up to $m (m-1)/2$ possible moves), (3) one singleton lineage and one vintaged lineage coalesce (up to $m$ possible moves), or (4) two vintaged lineages coalesce (only one possibility because for vintages, the sampling information is irrelevant). Each pair coalesces with the same probability and the transition probabilities at coalescent times are thus given by
\begin{align}
\label{transition}
P\Big[(\bfa{(t_i)}, &b(t_i))\Big| (\bfa{(t_i-)},b(t_i-))\Big]\\
& =\left\{
\begin{array}{ll}
\frac{\mathlarger\prod_{j=1}^m \dbinom{a_j(t_i-)}{a_j(t_i-)-a_j(t_i)}}{\dbinom{\sum_{j=1}^m a_j(t_i-)+|b(t_i-)|}{2}} \,\ \,\ \,\ \,\ & \text{if} \ \,\  (\bfa{(t_i)},b(t_i)) \prec  (\bfa{(t_i-)},b(t_i-))\nonumber\\
\,\ \\
0 \,\ \,\ \,\ \,\  \,\ \,\ \,  \,\ \,\  & \text{otherwise}
\end{array}
\right.
\end{align}
where $(\bfa{(t_i)},b(t_i)) \prec  (\bfa{(t_i-)},b(t_i-))$ means that $(\bfa{(t_i)},b(t_i))$ can be obtained by merging two lineages of $(\bfa{(t_i-)},b(t_i-))$ and $|b|$ denotes the cardinality of the set $b$.

Observe that the quantity $\sum_{j=1}^m a_j(t_i-)+|b(t_i-)|$ appearing in \eqref{transition} corresponds to the total number of extant lineages just before the event at $t_i$. Furthermore, since only two lineages coalesce at time $t_i$, at most two terms in the product appearing in the numerator of \eqref{transition} are not equal to one. Finally, if $m=1$, \eqref{transition} degenerates into the transition probabilities of the Tajima isochronous $n$-coalescent; on the other hand, if $m=n$, the process degenerates into the Kingman heterochronous $n$-coalescent since all singletons are uniquely labeled by their sampling times. Figure~\ref{fig:hetTaj} shows a possible realization from the Tajima heterochronous $n$-coalescent. Notice that in applications, the number of observations collected at any given time instance is generally larger than one, and hence, the heterochronous Tajima model would have a smaller state space than the Kingman model on heterochronous data. We investigate this assertion by quantifying how much bigger the state space of the Kingman heterochronous coalescent is compared to that of the Tajima heterochronous coalescent for a given dataset. We employ a sequential importance sampling to tackle this combinatorial question, extending the methodology of \cite{cap19}. Details can be found in the Supplementary material. The results suggest that, while it is true that the cardinalities of the two latent spaces are closer when there are more sampling groups, the difference between the cardinalities can be very significant when the entries of {\bfn}  are large.

To define the distribution of the holding times, we introduce the following notation. We denote the intervals that end with a coalescent event at $t_{k}$ by $I_{0,k}$ and the intervals that end with a sampling time within the interval $(t_{k+1},t_{k})$ as $I_{i,k}$ where $i\geq 1$ is an index tracking the sampling events in $(t_{k+1},t_{k})$. More specifically, for every $k\in \{2,\ldots,n\}$, we define 
\begin{equation}
I_{0,k}=[\max\{t_{k+1},s_{j}\},t_{k}),\quad \text{ where the maximum is taken over all } s_{j}<t_{k},
\end{equation}
and for every $i\geq 1$ we set
\begin{equation}
I_{i,k}=[\max\{t_{k+1},s_{j-i}\},s_{j-i+1}) \text{ with the max taken over all } s_{j-i+1}>t_{k+1} \text{ and } s_{j}<t_{k}.
\end{equation}
We also let $n_{i,k}$ denote the number of extant lineages during the time interval $I_{i,k}$. For example, in Figure~\ref{fig:hetTaj}, in $(t_9, t_8)$ we have $I_{0,8}=[s_2,t_8)$, $I_{1,8}=[t_9,s_2)$ and no $I_{i,8}$ for $i\geq 2$.
The vector of coalescent times \bft{} is a random vector whose density with respect to Lebesgue measure on $\mathbb{R}_+^{n-1}$ can be factorized as the product of the conditional densities of $t_{k-1}$ knowing $t_k$, which reads: for $k=3,...,n+1$,
\begin{equation}
\label{prior_time}
p(t_{k-1}\mid t_{k},\mathbf{s},\mathbf{n},N_{e}(t))=\frac{C_{0,k-1}}{N_{e}(t_{k-1})} \exp \left\lbrace - \int_{I_{0,k-1}} \frac{C_{0,k-1}}{N_{e}(t)}\di t+\sum^{m}_{i=1}\int_{I_{i,k-1}} \frac{C_{i,k-1}}{N_{e}(t)}\di t\right\rbrace,
\end{equation}
where $t_{n+1}=0$ by convention, $C_{i,k}:=\binom{n_{i,k}}{2}$, and the integral over $I_{i,k-1}$ is zero if there are less than $i$ sampling times between $t_k$ and $t_{k-1}$. The distribution of the holding times defined above corresponds to the same distribution of holding times in the heterochronous Kingman coalescent \citep{rod99}. Although the heterochronous Tajima coalescent takes value on a different state space, it remains true that every pair of extant lineages coalesces at equal rate.


Finally, given \bfn{}, \bfs{} and $\bft{}$, a complete realization of the Tajima heterochronous $n$-coalescent chain can be uniquely identified with an unlabeled binary ranked tree shape $g$ of $\mathbf{n}=(n_{1},\ldots,n_{m})$ samples at $(s_{1},\ldots,s_{m})$ with its $n-1$  coalescent transitions, so that
\begin{equation}
\label{prior_rts}
P(g \mid \bft{},\bfs{},\bfn)=\prod_{i=2}^{n} P\Big[(\bfa{(t_i)},b(t_i))\,\Big|\, (\bfa{(t_i-)},b(t_i-))\Big].
\end{equation}
 Equation~\eqref{prior_rts} gives the prior probability of the tree topology under the Tajima heterochronous $n$-coalescent. Putting together \eqref{prior_time} and \eqref{prior_rts}, we obtain a prior $\pi(\bfg{}\mid \bfs{},\bfn{}, N_e)$
\begin{equation}
\label{prior}
\pi(\bfg{}\mid \bfs{},\bfn{}, N_e)=P(g \mid \bft{},\bfs{},\bfn) \prod_{k=3}^{n+1}p(t_{k-1}\mid t_{k},\bfs{},\bfn,N_{e}).
\end{equation}


\section{Data and Likelihood}
\label{lik}

\subsection{Infinite Sites Model and the Perfect Phylogeny}
\label{data}
We assume that the observed data \bfY{} consists of $n$ sequences at $z$ polymorphic (mutating) sites at a non-recombining contiguous segment of DNA of organisms with a low mutation rate. Under these assumptions, a widely studied mutation model is the \textit{infinite sites model} (ISM) \citep{kim69,wat75} with Poissonian mutation, which corresponds to a Poisson point process with rate $\mu$ on the branches of \bfg{} such 
that every mutation occurs at a different site and 
no mutations are hidden by a second mutation affecting the same site.

\begin{figure}[t]
	\includegraphics[clip, width=1\textwidth]{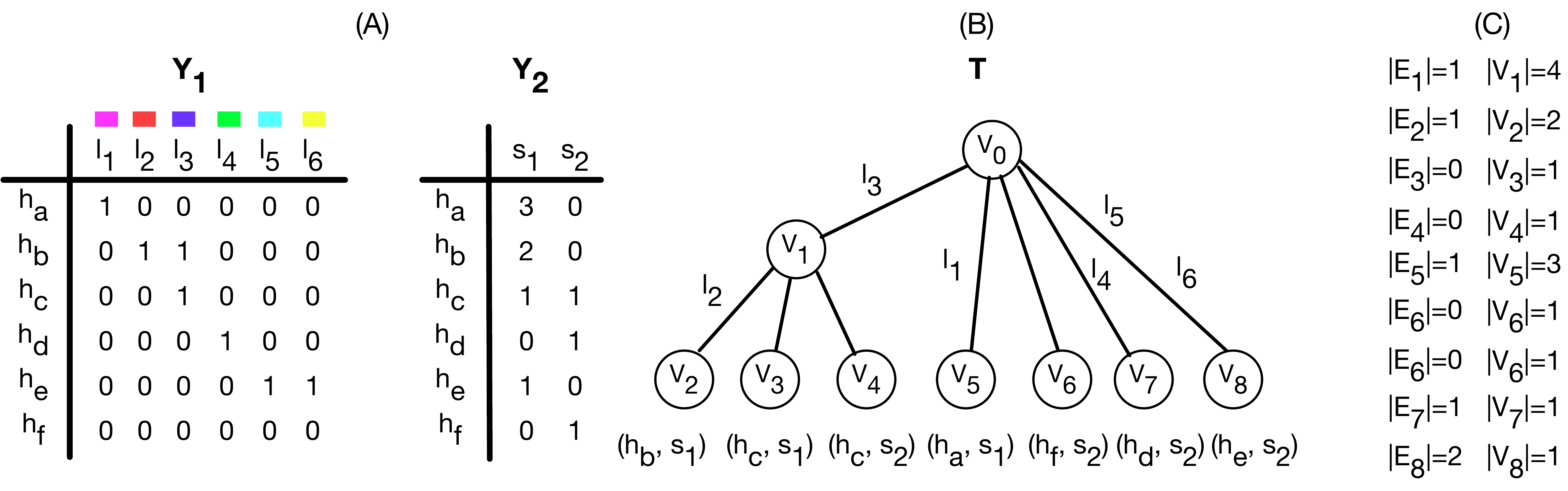}
	\caption{\small{\textbf{Incidence matrix, frequency matrix and perfect phylogeny representation}. Panel (A): data is summarized as an incidence matrix $\bfY{}_1$ ($h$ denotes the haplotypes, $l$ the segregating sites, the colors correspond to those depicted in Figure \ref{fig:coalescent}) and a matrix of frequencies $\bfY{}_2$ ($s$ denotes the sampling group). Panel (B): \bfT{}  denotes the perfect phylogeny corresponding to $\bfY_1$ and $\bfY_2$; each of the 6 polymorphic sites labels exactly one edge. When an edge has multiple labels, the order of the labels is irrelevant. Each leaf node is labeled by a pair (haplotype, sampling time), with each haplotype possibly labeling more than  one leaf nodes. Panel (C): $|E_{i}|$ corresponds to the number of mutations along the edge subtending node $V_{i}$ in (B) and $|V_{i}|$ corresponds to the number of sequences descending from $V_{i}$ in (B), see the text for details.}}
	\label{perfect_phylo}
\end{figure}

An important consequence of the ISM is that \bfY{} can be represented as an incidence matrix $\bfY{}_1$ and a frequency counts matrix $\bfY{}_2$. $\bfY{}_1$ is a $k \times z$ matrix with $0$-$1$ entries, where $0$ indicates the ancestral type and $1$ indicates the mutant type; $k$ is the number of unique sequences (or haplotypes) observed in the sample, and the columns correspond to polymorphic sites.  $\bfY{}_2$ is a $k \times m$ count matrix where the $(i,j)$th entry denotes  how many haplotype $i$ sequences belonging to group $s_j$ are sampled. For example, the $n = 10$ sequences defined by the realizations of the ancestral and mutation processes depicted in Figure~\ref{fig:coalescent} can be summarized into $\bfY _1$ and $\bfY_2$ displayed in Figure~\ref{perfect_phylo}(A). Note that we make the implicit assumption that we know which state is ancestral at each segregating site. However, this assumption can be relaxed, see \citep{grif95}, although the computational cost will substantially increase.

$\bfY{}_1$ and $\bfY_2$ can alternatively be represented graphically as an \textit{augmented perfect phylogeny} $\bfT$. Our likelihood algorithm exploits this graphical representation of the data. The augmented perfect phylogeny representation is an extension of the \textit{gene tree} or \textit{perfect phylogeny} \citep{gus91,gri94b,pal19} to the heterochronous case. The standard perfect phylogeny definition leaves out the information carried by $\bfY_2$. In the augmented perfect phylogeny $\bfT=(\bfV,\bfE)$, $\bfV{}$ is the set of nodes of \bfT{}, and $\bfE{}$ is the set of weighted edges. We define \bfT{} as follows:
\begin{enumerate}
	\item Each haplotype labels at least one leaf in $\bfT$. If a haplotype is observed at $k$ different sampling times, then $k$ leaves in $\bfT$ will be labeled by the same haplotype. The pair (haplotype label, sampling group) uniquely labels each leaf node.
	\item Each of the $z$ polymorphic sites labels exactly one edge. When multiple sites label the same edge, the order of the labels along the edge is arbitrary. Some external edges (edges subtending leaves) may not be labeled, indicating that they do not carry additional mutations to their parent node.
	\item For any pair (haplotype $h_{k}$, sampling group), the labels of the edges along the unique path from the root to the leaf $h_{k}$ specify all the sites where $h_{k}$ has the mutant type.
\end{enumerate}

Figure~\ref{perfect_phylo}(B)  plots \bfT{} corresponding to $\bfY_1$ and $\bfY_2$ displayed in Figure~\ref{perfect_phylo}(A). Observe that \bfT{} includes sampling information in the leaf labels. In the example, $h_C$ labels two leaves because it is observed at times $s_1$ and $s_2$. The corresponding edges $E_3$ and $E_4$ are unlabeled, \textit{i.e.}, no mutations are allocated to those edges because the underlying nodes carry identical sequences (same haplotype). We ``augment" Gusfield's perfect phylogeny because the sampling information is crucial in the likelihood calculation.

\bfT{} implicitly carries some quantitative information that can be quickly summarized. We denote the number of observed sequences subtended by an internal node $V$ by $|V|$. If $V$ is a leaf node, $|V|$ denotes the frequency of the haplotype $h$ observed at the corresponding sampling time~$s$. Similarly, we denote the number of mutation labels assigned to an edge $E$ by $|E|$. If no mutations are assigned to $E$, then $|E|=0$. For parsimony, the edge that connects node $V_i$ to its parent node is denoted by $E_i$. See Figure~\ref{perfect_phylo}(C) for an example.


 \cite{gus91} gives an algorithm to construct the perfect phylogeny \bfT'  in linear time.  Constructing \bfT{} from \bfT' is straightforward since all we need is to incorporate the sampling information and add leaf nodes if a haplotype is observed at multiple sampling times.
If we drew $\bfT'$ from the data in Figure~\ref{perfect_phylo}, it would not have node $V_4$, but only a single node $V_3$ labeled by haplotype $h_C$. A description of the algorithm can be found in the supplementary material.

\subsection{Likelihood}


The crucial step needed to compute the likelihood of a Tajima genealogy $\bfg{}$ is to sum over all possible allocations of mutations to its branches. This can be efficiently done by exploiting the augmented perfect phylogeny representation of the data \bfT{} and by first mapping nodes of \bfT{} to subtrees of \bfg. We stress that the need for an allocation step arises only when working with Tajima genealogies. In Kingman's coalescent, tree leaves are labeled by the sequences to which they correspond, and so there is a unique possible allocation. In Tajima's coalescent, leaves are unlabeled, creating potential symmetries in the tree, and so we have to scan all the possible ways in which the observed sequences may be allocated to \bfg{}.

\subsubsection{Allocations}
\label{sec:alloc}

\begin{SCfigure}[2][t]
	\hspace{-0.2cm}\includegraphics[width=.5\textwidth]{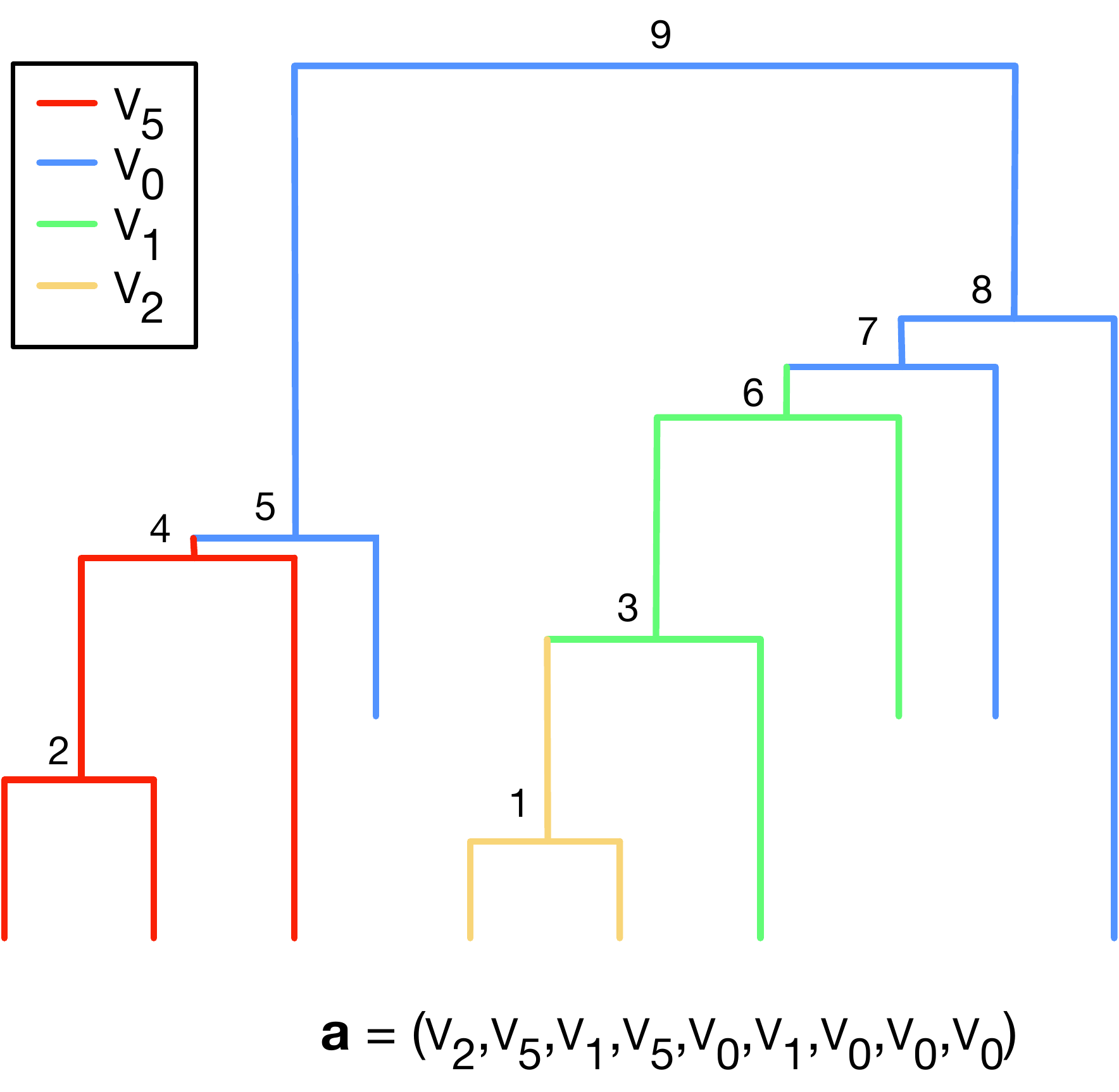}
	\caption{\small{\textbf{A possible allocation of non-singleton nodes of \bfV{} to subtrees of \bfg{}}. For a given allocation \bfa{} (bottom figure), we display how subtrees in \bfg{} (identified by the vintage tag at their root -- black number in the top figure) are allocated to the nodes of \bfT{}. Each color depicts an allocation of a subtree to a node: $V_5$ (red), $V_0$ (blue), $V_1$(green) and $V_2$ (yellow). 
	\hspace{1cm}}}
	\label{sub_all}
\end{SCfigure}

Let \bfa{} denote a possible mapping of nodes of  $\textbf{T}$ to subtrees of $\bfg$. \bfa{} is encoded as a vector of length $n-1$, where the $i$-th entry gives the node in $\bfT$ which is mapped to the subtree with vintage $i$, $\bfg_i$ (including the branch that subtends vintage $i$). Our algorithm first maps all \textit{non-singleton} nodes $\mathbf{V}$ of $\mathbf{T}$ to subtrees of $\bfg{}$, that is, only nodes such that $|V|>1$ are entries of \bfa{}. Singleton nodes in $\mathbf{T}$ ($V\in \mathbf{V}$ such that $|V|=1$) are treated separately and are initially excluded from the allocation step. For example, Figure \ref{sub_all} shows a possible vector \bfa{} whose entries are the non-singleton nodes $V_{0},V_{1},V_{2},$ and $V_{5}$ of $\bfT$ of Figure \ref{perfect_phylo}. We note that
nodes can appear more than once in \bfa{}, meaning that they can be mapped to more than one subtree. 
On the other hand, a single node $V_i$ is not necessarily mapped to all the vintages, leaves and internal branches of $\bfg_j$; different nodes may be mapped to some subtrees of $\bfg_j$ (including external branches), leading to a situation where $V_i$ is mapped to only a subset of the vintages and branches constituting $\bfg_j$. For example, in Figure \ref{sub_all}, $V_1$ is mapped to $\bfg_6$ and $\bfg_3$, but  $V_2$ is mapped to $\bfg_1$, a subtree of both $\bfg_6$ and $\bfg_3$; hence $V_{1}$ is only mapped to the green part of $\bfg_{6}$ and $\bfg_{3}$ as depicted in the Figure.

The precise mapping of nodes in  $\bfT$ to subtrees of $\bfg$ described below is needed to allocate mutations in \bfT{} to branches of $\bfg$. We will explain the allocation of mutations on $\bfg$ for a given $\bfa{}$ in the next subsection. 

We now define an algorithm to efficiently find all possible mappings \bfa{} for a given $\bfg{}$. We encode the set of all possible \bfa{}, as an $\#\bfa \times (n-1)$ matrix $\mathbf{A}$, where each row is a possible $\bfa{}$ ($n-1$ columns) and the number of rows $\#\bfa$ is equal to the number of possible allocations. To generate $\mathbf{A}$, the algorithm proceeds recursively from top to bottom in \bfg{}, by sweeping through subtrees in \bfg{} and matching them to nodes in $\mathbf{T}$ according to parent-offspring relationships and number of descendants in both \bfT{} and \bfg{}. To be more precise, the algorithm is initialized by setting the $1 \times (n-1)$ $\mathbf{A}$ matrix to $\mathbf{A}=(V_0,\ldots,V_0)$, \textit{i.e.}, $V_0$ is mapped to all subtrees in $\bfg{}$. The algorithm proceeds iteratively, adding and removing rows from $A$, iterating over an index $i$ 
going from $n-2$ to $1$.
The first step is to define $A(i)$, the set of node allocations in the $i$-th column of $\mathbf{A}$. Then for all $V \in A(i)$, the algorithm iterates through the following steps: define $T_{V}$ as the set of child nodes of $V$ that have $|\bfg_{i}|$ descendants. If the number of child nodes of $V$ is at least $3$, $V$ is also included in $T_{V}$. 
If $T_V=\emptyset$, for example if $V$ is a leaf node, the algorithm does nothing. If $|T_{V}|=1$, the algorithm replaces $V$ by the element of $T_{V}$ in the columns $I$ of $\textbf{A}$ corresponding to all subtrees of $\bfg_{i}$. 
If $|T_{V}|>1$,  the matrix $\mathbf{A}$ is augmented by stacking $|T_{V}|-1$ copies of $\mathbf{A}_{V}(,I)$, the submatrix of $\mathbf{A}$ obtained by extracting all the row vectors whose $I$-th elements are $V$.  The original submatrix $\mathbf{A}_{V}(,I)$ is referred to as $\mathbf{A}^{(1)}_{V}(,I)$, and $\mathbf{A}^{(2)}_{V}(,I),\ldots,\mathbf{A}^{(|T_{V}|)}_{V}(,I)$ denote its copies. Lastly, the algorithm replaces $V$ by the first element of $T_{V}$ in $\mathbf{A}^{(1)}_{V}(,I)$, by the second element of $T_{V}$ in $\mathbf{A}^{(2)}_{V}(,I)$ and so on, until the last element of $T_{V}$ is substituted in  $\mathbf{A}^{(|T_{V}|)}_{V}(,I)$.

The simple rule described above is fast to compute but it leads to incorrect allocations because nodes may be mapped a redundant number of times. For example, it is easy to see that implementing the algorithm above, we could define an allocation \bfa{} where node $V_2$ is allocated to all subtrees of size two; however, $V_2$ should be allocated at most once. This issue can be avoided by noting that internal nodes in \bfV{} should appear in each \bfa{} a number of times equal to their number of child nodes minus one, while leaf nodes, say $V' \in \bfV{}$, should appear $|V'|-1$ times. Hence, we complete each iteration by eliminating rows of $\mathbf{A}$ where this rule is violated. A second elimination rule is needed to account for the constraints imposed by the sampling time information: 
rows are eliminated when their assignments involve nodes labeled by a sampling time $s'$ ``matched'' to subtrees of \bfg{} that have leaf branches terminating at a different sampling time.  Algorithm~\ref{all} in the Supplementary material summarizes the above description.

\begin{figure}[t]
\centering\includegraphics[clip, scale=0.25]{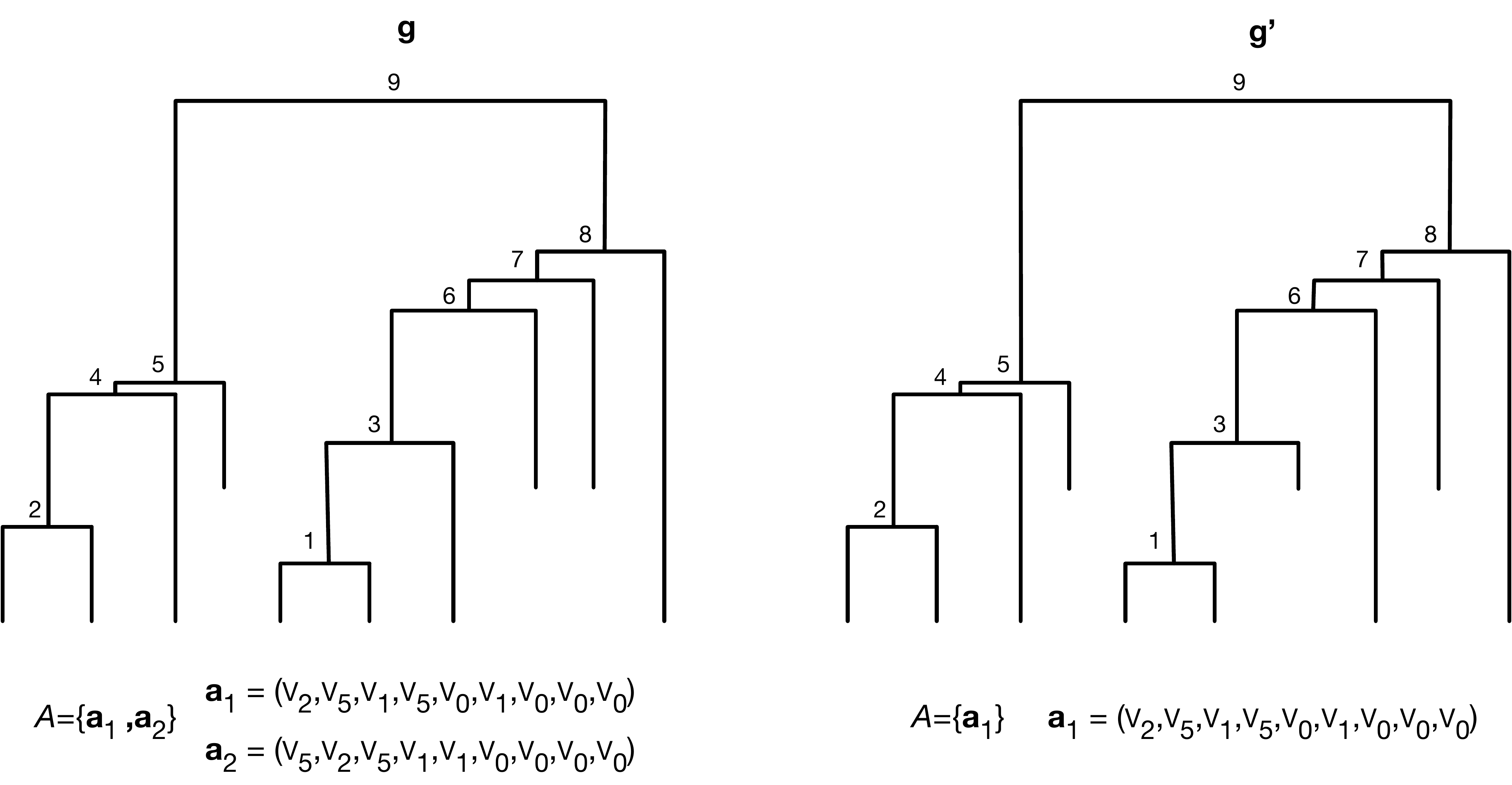}
	\caption{\small{\textbf{Example of allocations for two distinct genealogies}. Two possible samples from the Tajima heterochronous $n$-coalescent. Below we list all the possible allocations of nodes of \bfT{} to \bfg{} and \bfg{}'. The two genealogies differ solely by the inversion of the coalescent events $3$ and $6$. This change gives rise to differences in the possible allocations: for example, $V_5$ can be mapped to the subtree defined by node $3$ in \bfg{} but not in \bfg{}'.}}
	\label{allocations}
\end{figure}

Figure~\ref{allocations} gives examples of possible allocations of \bfT{} to two different genealogies \bfg{} and \bfg{}'. The second genealogy \bfg' differs from \bfg{} by the order of the coalescent events $3$ and $6$ which are inverted.  \bfg{} and \bfg{}' share the common allocation $\bfa_1=(V_2,V_5,V_1,V_5,V_0,V_1,V_0,V_0,V_0)$; however,  \bfg{} has a second possible allocation $\bfa_2=(V_5,V_2,V_5,V_1,V_1,V_0,V_0,V_0,V_0)$ that it is not compatible with $\bfg{}$'. This difference is due to the fact that $V_5$ has three descendants belonging to sampling group $s_1$, while \bfg{} has two subtrees with $3$ leaves sampled at $s_1$, and \bfg{}' has only one. We note that singleton nodes also need to be allocated, both in $\bfa_1$ and $\bfa_2$. We will elaborate on this point in the next subsection.

\subsubsection{Likelihood Calculations}
\label{fastlik}
To calculate the likelihood, we assume the
ISM of mutations and that mutations occur according to a Poisson point process with rate $\mu$ on the branches of \bfg{}, where $\mu$ is the total mutation rate. To compute the likelihood we need to
map mutations in \bfT{} to branches of \bfg{} and this is done for each mapping $\bfa_{i}$ of non-singleton nodes of \bfT{} to subtrees of \bfg{} . For every $V$ in \bfT{} such that $|V|>1$, we define
$\bfE_{V}$ as the set formed by the edges in $\bfT$ that subtend singleton children of $V$ and, with the exception of $V=V_{0}$, $\bfE_{V}$ in addition includes the edge that subtends $V$. For the example in Figure \ref{perfect_phylo}(B), $\bfE_{V_{1}}=\{E_{1},E_{3},E_{4}\}$. Let $\bfV^*$ be the set of all $V \in \bfV$ such that $|V|>1$,  
then the likelihood function is defined as
\begin{align}
\label{lik_2}
P(\textbf{Y}\mid \bfg, N_e,\mu)&=\sum_{i=1}^{\#\bfa} P(\textbf{Y}, \bfa{}_i \mid \bfg, N_e, \mu)\nonumber\\
&= \sum_{i=1}^{\#\bfa} \prod_{V \in \bfV^*} P(V, \bfE_{V}, \bfa{}_i  \mid \bfg,  N_e, \mu),
\end{align}
where we recall that $\#\bfa$ is the number of possible allocations, and 
 $P(V, \bfE_{V}, \bfa{}_i \mid  \bfg, N_e, \mu)$ is the probability of observing the mutations of the $\bfE_{V}$ edges along the corresponding branches of $\bfg{}$ defined by the mapping $\bfa_i$ as follows.

If $V$ has no singleton child nodes, then $\bfE_{V}=\{E\}$ and 
\begin{equation}
\label{intern_mut}
P(V, \{E\}, \bfa_{i} \mid  \bfg, N_e, \mu)  \propto (\mu l)^{|E|} e^{-\mu \mathcal{T}},
\end{equation}
where $l$ is the length of the branch in $\bfg$ that subtends $\bfg_{j}$, $j$ is the largest index such that $\bfa_{i,j}=V$, and $\mathcal{T}$ denotes the length of the subtree  in $\bfg$ to which $V$ is mapped in $\bfa_{i}$ (as described in Subsection \ref{sec:alloc}). For example, considering $V_2$ in Figure~\ref{sub_all}, we have $\mathcal{T}_2= 2 t_n +(t_{n-2}-t_n)$ and $l=(t_{n-2}-t_n)$ is the length of the branch connecting vintage $1$ to vintage $3$. 

If node $V$ has singleton child nodes, 
\begin{equation}
\label{leaf_all}
 P(V, \{E, E_{ch_1}, \ldots, E_{ch_k} \}, \bfa_{i} \mid  \bfg,N_e, \mu) \propto (\mu l)^{|E|} e^{-\mu \mathcal{T}} \sum_{\textbf{R} \in \Pi(\bfE_{V})} \prod_{j=1}^k (\mu l_{R_j})^{|E_{ch_j}|},
\end{equation}
where the first term on the r.h.s is defined exactly as the quantity on the r.h.s. of \eqref{intern_mut}, while the second term corresponds to the probability of all possible different matchings between $R_{1},\ldots,R_{k}$, the first $k$ indexes such that $\bfa_{i,R_{j}}=V$, and $|E_{ch_1}|,|E_{ch_2}|,\ldots,|E_{ch_k}|$, the $k$ numbers of mutations observed on the edges $E_{ch_1}, \ldots, E_{ch_k}$ leading to the child nodes of $V$.
In this expression, 
 $\Pi(\bfE_{V})$ is the set of all possible such matchings \textbf{R}.

Before defining $\Pi(\bfE_{V})$ more precisely, we make two observations. First, not all matchings are possible since not all leaf branches terminate at the same time (heterochronous sampling). Second, it is enough to consider the allocations that contribute to distinct likelihood values, \textit{i.e.} allocations for which the underlying samples are ``distinguishable" in the sense that they have a different number of mutations. 

We define $\Pi(\bfE_{V})$  as the set of all possible ``distinct matchings of number of observed singleton mutations to singleton branches", that is, allocations which lead to a distinct likelihood values. To construct $ \Pi(\bfE_{V})$, we first partition the singleton edges  $E_{ch_1}, \ldots, E_{ch_k}$ according to the sampling times of the corresponding nodes $V_{ch_1}, \ldots, V_{ch_k}$. Let $k_i$ be the number of nodes in $\{V_{ch_1}, \ldots, V_{ch_k}\}$ with sampling time $s_i$, \textit{i.e.}, the size of each subset of the partition. We then further partition these subsets by grouping together the edges carrying the same number of mutations (defined as $|E_{ch_1}|, \ldots, |E_{ch_k}|$). For each given sampling time $s_j$, let $k^{(1)}_j, \ldots, k^{(m_j)}_j$ denote the cardinalities of the $m_j$ sub-subsets obtained by this procedure, so that $k_j=\sum_{h=1}^{m_j} k^{(h)}_j$. The cardinality of $\Pi(\bfE_{V})$ is then
 \begin{equation}
 \label{perm}
 |\Pi(\bfE_{V})|=\prod_{j=1}^m \frac{k_j!}{k^{(1)}_j! \dots k^{(m_j)}_j!},
 \end{equation}
where the product in \eqref{perm} is the number of permutations with repetition of the different edges that are compatible with the data in terms of sampling times and numbers of mutations carried. Note that Equation~\eqref{perm} is not the same as Equation~(6)  in \cite{pal19} because here we account for the different sampling groups. It degenerates into Equation~(6) in \cite{pal19} in the isochronous case.

%

Lastly, we note that knowing a priori the full matrix $\mathbf{A}$ allows to compute efficiently the likelihood \eqref{lik_2} via a sum-product algorithm. Indeed, for each $V\in \bfV^*$ there may be several rows $\bfa$ of  $\mathbf{A}$ such that $P(V, \bfE_{V}, \bfa{}  \mid \bfg,  N_e, \mu)$ is the same, due to the fact that $V$ is mapped to the same subtree in all these allocations. For such a $V$, one could compute the likelihood corresponding to these $r$ allocations $\bfa'_1,\ldots,\bfa'_r$ in the following way:
\begin{align}
\label{eq:sumprod}
\sum_{i=1}^{r} & \prod_{V \in \bfV^*} P(V, \bfE_{V}, \bfa'_i  \mid \bfg,  N_e, \mu)\nonumber \\
& =P(V, \bfE_{V}, \bfa'_1  \mid \bfg,  N_e, \mu)\sum_{i=1}^{r} \prod_{V' \in \bfV^*\setminus \{V\}} P(V', \bfE_{V'}, \bfa'_i  \mid \bfg,  N_e, \mu).
\end{align}
The exact sum-product formulation of \eqref{lik_2} is specific to the observed \bfY{} and $\mathbf{A}$. 

\section{Bayesian Model and MCMC inference}
\label{mcmc}
In Section~\ref{sec:hettaj} we have introduced a probability distribution on genealogies, and in Section~\ref{lik} we have expounded how to compute the likelihood of heterochronous data \bfY{} generated by a Poisson process of mutations superimposed on this genealogy. We finally need to specify a prior distribution on $\log N_e$ (the logarithm is used to ensure that $N_e(t)> 0$ for $t>0$ to complete our Bayesian model. In this paper, we follow \citep{pal13} and place a Gaussian process (GP) prior on $(\log(N_e(t)))_{t\geq 0}$ (the logarithm is used to ensure that $N_e(t)\geq 0$ for $t>0$). We thus have:
\begin{align}
\label{model}
\bfY{}\mid \bfg{},\mu, N_e,\bfn{},\bfs{} & \sim \text{Poisson process} \nonumber \\
 \bfg \mid N_e,\bfs{}, \bfn{} & \sim \text{Tajima heterochronous $n$-coalescent}\\
\log N_e \mid \tau&  \sim \text{GP} (0, C(\tau)) \nonumber\\
\tau &\sim \Gam(\alpha, \beta) \nonumber
\end{align}
where $C(\tau)$ is the covariance function of the Gaussian process. As in \cite{pal13}, for computational convenience we use Brownian motion with covariance elements\\
$
\mathrm{Cov} (\log(N_e(t)), \log(N_e(t'))= \tau \min (t, t')
$
for any $t,t' >0$ as our GP prior.  From \eqref{model}, the posterior distribution can be written as
\begin{equation}
\label{posterior}
\hspace{-1cm}\pi(\log N_e, \tau, \bfg{}| \bfY{}, \mu) \propto P (\bfY | \bfg{}, \log N_e, \mu) \pi(\bfg{}|\log N_e) \pi(\log N_e|\tau) \pi (\tau),
\end{equation}
which we approximate via MCMC methods. Full conditionals are not available, and so we use Metropolis-within-Gibbs updates. At each MCMC iteration, we jointly update  $(\log N_e, \tau)$ via a Split Hamiltonian Monte Carlo (HMC) \citep{shah14} suitably adapted to phylodynamics inference by \cite{lan15}; then we update the topology $g$ and $\bft{}$. We propose two Metropolis steps to update $g$ and $\bft{}$. The latter may also be combined in a single step. The transitions for $g$ and \bft{} are tailored to the Tajima $n$-coalescent genealogies. To update $g$, we employ the scheme in \cite{pal19}, with two local proposals that either swap two consecutive coalescent events or swap two  offspring, each descending from two different and consecutive coalescent events \citep[Figure 4]{pal19}. To update $\bft{}$, we propose a new sampler that accounts for the observed sampling times constraints, 
an issue specific to heterochronous samples under the ISM assumption, which we detail in the next subsection.

\subsection{Constraints imposed by the ISM hypothesis}
\label{sec:ISMconstr}

Under the ISM hypothesis, mutations partition the observed sequences into two sets: the sequences that carry the mutations and the sequences that do not. This recursive partitioning of the sequences is graphically represented by \bfT{}. As a consequence, not all topologies $g$ and not all vectors \bft{} are compatible with the data, i.e. have posterior probability or density greater than 0. The combinatorial constraints imposed by the ISM on the space of topologies are discussed in detail in \cite{cap19}.

The constraints on \bft{} solely arise in the heterochronous case. 
First note that the definition of the Tajima heterochronous $n$-coalescent implies that there can be at most $n_1-1$ coalescence events before $s_2$,  at most $n_1+n_2-1$ events before $s_3$, and so on. 
Moreover, if there are shared mutations between some (but not all) samples with different sampling times, the maximum number of coalescent events between the involved sampling times is further restricted. In the example of Figure~\ref{perfect_phylo}(A), there is a shared mutation $l_{3}$ between 3 samples with sampling time $s_{1}$ and a sample with sampling time $s_{2}>s_{1}$. Out of the 7 samples obtained at time $s_{1}$, the 3 samples that share the $l_3$ mutation could coalesce first some time between $s_{1}$ and $s_{2}$ (at most $2$ coalescent events among the 4 sequences descending from node $V_{1}$), but they need to coalesce with the sample at time $s_2$ in node $V_1$ before they coalesce with the other 4 samples collected at time $s_{1}$ (those can coalesce at most 3 times between $s_{1}$ and $s_{2}$). Therefore, there are at most five coalescent events before $s_{2}$.


To encode the constraints imposed by the sampling information, we define a vector \textbf{c} of length $m$, where the $i$th entry denotes the maximum number of coalescent events that can happen (strictly) before time $s_{i}$ for given \bfY{}, \bfs{} and \bfn{}; note that \textbf{c} is not the number of coalescent events in a given interval. Trivially $c_1$=0 because there are no samples. In the example of Figure \ref{perfect_phylo}(A), we have $\textbf{c}=(0,5)$. Note that $c_2$ is $5$ and not $n_1-1=6$. In the online supplementary material, we provide a greedy search algorithm to define \textbf{c}.

\subsection{Coalescent times updates}

Let $\Delta\mathbf{t}:=(t_{n}-t_{n+1}, \ldots, t_{2}-t_{3})$ be the vector of intercoalescence times, and $(\Delta t_i)_{i \in I}$ the subvector of  elements of $\Delta\mathbf{t}$ at positions $I \subseteq\{1,\ldots,n-1\}$. The proposal is generated in three steps. First, we uniformly sample the number of intercoalescent times proposal moves -- \textit{i.e.}, the cardinality of $I$, then we uniformly choose which times to modify -- \textit{i.e.},  we define $I$, and lastly, we sample the proposals $(\Delta t_i)'_{i \in I}$. The first two steps balance between fast exploration of the coalescent times state-space and a high acceptance probability 
-- few changes are expected to lead to higher acceptance rates while many changes are expected to lead to faster exploration of the state space. 
In our implementation, we limit the maximum possible number of intercoalescent times moves to a fixed number $Z \ll n-1$. 
Lastly, we sample new states $(\Delta t_i)'$, for $i \in I$ 
from a truncated normal with mean $\Delta t_i$ and standard deviation $ \sigma \Delta t_i$. The left tail is truncated by a parameter $lo_i$, and the right tail is left unbounded. Three reasons motivate this choice:  it has positive support, it can be centered and scaled around the current $\Delta t_i$ using a single parameter $\sigma$, and we can set the lower bound $lo_i$ to ensure that only compatible times $\bft{}'$ are proposed. 
To set the values of $lo_i$, we rely on $\textbf{c}$, the vector that specifies the maximum number of coalescent events possible before each sampling time. We note that the elements of $\textbf{c}$ can be used to index coalescent times. In particular, $t_{n-c_i}$ denotes the time of the $(c_i+1)$th coalescent event. For example in Figure \ref{fig:hetTaj}, $t_{n-c_1}=t_{10}$ is the first coalescent event, and $t_{n-c_2}=t_5$  is the sixth coalescent event. Given $\textbf{c}$, $lo_i$ is set to
\begin{equation}
\label{eq:lower_time}
lo_i= \max_{j=1,\ldots,m} \big\{0,\{[s_j-(t_{n-c_j}-\Delta t_i)] \mathbbm{1}(i \leq c_j+1)\}\big\},
\end{equation}
where 
$\mathbbm{1}(i \leq c_j+1)$ is an indicator function.
Equation~\eqref{eq:lower_time} ensures the proposal 
$t'_{n-c_j} \geq s_j$ for all $j$. Indeed, note that $t_{n-c_j}=\sum_{k=1}^{c_j+1}\Delta t_k$. Hence,
if $(t_{n-c_j}-\Delta t_i)-s_j>0$ for any given $j$ such that $i \leq c_j+1$, then the proposed value of $(\Delta t_i)'$ could be zero and still $t'_{n-c_j}$ would be a compatible time. In this case, we do not need to impose any restriction on the lower bound of the truncated normal. On the other hand, if the vector considered in \eqref{eq:lower_time} has one or more positive values, the proposed value $(\Delta t_i)'$ should be large enough to ensure that
for all sampling times $s_j$, there will never be more than $c_j$ coalescent events before $s_j$. In other words, we truncate the proposal distribution support to ensure the compatibility of \bft'. We discuss how to set $Z$ and $\sigma$ in Section~\ref{sim}.


The transition density of coalescent times is given by
\begin{equation}
\label{timeMH}
k(\bft{},\bft')= \frac{1}{Z} \binom{n-1}{|I|}^{-1} \prod_{i \in I}\text{Truncated} \, \Norm (\Delta t_i, \sigma \, \Delta t_i, lo_i, \infty),
\end{equation}
with $\text{Truncated} \, \Norm(\Delta t_i, \sigma \, \Delta t_i, lo_i, \infty)$ denoting a truncated normal density function with mean $\Delta t_i$, standard deviation $\sigma \Delta t_i$, lower bound $lo_i$ and upper bound $\infty$.

\subsection{Multiple loci, unknown mutation rate, and unknown ancestral state}

\label{sec:extensions}
\textbf{Multiple loci.} Thus far, we have assumed data observed at a single linked locus (without recombination). We now extend the methodology to multiple independent loci, assuming a constant mutation rate across loci. Let $\bfY^{(1)}, \ldots, \bfY^{(L)}$ be the observed data at $L$ independent loci. The independence assumption implies that there is an underlying genealogy at each locus denoted as $\bfg^{(1)}, \ldots, \bfg^{(L)}$, all resulting from a single population with population size parameter $N_e$. The posterior distribution is now:
\begin{align}
\label{eq:posterior_multi}
\hspace{-1cm}\pi(\log N_e, \tau, (\bfg{}^{(i)})_{1:L})| (\bfY{}^{(i)})_{1:L}), \mu) \propto& \prod_{i=1}^L \Big[P (\bfY{}^{(i)} | \bfg{}^{(i)}, \log N_e, \mu)  \pi(\bfg{}^{(i)}|\log N_e)\Big] \nonumber \\
 & \times \pi(\log N_e|\tau) \pi (\tau).
\end{align}
To sample from posterior \eqref{eq:posterior_multi} we employ the same MCMC described in the previous sections. Now, every iteration requires $L$ MH steps to update $(g{}^{(i)})_{1:L}$, $L$ steps to update $(\bft{}^{(i)})_{1:L}$, and one step to update $\log N_e$.

\textbf{Unknown mutation rate.} Observed samples at different time points provide information about the rate of mutation. Heterochronous data allow the joint estimation of the mutation rate and the effective sample size \citep{dru02}. That is, we can approximate the posterior $\pi(\log N_e, \tau, \bfg{},\mu| \bfY{})$,
 by placing a prior distribution on $\mu$. Common priors in this context include a Gamma distribution, or a uniform distribution with a given support \citep{dru15}. The Gibbs sampler targeting the posterior includes an additional full conditional $\pi(\mu| \bfY{},\log N_e, \tau, \bfg{})$, which we sample from using an additional Metropolis-within-Gibbs step. Note that despite the fact that we can sample from $\pi(\mu| \bfY{},\log N_e, \tau, \bft{})$ (\textit{i.e} marginalizing $g$ and we place a Gamma prior on $\mu$), this collapsed-step cannot be employed in the Gibbs sampler because it changes the stationary distribution \citep{van15}.

\textbf{Unknown ancestral state.} In the case of unknown ancestral state, one needs to sum over all possible compatible ancestral states that can explain the data \citep{grif95}.

\section{Simulations}
\label{sim}

\begin{figure}
\includegraphics[scale=0.5]{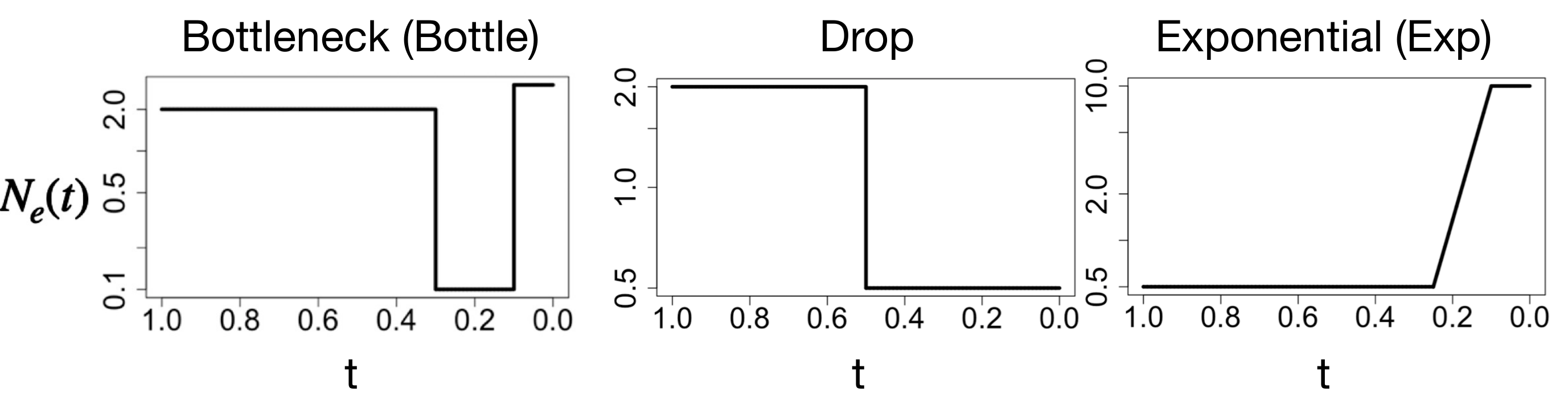}
\caption{\small{\textbf{Simulation: true effective population size trajectories considered (on a log scale).}}}
\label{fig:sim_details}
\end{figure}

We explore the ability of our procedure to reconstruct $N_e$ in simulation 
across a range of demographic scenarios which 
capture realistic and challenging population size trajectories 
encountered in applications. The code for simulations and inference is 
implemented in a \texttt{R }package. The validity of the algorithms' implementation is discussed in the supplementary material.

\textit{Simulation setup.} Given \bfn{}, \bfs{}, and $N_e$, we simulate genealogies under the Tajima heterochronous $n$-coalescent (Section~\ref{sec:hettaj}). Given a realized $\bfg{}$ and fixed $\mu$, we draw $M$ mutations from a Poisson distribution with parameter $\mu L$ ($L$ is the length of the tree \bfg{}: the sum of all branch lengths of $\bfg{}$) and place them independently and uniformly at random along the branches of the timed genealogy.  $\bfY_1$, $\bfY_2$ and \bfT{} are then constructed as described in Section~\ref{data}. We simulate genealogies with three population scenarios: a bottleneck  (``bottle''), an instantaneous drop (``drop''), and two periods of constant population size with exponential growth in between (``exp''). Figure \ref{fig:sim_details} sketches the trajectories used; details are given in the supplementary material. For each scenario, we generated genealogies with three numbers of leaves  ($n =14,\, 35,\, 70$) and different $\bfn, \bfs{}$ as summarized in Table~\ref{sim_summary}. The mutation parameter is varied to analyze the effect of the number of segregating sites on the quality of the estimation, but  in this section it is assumed to be known. Results for the joint estimation of $N_e$ and $\mu$ for a subset of the datasets analyzed are available in the supplementary material. In Table \ref{sim_summary}, we also provide estimates of the number of Tajima and Kingman trees with positive likelihoods for the corresponding simulated data set, respectively denoted $\widehat{|\mathcal{G}^T|}*$ and $\widehat{|\mathcal{G}^K|}*$. These estimates were obtained with a sequential importance sampling algorithm described in the supplementary material. Our estimates lacked numerical precision in the ``exp" scenario with $n=70$  and are not shown.

\begin{table} \centering
	\renewcommand{\arraystretch}{0.7} 
	\caption{\textbf{Summary of parameter values used in simulations.} List of parameters $\bfn, \bfs{}, \mu$ and demographic scenarios $N_e$ used to simulate data. We report the realized number $M$ of mutations for each of the 9 data sets ($\#$ mutations), and the estimated cardinalities of the spaces of Kingman ($\widehat{|\mathcal{G}^K|}*)$ and Tajima ($\widehat{|\mathcal{G}^T|}*$) trees that are compatible with the simulated data obtained through the sequential importance algorithms described in the supplementary material (N/A is used when the estimates obtained lacked numerical precision)}
	\label{sim_summary}

			\scalebox{0.7}{
	\begin{tabular}{@{\extracolsep{5pt}}  l  l |ccc}
&		& $n=14$ & $n=35$ & $n=70$  \\
		\hline \\[-1.8ex]
	\multirow{6}{*}{\rotatebox[origin=c]{90}{Bottleneck}}&	\bfn{} &	(5,5,4) & (10,10,10,5) &  (10,10,10,10,5,10,5,5,5)\\
	&	\bfs{} &	(0,.11,.32) & (0,0.045,0.11,0.32) & (0,0.045,0.075,0.11,0.2,0.25,0.31, 0.35,0.45) \\
	&	$\mu$ &	15 & 30 & 18 \\
	&	$\#$ mutations &	122 & 186 & 252  \\
	& $\widehat{|\mathcal{G}^K|}*$ & $3\times10^9 \pm 1.6\times10^7$&$1.1\times 10^{45} \pm 3.6\times 10^{42}$& $8.2\times 10^{88} \pm 4.3\times 10^{87}$\\
	& $\widehat{|\mathcal{G}^T|}*$  &$7\times 10^6 \pm 5.1\times 10^4$ & $6.9\times 10^{34} \pm 3.4\times 10^{33}$& $2.4\times 10^{80} \pm 1.6\times 10^{79}$\\
		\hline
		\multirow{6}{*}{\rotatebox[origin=c]{90}{Drop}}&	\bfn{} &	(8,3,3) & (10,10,10,5) &  (15,10,10,15,10,5,5)\\
	&	\bfs{} &	(0,0.4,0.6) & (0,0.2,0.4,0.6) & (0,0.1,0.2,0.4,0.47,0.6,0.8) \\
	&	$\mu$ &	12 & 12 & 12  \\
	&	$\#$ mutations &	121 & 127 & 190  \\
		& $\widehat{|\mathcal{G}^K|}*$  & $3.1\times10^5 \pm 3.2\times10^3$&$2.5\times 10^{31} \pm 4\times 10^{29}$& $2.5\times 10^{89} \pm 1\times 10^{88}$\\
	& $\widehat{|\mathcal{G}^T|}*$  &$4.2\times 10^4 \pm 4.1\times 10^2$ & $2.2\times 10^{26} \pm 4.8\times 10^{24}$& $2.8\times 10^{81} \pm 2.1\times 10^{80}$\\
		\hline
	\multirow{6}{*}{\rotatebox[origin=c]{90}{Exp}}&		\bfn{} &	(14) & (20,5,5,5) &  (20,15,10,10,10,5)\\
	&	\bfs{} &	(0) & (0,0.11,0.16,0.255) & (0,0.05,0.07,0.11,0.21,0.26) \\
	&	$\mu$ &	15 & 22 & 22 \\
	&	$\#$ mutations &	66 & 174 & 254  \\
			& $\widehat{|\mathcal{G}^K|}*$  & $1.6\times10^7 \pm 1.3\times10^5$&$2.1\times 10^{29} \pm 1.2\times 10^{28}$& N/A\\
	& $\widehat{|\mathcal{G}^T|}*$  &$9.8\times 10^5 \pm 8.1\times 10^3$ & $6.2\times 10^{28} \pm 3.1\times 10^{27}$& N/A\\
		\hline
	\end{tabular}}
\end{table}
We empirically assess the accuracy of our estimates with three commonly used criteria. The first one is the sum of relative errors (SRE), 
$SRE=\sum^{k}_{i=1}\frac{|\widehat{N}_e(v_{i})-{N_e(v_{i})}|}{{N_e(v_{i})}},$
where $(v_1, \ldots, v_k)$ is a regular grid of $k$ time points, $\widehat{N}_e(v_{i})$ is the posterior median of $N_e$  at time $v_{i}$ 
and $N_e(v_i)$ is the value of the true trajectory at time $v_i$. The second criterion is the mean relative width, defined by
$MRW=\frac 1 k \sum^{k}_{i=1}\frac{|\hat{N}_{97.5}(v_{i})-\hat{N}_{2.5}(v_{i})|}{N(v_{i})},$
where $\hat{N}_{97.5}(v_{i})$ and $\hat{N}_{2.5}(v_{i})$ are respectively the $97.5\%$  and $2.5\%$ quantiles of the posterior distribution of $N(v_{i})$. Lastly, we consider the envelope measure defined by
$ENV= \frac 1 k \sum^{k}_{i=1}\mathbf{1}_{\{\hat{N}_{2.5}(v_{i})\leq N_e(v_{i}) \leq  \hat{N}_{97.5}(v_{i})\}},$
which measures the proportion of the curve that is covered by the 95\% credible region. In this simulation study we fix $k=100$, $v_1=0$ and $v_k=.6 \, t_2$.


\textit{MCMC tuning parameters.} The posterior approximation is sensitive to the initial values of \bfg{}, $N_e$, and the MCMC parameters. We initialize \bfg{} with the serial UPGMA \citep{dru00}. In addition to the usual MCMC parameters such as chain length, burnin and thinning, there are three parameters specific to our method: the HMC step size $\epsilon$, the maximum number of intercoalescent times proposals ($Z$), and the standard deviation $\sigma$ that parametrizes the transition kernel $k(\bft{},\bft{'})$. While all three parameters contribute to the mixing of the Markov chain and acceptance rates, in our experience,  $\epsilon$ and $\sigma$ are the most influential. 
In settings similar to the ones analyzed here (time scale, type of trajectory patterns, and mutation rate), parameter values $\epsilon \in [0.03, 0.09]$, $Z \in \{1,2,3\}$, and $\sigma \in [0.01,0.03]$ lead to a similar mixing of the Markov chain and accuracy (w.r.t the metrics considered). We based these guidelines on extensive simulation studies on the nice datasets considered, which we believe to be representative of a broad set of settings encountered in applications. In our simulations, we set $\epsilon=0.07$, $Z=2$, and $\sigma=0.02$.


\textit{Comparison to other methods.} To our knowledge, there is no publicly available software implementing Bayesian nonparametric inference for  $N_e$ under the ISM and variable population size. To test the performance of our model, we implemented a function for computing the likelihood of Kingman's genealogies for labeled data. For posterior approximation via MCMC, we used the Markovian proposal on the space of ranked labeled topologies of \cite{mark00} (recall that the Tajima implementation uses the same proposal but on the space of ranked unlabeled topologies). The kernels used to update \bft{} and $\log (N_e)$ are shared between the two implementations. We generated two realizations of the Markov chains used in the MCMC scheme to approximate the posterior distributions under the Kingman and Tajima models: one 
under a fixed time budget ($72$ hours) and one under a fixed number of iterations (one million). For parsimony, the results of the latter study are given in the supplementary material. We use the mean effective sample size (ESS) of \bft{} and of $\log (N_e)$ as an empirical assessment of convergence (for implementation we use the \texttt{R} package \texttt{coda} \citep{coda}).

We also compare our results to an oracle estimator that infers $N_e$ from the true \bfg{}. The oracle estimation is obtained using the method of \cite{pal12}, which is equivalent to model \eqref{model} removing the randomness on $g$ and \bft{}.  A comparison with two other methodologies implemented in  BEAST \citep{drummond2012bayesian} is included in the supplementary material. We do not include the results in the main manuscript because these methodologies assume a different mutation model, a different prior on $N_e$, and a different MCMC scheme. The comparisons with BEAST should be mostly interpreted as validity checks of our implementations.

\textit{Results.} The difference in cardinality between the two spaces varies considerably across data sets. In the ``bottle" $n=70$ data set, the cardinality of Kingman trees is about $10^8$ times larger than that of Tajima trees, however in the ``exp" $n=35$ simulation, the ratio between the cardinalities is approximately $10$. Table \ref{tab:sim_time_eff} summarizes the mean ESS for the $9$ simulated data sets achieved with Tajima and Kingman. The high ESSs suggest convergence of the MCs. This is confirmed by the visual inspections of the trace plots (Supplementary material).

Tajima has the highest ESS for $\log N_e$ in $5$ out of $9$ instances, Kingman in $3$, and there is one tie. In $6$ out of $9$ instances, Tajima has the highest ESS for $\bft{}$, Kingman in $2$, and there is one tie. We interpret this result as evidence that the Tajima chain is more efficient. However, we invite caution: first of all, ESS is only a proxy for convergence; besides, the results obtained for a fixed number of iterations suggest a more even performance (Tables \ref{tab:sim_ite_eff} and \ref{tab:sim_ite} in supplementary material).

\begin{figure}[!ht]
	\begin{center}
		\includegraphics[scale=0.55]{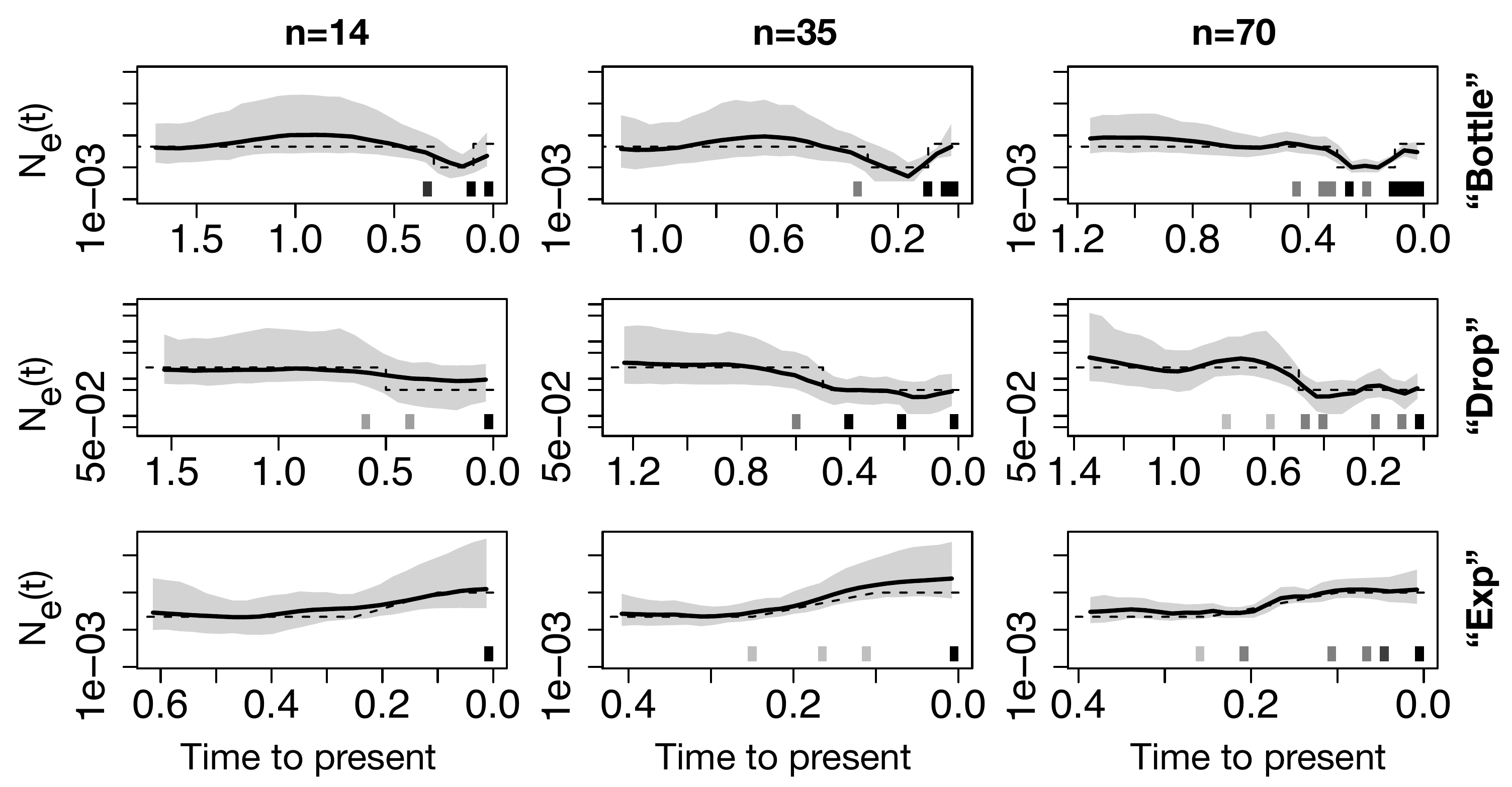}
	\end{center}
	\caption{\small{\textbf{Simulation: effective population size posterior medians from different trajectories and sample sizes for the Tajima-based model.}
$N_e$ posterior distribution from simulated data with three population size trajectories (rows) - bottleneck (``Bottle"), exponential growth (``Exp") and instantaneous fall (``Drop") -  different sample sizes (columns) - $n=14$, $n=35$ and $n=70$. Posterior medians are depicted as solid black lines and 95\% Bayesian credible intervals are depicted by shaded areas. \bfn{} and \bfs{} are depicted by the heat maps at the bottom of each panel: the squares along the time axis indicate the sampling time, while the intensity of the black color depicts the number of samples. More details are given in Table~\ref{sim_summary}.}}
	\label{fig:sim1}
\end{figure}

The results of the nine curves estimated with our method are plotted in Figure~\ref{fig:sim1}. The supplementary material includes the plots for Kingman (Figure \ref{fig:sim_kingman}) and the BEAST-based methodologies (Figure \ref{fig:sim2}). True trajectories are depicted as dashed lines, posterior medians as black lines, and $95\%$ credible regions as gray shaded areas. Note that the $y$ axis is logarithmic. Table \ref{tab:simeff_time} summarizes SRE, MRW, ENV, and the mean ESS for the $9$ simulated data sets achieved with Tajima, Kingman, and ``Oracle" for the fixed computational budget runs in all three scenarios. As $n$ increases, posterior medians track the true trajectories more closely. It is well known in the literature that abrupt population size changes are the most difficult to recover. The ``drop" and ``bottleneck" scenarios are less accurate for $n=14$, as exhibited by the wider credible region. We recover the bottleneck (panel first row and first column), but we do not recover the instantaneous drop (panel first row and third column).

\begin{SCtable}[2][t]
	
	\label{tab:sim_time_eff}
	\caption{\small{\textbf{Simulation: mean effective sample sizes of  \bft{}  (ESS \bft{}), and $\log N_e$ (ESS $\log N_e$) of  Tajima and Kingman for a fixed computational budget.}
			Mean ESS for three population trajectories (Bottle, Exp, Drop) and three sample sizes ($n=14,35,70$). Bold marks the method with the best performance (excluding the ``oracle") or within $10\%$ of the best performance. The MCMC was run for $72$ hours for both models.}}
	\scalebox{0.75}{
		\vspace{1cm}
		\renewcommand{\arraystretch}{0.7} 
		\begin{tabular}{@{\extracolsep{5pt}} cc|cc|cc}
			\hline \\[-1.8ex]
			& & \multicolumn{2}{c}{ESS \bft} & \multicolumn{2}{c}{ESS $\log N_e(t)$}\\
			Label & n & Tajima & Kingman & Tajima & Kingman \\ 
			\hline \\[-1.8ex] 
			\multirow{3}{*}{\rotatebox[origin=c]{90}{Bottle}} & 14 & \textbf{557.82} & 211.74 & \textbf{191.07} & 104.24 \\ 
			& 35 & 1237.53 & \textbf{2076.18} & 129.06 &\textbf{249.52} \\ 
			& 70 & \textbf{1717} & 1167 & \textbf{173.67} & 136.05 \\ 
			\multirow{3}{*}{\rotatebox[origin=c]{90}{Drop}} & 14 & \textbf{96.3} & 82.26 & 488.88 & \textbf{1027.6} \\ 
			& 35 & 599.98 & \textbf{2435.09} & \textbf{281.89} & \textbf{270.86} \\ 
			& 70 & \textbf{1450} & 1272 & \textbf{153.8} & 128.43 \\ 
			\multirow{3}{*}{\rotatebox[origin=c]{90}{Exp}}& 14 & \textbf{292.69} & \textbf{308.23} & \textbf{384.57} & 74.9 \\ 
			& 35 & \textbf{2775} & 2236.63 & \textbf{133.57} & 113.23 \\ 
			& 70 & \textbf{1406} & 667.76 & 68.5 & \textbf{80.61} \\ 
			\hline \\[-1.8ex]
	\end{tabular} }
	
\end{SCtable}

\begin{table}[!ht] \centering
		\renewcommand{\arraystretch}{0.7} 
	\caption{\small{\textbf{Simulation: performance comparison between Tajima, Kingman and Oracle models for a fixed computational budget.}
			Envelope (ENV), sum of relative errors (SRE), and mean relative width (MRW) for three population trajectories (Bottle, Exp, Drop) and three sample sizes ($n=14,35,70$). Tajima (our model), Kingman  (Kingman $n$-coalescent), Oracle \citep{pal12} (known \bfg). Bold depicts the method with the best performance (excluding the ``oracle") or within $10\%$ of the best performance. The MCMC was run for $72$ hours for both models. }}
					\label{tab:simeff_time}
	\vspace{0.25cm}
\scalebox{0.75}{
\begin{tabular}{@{\extracolsep{5pt}} cc|c|cc|c|cc|c|cc}
	\hline \\[-1.8ex]
	& &\multicolumn{3}{c}{$\%$ENV} & \multicolumn{3}{c}{SRE} & \multicolumn{3}{c}{MRW} \\
Label & n & Oracle & Tajima & Kingman & Oracle & Tajima & Kingman & Oracle & Tajima & Kingman  \\ 
\hline \\[-1.8ex] 
\multirow{3}{*}{\rotatebox[origin=c]{90}{Bottle}} & 14 & 100 & 100 & 100 & 408.11 & 175.66 & \textbf{123} & 20164.85 & 2298.28 & \textbf{6241.1}  \\ 
 & 35 & 99 & 96 & 96 & 155.81 & 148.33 & \textbf{78.81} & 203.52 & 1385.86 & \textbf{148.73}  \\
 & 70 & 98 & 88 & 82 & 121.34 & 124.55 & \textbf{98.84} & 23.33 & 22.8 & \textbf{17.12}  \\ 
\multirow{3}{*}{\rotatebox[origin=c]{90}{Drop}} & 14 & 100 & 100 & 100 & 28.78 & \textbf{36.47} & \textbf{38.21} & 10.54 & 8.8 & \textbf{6.24}  \\ 
 & 35 & 99 & 96 & 93 & 21.27 & \textbf{31.73} & 67.69 & 2.96 & \textbf{6.02}  & 24.78 \\ 
& 70 & 99 & 92 & 98 & 17.1 & \textbf{29.09} & 34.41 & 2.13 & \textbf{3.66}  & 4.86  \\ 
\multirow{3}{*}{\rotatebox[origin=c]{90}{Exp}}& 14 & 100 & 100 & 100 & 35.94 & \textbf{50.91} & \textbf{53.48} & 16.56 & \textbf{19.33} & 1163.38  \\ 
 & 35 & 100 & 100 & 100 & 35.58 & \textbf{112.5} & \textbf{114.42} & 11.41 & \textbf{116.97}  & 148.147 \\ 
 & 70 & 100 & 100 & 100 & 30.71 & 43.16 & \textbf{37.31} & 3.64 & 3.97 & \textbf{2.75}  \\ 
	\hline \\[-1.8ex]
\end{tabular} }

\end{table}

Table~\ref{tab:simeff_time} quantifies the analysis of Figure~\ref{fig:sim1}. First, no method unequivocally outperforms the others. The methods have identical performance for the ENV metric, according SRE and MRW metrics, our method has a superior performance  in the ``drop" and ``exp" scenarios, while Kingman-based inference is superior in the ``Bottle" scenario. We note that the Tajima methodology is the one that more closely tracks the ``Oracle" results (in eight out of nine cases, Tajima has the closest SRE and MRW to the Oracle). We consider this a positive feature given that  ``Oracle" posterior does not account for the uncertainty in \bfg{}.  Surprisingly, both Tajima and Kingman outperform the ``Oracle" methodology in certain examples.

\section{North American Bison data}
\label{app}



Recent advances in molecular and sequencing technologies allow recovering genetic material from ancient specimens \citep{paa04}. In this section, we analyze modern and ancient bison sequences. These mammals offer a case study of a population experiencing population growth followed by a decline. It was a long-standing question whether the drop was instigated by human intervention or by environmental changes. \cite{sha04} first reconstructed the genetic history of Beringian bisons. Their estimate for the start of the decline supports the environmental hypothesis. In particular, they suggest that the decline may be due to environmental events preceding the last glacial maxima (LGM). This data-set has been the subject of extensive research in the past decade.

We analyze new bison data recently described by \cite{fro17}. 
We fit our coalescent model to these sequences and estimate population size dynamics. To our knowledge, there is no phylodynamics analysis of this data set in the literature. Two motivations underlie this study: first, \cite{sha04} sequences include $602$ base pairs from the mitochondrial control region, while \cite{fro17} provide the full mitochondrial genome ($16322$ base pairs after alignment); second, we are interested in testing whether the previously published overwhelming evidence in favor of the environmentally induced population decline is confirmed by this new data. We analyzed $38$ sequences ($10$ modern, $28$ ancient). Details on the data set and on the implementation of our method and a BEAST-based alternative (GMRF \cite{min08}) are given in the supplementary material.

\begin{figure}[!t]
	\begin{center}
		\includegraphics[scale=0.55]{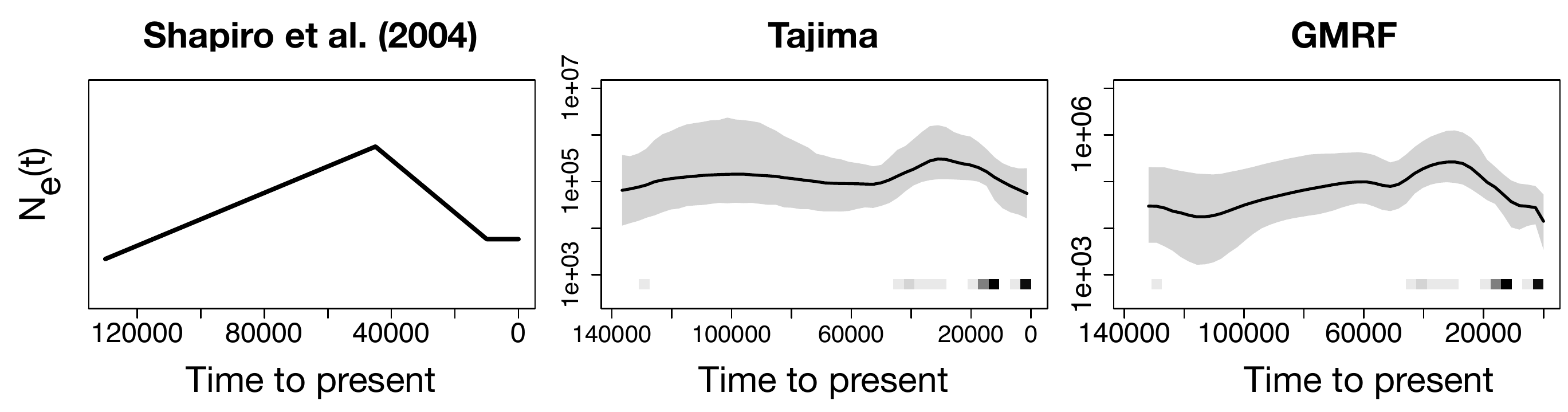}
	\end{center}
	\caption{\small{\textbf{Bison in North America: effective population size ``expected trajectory" and posterior median estimates from \cite{fro17} data set.}
			The first panel depicts a sketch of a ``consensus population trajectory" obtained from the phylodynamics study of the data of \cite{sha04} in \cite{fau20}. The second and third panels display estimated posterior medians of $N_e$ (as black curves) obtained from $n=38$ ancient and modern sequences from North America specimens in \cite{fro17} data. The second panel corresponds to our method, and the third panel to GMRF. The posterior medians are depicted as solid black curves, and the 95\% Bayesian credible regions are depicted by shaded areas. \bfn{} and \bfs{} are depicted by the heat maps at the bottom of the last two panels: the squares along the time axis depicts the sampling time, while the intensity of the black color depict the number of samples.}}
	\label{fig:bison}
\end{figure}

The first panel of Figure~\ref{fig:bison} plots a summary of the effective population size pattern recovered by a recent analysis of \cite{sha04} data by \cite{fau20}. While the precise timings and the trajectory details differ from method to method, the broad patterns are consistent. 
The population peak is estimated to be between 41.6 and 47.3 kya. The timing of the start of the decline is the main feature of interest. We plot the posterior medians (black lines) of $N_e$ along with the $95\%$ credible regions (gray area) obtained from posterior samples by sampling Tajima's trees (``Tajima", second panel) and  Kingman's trees (``GMRF", third panel).

Both our method and GMRF recover the pattern described in the first panel. We detect the population decline only up to about $60$kya ago. Afterward the median trajectory is relatively flat while the credible regions are wide. This can be explained by the fact that we have no samples from  $42$kya to $128.5$kya. On the other hand, GMRF detects more clearly the population decline. 
The GMRF median time estimate of the population peak is $29.6$~kya, while the median time estimate for our method is $29.7$ kya. 
Thus, the estimates of the main event of interest, the population decline, are practically identical. The difference between the estimates obtained analyzing $2017$ data differ substantially from the estimates of a population peak between 41.6 and 47.3 kya obtained analyzing the $2004$ data.

The LGM in the Northern hemisphere reached its peak between $26.5$ and $19$ kya \citep{cla09}. Hence, the analysis of the $2017$ data still supports the hypothesis of a decline initiated before the LGM. 
However, our estimates suggest an initial decline much closer to the LGM peak than the analysis of the $2004$ data. Human arrival in North America via the Berigian bridge route should have happened around $14-16$ kya \citep{lla16}. Therefore, despite the mismatch of the timing, the human-induced decline hypothesis has little evidence also according to our analysis of this new dataset. 

\section{SARS-CoV-2}
\label{covid}

\begin{figure}[!t]
	\begin{center}
		\includegraphics[scale=0.55]{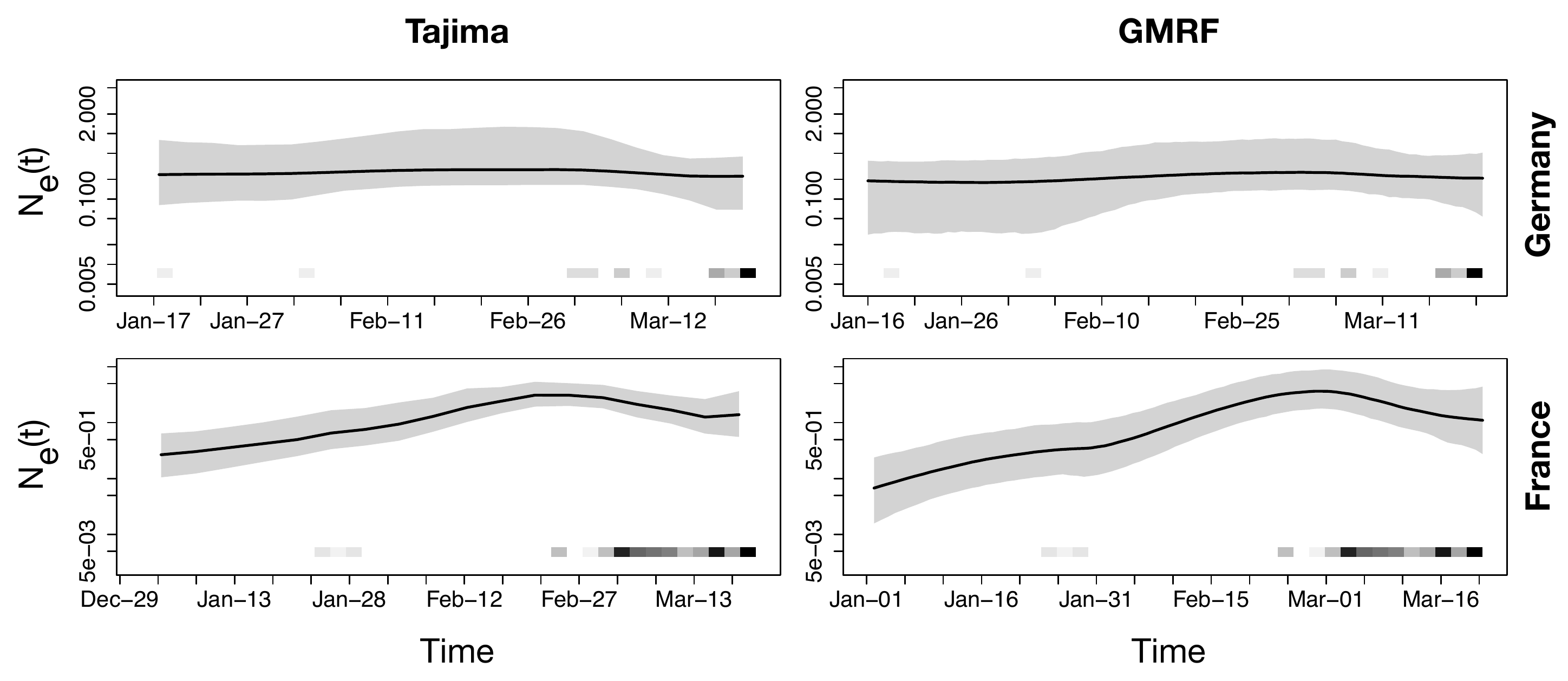}
	\end{center}
	\caption{\small{\textbf{2019-2020 SARS-CoV-2: $N_e$  posterior median estimates from SARS-CoV-2 GISAID data sets from France and Germany.}
			Black curves in the first row panels depict estimated posterior medians of $(N_e(t))_{t\geq 0}$ obtained from $n=32$ viral samples from Germany. Black curves in the second row panels depict estimated posterior medians of $N_e$ obtained from $n=123$ viral samples from France. The left column corresponds to our method's results, and the second column to GMRF results. The posterior medians are depicted as solid black lines, and the 95\% Bayesian credible regions are depicted by shaded areas. \bfn{} and \bfs{} are depicted by the heat maps at the bottom of the last two panels: the squares along the time axis depict the sampling time, while the intensity of the black color depicts the number of samples.}}
	\label{fig:covid}
\end{figure}

SARS-CoV-2 is the virus causing the pandemic of novel coronavirus disease in 2019-2020 and it is of interest to explore the utility of viral molecular sequences for surveillance during the outbreak of the epidemic. Here, we analyze $123$ whole genome sequences collected in France, and $32$ sequences collected in Germany that were made publicly available in the GISAID EpiCov database \citep{shu2017gisaid}. Details on the data sets and on the implementation of our method and the BEAST-based alternative (GMRF \cite{min08}) are given in the supplementary material.

We show the estimates of effective population size with our method in the first column of Figure~\ref{fig:covid} and with BEAST in the second column. Results for Germany correspond to the first row and for France to the second row. Both analyses of the French dataset exhibit exponential growth from mid-December of $2019$ to the end of February (Tajima estimate of median population peak is 2020/02/29, GMRF estimate is 2020/03/1). Following the exponential growth, both methods suggest a decline. Both analyses of the German dataset recover nearly constant trajectories, possibly due to sampling time concentration in mid-march and spatial sampling concentration in Duesseldorf (see online supplementary material for details).

A final remark. Our estimates should be interpreted as estimates of genetic diversity over time and not as number of infections. Our model ignores recombination, population structure, and selection. Viruses tend to exhibit antigenic drifts, selective sweeps, and cluster spatially following migration events \citep{ram08}. All these aspects may hinder using the models employed in this section to analyze large-scale viral population size dynamics.

\section{Discussion}
\label{concl}

We have studied an alternative to the Kingman $n$-coalescent to do Bayesian nonparametric inference of population size trajectory from heterochronous DNA sequences collected at non-recombining loci. The process, called Tajima heterochronous $n$-coalescent, allows for the analysis of serially sampled sequences. We proposed a fast algorithm to compute the likelihood function in this new model. Our research provides further evidence that using this lower-resolution coalescent process could help solve the scalability issues that prevent using the standard coalescent in the large datasets that are now being collected.

More research is needed to make this process an attractive alternative to the Kingman $n$-coalescent in scientific applications. The current methodology has some limiting assumptions, particularly the fact that it is based on the ISM mutation model. We deem moving away from the ISM a priority for future work. However, in practice this framework already covers an extensive range of possible applications. Another important future direction includes modeling recombination. Indeed, it is well understood that the information at a single locus saturates quickly as $n$ increases when the whole sample is taken at the same point in time. This is less so when dealing with longitudinal data, since the addition of elder samples restarts the genealogical process at different times and allows us to explore effective population size fluctuations deeper in the past. Furthermore, we have shown that our methodology could be applied to data collected at multiple independent loci. 

\bibliographystyle{agsm}
\begin{spacing}{1}
	\bibliography{biblio_postdoc}
\end{spacing}

%
%
%
%
%

\newpage

\begin{center}
	{\large\bf SUPPLEMENTARY MATERIAL}
\end{center}

\textbf{Examples of Likelihood under Kingman vs. Likelihood under Tajima.}

\begin{figure}[!b]
	\centering
	\includegraphics[scale=0.55]{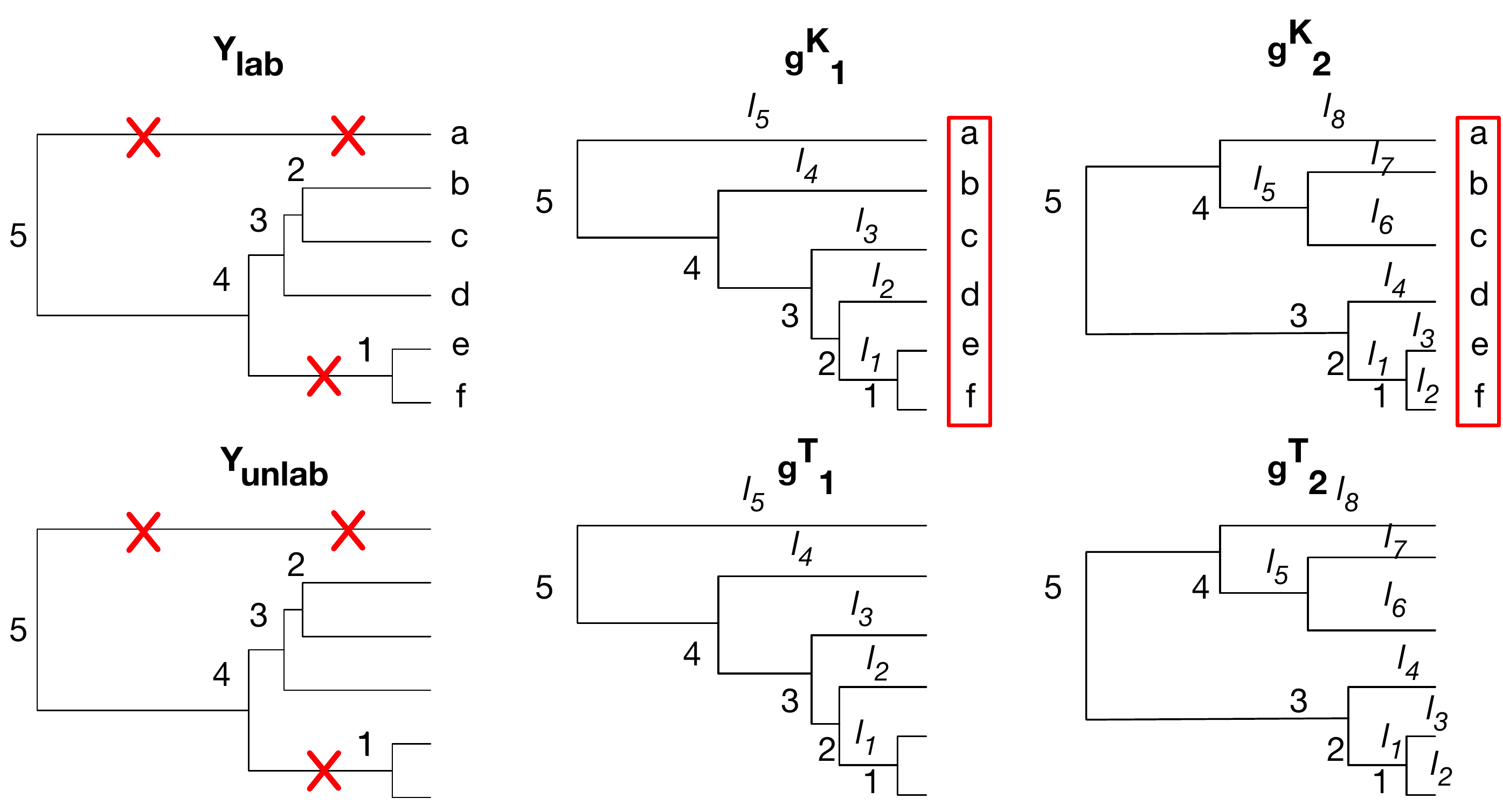}
	\caption{\small{\textbf{Dataset considered in the introduction of the manuscript (first column), two Kingman trees $g_1^K$ and $g_2^K$ compatible with the dataset, and two Tajima trees $g_1^T$ and $g_2^T$ compatible with the dataset.}  First column first row panel depicts a realization of a genealogy of $n=6$ samples (tips) with 3 mutations superimposed (marked as X). The corresponding labeled data $\bfY_{lab}$ would indicate that $a$ carries two mutations and the pair $e,f$ carries one mutation. The first column second row panel depicts a realization in which sequence labels are dropped. The corresponding unlabeled data $\bfY_{unlab}$ would indicate that one sequence carries two unique mutations and a pair of sequences carry the same mutation. The other four trees are used in this section to demonstrate the likelihood calculations. The red boxes in $g_1^K$ and $g_2^K$ are there to highlight the fact that we have to consider all possible label permutations (except within a cherry). The labels $l_i$ associated with the branches will be used in the likelihood calculations and denote the length of the corresponding branch.}}
	\label{fig:liksupp}
\end{figure}

This section has the following goals: $i)$ to provide an analytical expression for the likelihood for a fixed genealogy, $ii)$ elaborate on how the likelihood conditionally on a Tajima tree and a Kingman tree differ, $iii)$ show what entails to drop the sequence labels of the dataset, \textit{i.e} we clarify the difference between dealing with a labeled dataset $\bfY_{lab}$ and an unlabeled dataset $\bfY_{unlab}$, $iv)$ show that the marginal likelihoods $P(\bfY_{lab}|N_e,\mu)$ and  $P(\bfY_{unlab}|N_e,\mu)$ differ by a constant factor, \textit{i.e} there is no loss of information when estimating $N_e$ through $\bfY_{unlab}$. We continue the analysis of the example discussed in the Introduction of the manuscript and depicted in Figure \ref{fig:liksupp} (first column). We compute the likelihood conditionally on the genealogies depicted in Figure \ref{fig:liksupp}. 

The labeled dataset $\bfY_{lab}$, obtained from the realization depicted in the upper-left corner of Figure \ref{fig:liksupp}, carries the information that sequence $a$ has two unique mutations, and sequences $e$ and $f$ share one mutation. The unlabeled dataset $\bfY_{unlab}$, obtained from the realization depicted in the lower left corner of Figure \ref{fig:liksupp}, carries the information that there is one sequence with two unique mutations and two sequences sharing a unique mutation. This is a consequence of the ISM mutation model, which assumes that once a mutation occurs on a branch of the genealogy, individuals descending from that branch carry that mutation, and any new mutation occurs at a site that has not mutated before. 
For the $\bfY_{lab}$ dataset, any Kingman tree topology in which individual $e$ merges first with another individual who is not $f$ , for example $b$, would have null Kingman likelihood.

The genealogy $g_1^T$ in Figure \ref{fig:liksupp} (second column and second row) is the unlabeled ranked genealogy obtained by removing the leaf labels of $g_1^K$. All Kingman genealogies like $g_1^K$ obtained by all possible permutations of the leaf labels (marked in the red box) belong to the same equivalence class with unlabeled ranked tree shape $g_{1}^{T}$. We note that permuting the two labels of a cherry does not create a new Kingman genealogy. Similarly, all Kingman genealogies like $g_2^K$ obtained by all possible permutations of the leaf labels (again, excluding the permutations in the cherries) belong to the same equivalence class with the unlabeled ranked tree shape $g_{2}^{T}$.

We will start with calculating the conditional likelihood of the unlabeled data given $g^{T}_{1}$, the vector of coalescent times $\mathbf{t}$ and mutation rate $\mu$. Let us assume w.l.o.g. that $\mu=1$ and let $\mathcal{T}$ denote the tree length. Branch lengths are defined on the trees in Figure \ref{fig:liksupp} (we do not use coalescent times for compactness of the notation). The likelihood of $\bfY_{unlab}$ conditionally on $g_1^T$ is
\begin{equation*}
	\label{eq:lik1_un}
	P(\bfY_{unlab}|g_1^T,\bft{},\mu) =\exp(-\mathcal{T})\Bigg[l_1\Bigg(\frac{l_2^2}{2!}+\frac{l_3^2}{2!}+\frac{l_4^2}{2!}+\frac{l_5^2}{2!}\Bigg)\Bigg],
\end{equation*}
since we know from fixing the underlying genealogy that the shared mutation is carried by the internal branch with vintage $1$, while the two unique mutations are carried by an (unvintaged) external branch which may thus be either one among those of lengths $l_2$, $l_3$, $l_4$ or $l_5$ (and we therefore have to sum over all possibilities).
Now,
\begin{equation*}
	P(\bfY_{lab}|g_1^T,\bft{},\mu) =\sum_{g^K \sim g^K_1}P(\bfY_{lab}|g^K,\bft{},\mu)P(g^{K} \mid g_{1}^{T}),
\end{equation*}
where the sum is taken over all Kingman genealogies in the same equivalence class of $g^{K}_{1}$ and $P(g^{K}\mid g^{T}_{1})=\frac{2^c}{n!}$, and $c$ is the number of cherries. For the example of Figure \ref{fig:liksupp}, $P(g^{K}\mid g^{T}_{1})=\frac{2}{6!}$. 
Now the sum of the likelihoods in the equivalence class of $g_1^K$ is given by
\begin{equation}
	\label{eq:lik1_lab}
	\sum_{g^K \sim g^K_1}P(\bfY_{lab}|g^K,\bft{},\mu) =6 \exp(-\mathcal{T})\Bigg[l_1\Bigg(\frac{l_2^2}{2!}+\frac{l_3^2}{2!}+\frac{l_4^2}{2!}+\frac{l_5^2}{2!}\Bigg)\Bigg].
\end{equation}
To derive Eq.~\eqref{eq:lik1_lab}, note that to have a positive likelihood, the pair $e$-$f$ can only label the external branches of subtree $1$. Then, if label $a$ is assigned to the leaf subtending subtree 5, we can then assign the rest of the labels $b,c$, and $d$ in 6 different ways. In this case, the Poisson likelihood of the labeled data given that $e$ and $f$ subtend subtree 1 and $a$ subtends subtree 5 is: $6 \exp(-\mathcal{T})l_{1}\frac{l^{2}_{5}}{2!}$. Alternatively, $a$ can be assigned to the leaf subtending subtrees $2,3$ or $4$ as well. Considering all these possibilities, we obtain Eq.~\eqref{eq:lik1_lab}.


Now, we consider the other topology in Figure \ref{fig:liksupp} and compute the corresponding likelihoods of unlabeled and labeled data, respectively:

\begin{equation*}
	P(\bfY_{unlab}|g_2^T,\bft{},\mu) =\exp(-\mathcal{T})\Bigg[l_1\Bigg(\frac{l_4^2}{2!}+\frac{l_6^2}{2!}+\frac{l_7^2}{2!}+\frac{l_8^2}{2!}\Bigg)+l_5\Bigg(\frac{l_2^2}{2!}+\frac{l_3^2}{2!}+\frac{l_4^2}{2!}+\frac{l_8^2}{2!}\Bigg)\Bigg],
\end{equation*}
\begin{equation}
	\label{eq:lik2_lab}
	\sum_{g^K \sim g^K_2}P(\bfY_{lab}|g^K,\bft{},\mu) =3 \exp(-\mathcal{T})\Bigg[l_1\Bigg(\frac{l_4^2}{2!}+\frac{l_6^2}{2!}+\frac{l_7^2}{2!}+\frac{l_8^2}{2!}\Bigg)+l_5\Bigg(\frac{l_2^2}{2!}+\frac{l_3^2}{2!}+\frac{l_4^2}{2!}+\frac{l_8^2}{2!}\Bigg)\Bigg].
\end{equation}
Observe that the constant multiplying the Poisson likelihoods in the expression on the right-hand side of the above equation is 3. This is due to the fact that there are now two cherries in the tree topology, for which permuting the labels leads to the same Kingman topology, and so for a fixed labeling of the external branches $e$, $f$, and $a$, we have to consider only three possible permutations of $b$, $c$, and $d$. 

Note that likelihood \eqref{eq:lik1_lab} includes a factor $6$ while \eqref{eq:lik2_lab} includes a factor $3$. This difference is reconciled when computing the marginal likelihood/posterior distributions because under the Tajima $n$-coalescent prior $P(g^T_1)/P(g^T_2)$=2.

\newpage
\textbf{Algorithm for Augmented Perfect Phylogeny.} The algorithm below uses Gusfield's perfect phylogeny as an input, duplicates nodes corresponding to haplotypes that are sampled at more than one sampling time, and returns the augmented perfect phylogeny~\bfT{}.

\begin{algorithm*}[!h]
	\caption{Define \bfT{}}
	\label{alg:pp}
	\begin{algorithmic}
		\State \textbf{Inputs:} \bfT{}' \citep{gus91}, \bfs{}, $\bfY_2$
		\State \textbf{Output:} \bfT{}
		\begin{enumerate}
			\item \textbf{For} $i=1$ to $k$ \textbf{do}
			
			\textbf{If} $h_{i}$ is observed at multiple sampling times (from $\bfY_2$):
			
			[let w.l.o.g. $r$ be the number of sampling groups in which $h_i$ is observed, and $ s_{i_1},\dots, s_{i_r} $ the corresponding sampling times]
			\begin{enumerate}
				\item Take the leaf node $V'$ in \bfT' labeled by $h_{i}$ (each haplotype labels a unique node in Gusfield \bfT')
				
				\item 	\textbf{If} $|E'|=0$: make $r-1$ copies of $V$ ($r-1$ nodes with edges connecting them to the same parent of $V$ with no edge labels). Then label each of these nodes uniquely by a pair $(h_i,s_{i_1}),\dots,(h_i,s_{i_r})$
				
				\textbf{Else if} $|E'|\geq 1$: create $r$ new nodes with unlabeled edges connecting them to $V'$. Then label each of these nodes in a unique way with a pair $(h_i,s_{i_1}),\dots,(h_i,s_{i_r})$
			\end{enumerate}
			
			\textbf{Else if} $h_{i}$ is observed at a single sampling time (from $\bfY_2$):
			\begin{enumerate}
				\item Identify $V'$ in \bfT' labeled $h_{i}$
				\item Label $V'$ with a pair ($h_{i}$, its corresponding sampling time)
			\end{enumerate}
			
			\item Return \bfT{}.
		\end{enumerate}
	\end{algorithmic}
\end{algorithm*}

\newpage
\textbf{Algorithm for Allocation Matrix.} The algorithm below uses \bfT{} and \bfs{} as an input and return the allocation matrix $A$. 

\begin{algorithm}
	\caption{Description of the algorithm to define the allocation matrix}
	\label{all}
	\begin{algorithmic}
		\State \textbf{Inputs:} \bfT{}, \bfs{}
		\State \textbf{Output:} $A$
		\begin{enumerate}
			\item Initialize $A=(V_0,\ldots,V_0)$
			\item \textbf{For} $i=n-2$ to $1$ \textbf{do}
			\begin{enumerate}
				\item Define $A(i)$ unique nodes in the $i$th column of $A$
				\item \textbf{For all} $V \in A(i)$ \textbf{do}
				\begin{enumerate}
					\item Define $T_V$, set of (non-singleton) child nodes of $V$ having $|g_i|$ descendants
					\item Include $V$ in $T_V$ if it has more than two child nodes
					\item Define $I$, set of vintages corresponding to all subtrees of $g_i$
					
					\item 	\textbf{If} $|T_V|=0$: do nothing
					
					\textbf{Else if} $|T_V|=1$: set column $A_V(\cdot,I)$ equal to $T_V$
					
					\textbf{Else if} $|T_V|>1$: copy $A_V(,I)$ $|T_V|-1$ times, attach the copies to $A$ and set each copy equal to one element of $T_V$
					
					\item Eliminate rows in $A$ where $V$ appears too frequently (rule in the paper)
					\item Eliminate rows not compatible with \bfs{} and \bft{}
				\end{enumerate}
			\end{enumerate}
			\item Return $A$.
		\end{enumerate}
	\end{algorithmic}
\end{algorithm}

\textbf{Algorithm for computing \textbf{c}.} In Section \ref{sec:ISMconstr} we discussed some constraints that are imposed by the ISM on $g$ and $\bft{}$. Under the ISM, we say that a vector \bft{} is not compatible if, conditionally on it, it is impossible to construct a topology $g$ having positive likelihood. This notion of compatibility of \bft{} arises solely in the heterochronous case. It has to do with the number of coalescent events that can happen before each given sampling time.  Our goal was to propose an MC sampler that samples only compatible \bft{}. To do this, we introduced a vector \textbf{c}, whose $i$th  entry denotes the maximum number of coalescent events that can happen (strictly) before time $s_{i}$ for a given \bfY{} and under the ISM. Here we explain how to compute \textbf{c} by a greedy search. The idea is simple: it is impossible to build a compatible topology~$g$ conditionally on an incompatible vector~\bft{}.  We initially assume that the ISM does not impose any constraints on \bft{} and check if we can build a compatible topology. If we can, the ISM does not impose constraints. Otherwise, we need to add some constraints. We continue iteratively until we manage to sample a compatible $g$. To do this process, we consider one sampling group at a time starting from $s_1$. We define a vector \textbf{add} of length $m$ whose $i$th entry is the number of coalescent events that happens before $s_i$. Note that if we are interested in sampling $g$ (ignoring branch information), \textbf{add} is the only time information we need. We can sample compatible $g$'s through a simple extension of an Algorithm~2 in \cite{cap19} (also used in the next subsection). We refer to that paper for details.

In the example of Figure \ref{perfect_phylo}(A), the algorithm proceeds as follows. We initialize $\textbf{c}=(0,6)$. Then we start the ``for cycle" at $i=2$ and set $\textbf{add}=(0,6)$: this assumes that $6$ coalescent events happens before $s_2$, \textit{i.e.} after $6$ coalescent events we sample all the remaining samples. We try to build a compatible topology under this assumption and we fail (see explanation in Section \ref{sec:ISMconstr}). Then we set $\textbf{add}=(0,5)$ and we try to build a compatible $g$. Now we succeed, hence we set $c_2=add_2$. If we had additional sampling times, we would move to the next sampling time. In this example, we stop and keep $\textbf{c}$ as the output.

\begin{algorithm}[!h]
	\caption{Define \textbf{c}}
	\label{all:c}
	\begin{algorithmic}
		\State \textbf{Inputs:} \bfT{}, \bfs{}
		\State \textbf{Output:} \textbf{c}
		\begin{enumerate}
			\item Initialize $\textbf{c}=(0,n_1-1,n_1+n_2-1,\dots,\sum_{i=1}^{m-1}n_i-1,n-1)$
			\item \textbf{For} $i=2$ to $m$ \textbf{do}
			\begin{enumerate}
				\item Set $\textbf{add}=(0,\ldots, add_{i}=\sum_{i=1}^{i}n_i-1,\ldots,add_{m}=\sum_{i=1}^{i}n_i-1)$ 
				\item Given \textbf{add}, try to sample a compatible topology $g$
				\item 	\textbf{If} $g$ compatible: set $c_i=add_{i}$
				
				\textbf{Else if} $g$ not compatible: set $add_{i}=add_{i}-1,\ldots, add_{m}=add_m-1$ and return to (b)
			\end{enumerate}
			\item Return $\textbf{c}$.
		\end{enumerate}
	\end{algorithmic}
\end{algorithm}

\pagebreak

\pagebreak
\textbf{Counting the number of compatible tree topologies under the ISM with heterochronous data:}

\begin{figure}[!b]
	\includegraphics[width=1.0\textwidth]{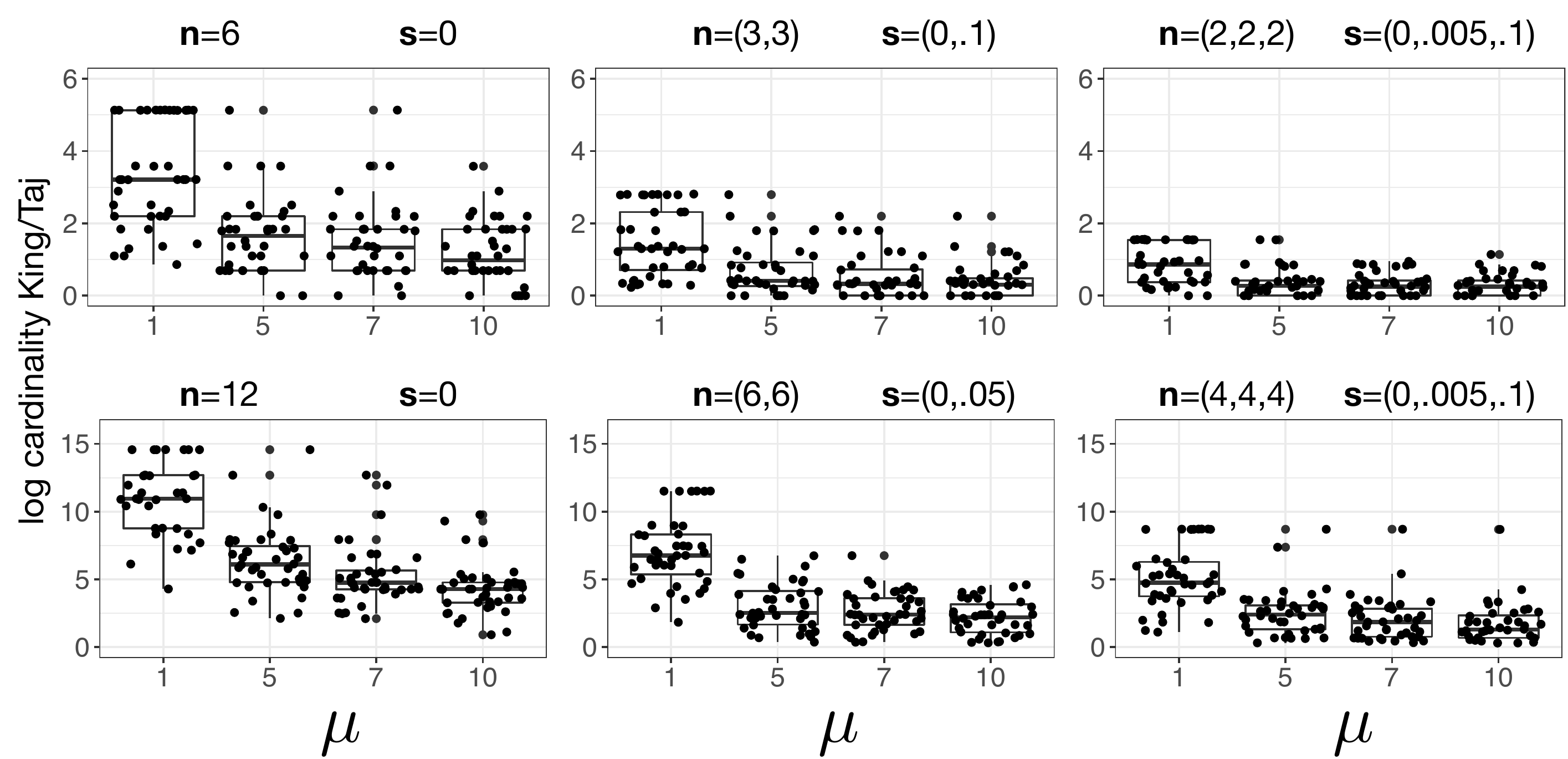}
	\caption{\small{\textbf{Multiresolution simulation study:} log ratio of estimated counts for varying $n$ and $\mu$.
			Rows correspond to the log ratio of cardinalities between Kingman and Tajima topologies for two distinct sample sizes. Columns represent different numbers of sampling groups $m$, and boxplots within each plot show results under different mutation rates. Boxplots are generated from $50$ independent simulations. Dots represent the SIS count estimates computed for $N= 5000$. Dots are spread over the box width for ease of visualization.}}
	\label{n_mu}
\end{figure}

We say that a genealogy is not compatible with the data if its likelihood is zero. 
Under the ISM, this happens when it is impossible 
to allocate the mutations on the genealogy and recover the observed dataset \bfY{}. \cite{cap19} provides sequential importance sampling (SIS) algorithms to estimate the cardinality of the space of coalescent trees (labeled and unlabeled) compatible with a given dataset. These algorithms allowed them to study how the cardinality of the space of compatible tree topologies varies as a function of $\mu, n$ and the coalescent process resolution (labeled or unlabeled) considered.

Roughly speaking, the difference between Tajima and Kingman state-space cardinalities decreases as the number of mutations increases \citep{cap19}. In Section $2$, we pointed out that if $m=n$, the Tajima heterochronous $n$-coalescent degenerates into the Kingman heterochronous $n$-coalescent.  A natural question is: how different are the cardinalities as a function of the total sample size, the sample size at each sampling time, and the number of sampling groups $m$? In this section, we do a simulation study and estimate the cardinalities of the spaces of Kingman and Tajima trees as a function of \bfn{} (and $n$), $m$, and $\mu$. 

Algorithms $1$ and $2$ in \cite{cap19} give a methodology to estimate the cardinalities when $m=1$. To extend these algorithms to the case $m>1$, we need to account for the fact that, conditionally on a given \bft{} and \bfs{}, not all samples may be available at a given time instance $t$, \textit{i.e.} if $t < s_i$, the sequences belonging to sampling group $i$ are not yet available for a coalescent event. 

Algorithms $1$ and $2$ in \cite{cap19} involve $n-1$ iterations. An iteration corresponds to one coalescent event. Each iteration includes a step designed to update the set of \textit{active nodes}, \textit{i.e.}, the nodes available for the next step; we refer to \cite{cap19} for more details. In the case of heterochronous data, we need to condition on \bft{} and \bfs{}: in each iteration, we add to the set of active nodes $(i)$ the nodes that are added in the original algorithms, $(ii)$ the nodes in \bfT{} that have been sampled after that coalescent event and before the next one. The rest of the algorithms proceeds exactly as in \cite{cap19}.

\textit{Data.}  We simulate $50$ incidence matrices for each of the following pairings: $[\bfn=6,\bfs=0]$, $[\bfn=(3,3),\bfs=(0,.1)]$, $[\bfn=(2,2,2),\bfs=(0,.05,.1)]$, $[\bfn=12,\bfs=0]$, $[\bfn=(6,6),\bfs=(0,.05)]$, $[\bfn=(4,4,4),\bfs=(0,.05,.1)]$, and $\mu$ in $(1,5,7,10)$, \textit{i.e.} we are considering two sample sizes and, for each sample size, three distinct sampling groups partitioning. For each simulated dataset, we estimate the cardinality of the two constrained topological spaces. Based on the results  in \cite{cap19},  we set the number of SIS samples to $N=5000$. 

\textit{Results.} Figure \ref{n_mu} shows the log ratio of the estimated cardinalities of Kingman topologies to Tajima topologies. The two rows differ for the sample size considered, while the columns differ for the number of sampling groups considered.  The figure shows that the cardinality of the space of Kingman trees is always larger than the cardinality of the space of Tajima trees. As in \cite{cap19}, we note that a high mutation rate will generally constrain the tree sample spaces more than a low mutation rate. Similarly, the more sampling groups we consider, the more negligible difference in cardinality between the two coalescent processes. However, we see that the difference increases as $n$ increases (see mean levels within a column). The case study suggests that modeling with lower resolution coalescent models still constitutes a relevant state space reduction when multiple samples are collected at a given time instance.

\vspace{1cm}
\textbf{Simulations details:}

\textit{Population size parameters:} We consider the following simulation scenarios:
We simulate genealogies with three population scenarios:

1. A bottleneck  (``bottleneck''):
\begin{equation}
	\label{bottle}
	N_e(t)=\begin{cases}
		3 & \qquad \hbox{if } t \in [0,0.1),\\[-0.2cm]
		0.1 & \qquad \hbox{if }t \in [0.1,0.3),\\[-0.2cm]
		2 & \qquad \hbox{if }t \in [0.3,\infty).
	\end{cases}
\end{equation}

2. An instantaneous drop (``drop''):
\begin{equation}
	\label{drop}
	N_e(t)=\begin{cases}
		0.5 & \qquad \hbox{if }t \in [0,0.5),\\[-0.2cm]
		2 & \qquad \hbox{if } t \in [0.5,\infty).
	\end{cases}
\end{equation}

3. Two periods of constant population size with an exponential growth in between (``exp''):
\begin{equation}
	\label{exp}
	N_e(t)=\begin{cases}
		10 & \qquad \hbox{if } t \in [0,0.1),\\[-0.2cm]
		10\,\exp(2-20\,t) & \qquad \hbox{if }t \in [0.1,0.25),\\[-0.2cm]
		0.5 & \qquad \hbox{if }t \in [0.25,\infty).
	\end{cases}
\end{equation}

\pagebreak
\textbf{Checking the validity of algorithms}

\begin{figure}
	\includegraphics[width=1.0\textwidth]{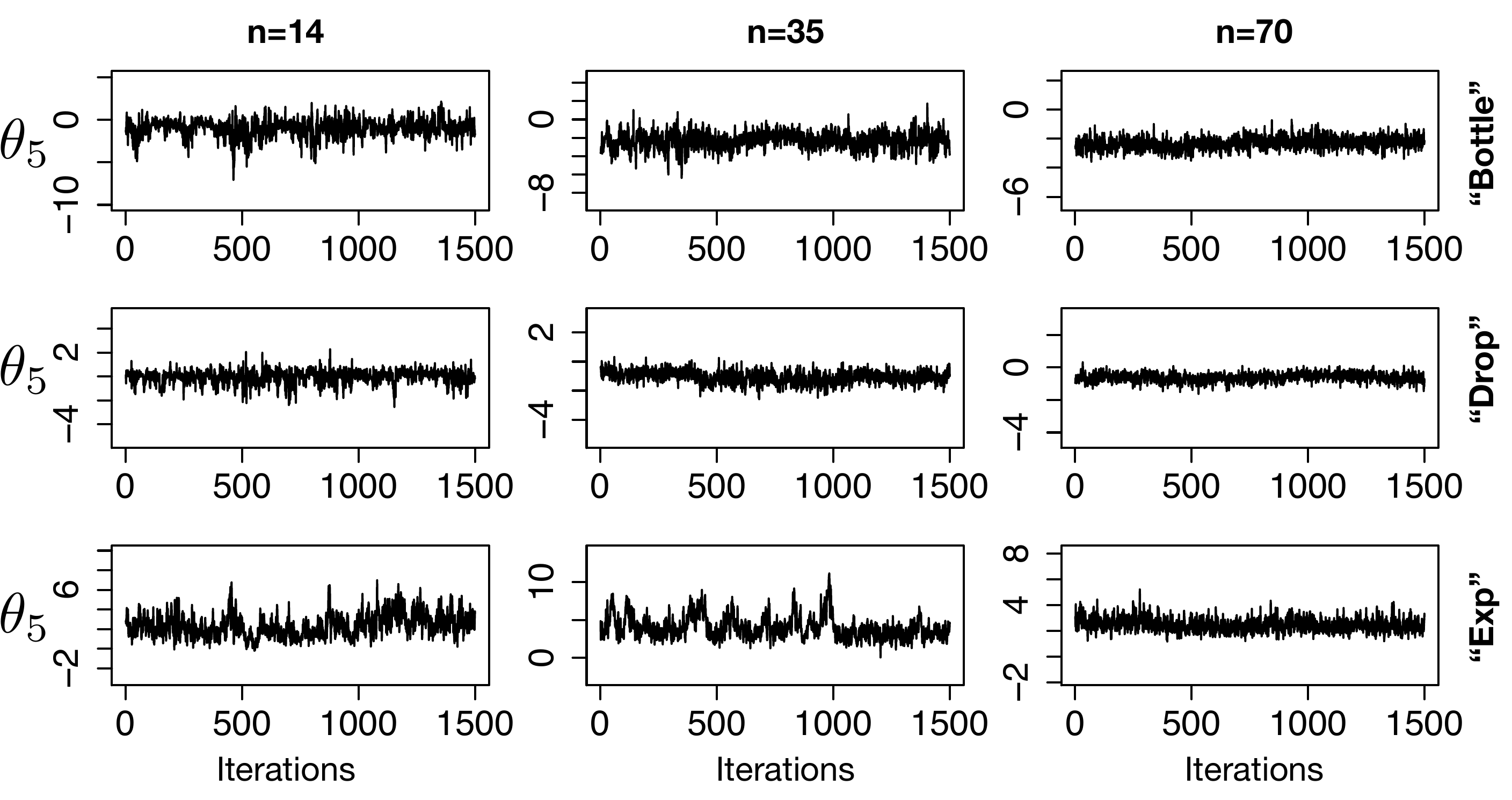}
	\caption{\small{\textbf{Trace plots of $\theta_{5}=\log N_e(t), t \in (x_{4},x_{5})$ from the Tajima model.}} The nine panels show the trace plot of $\log N_{e}(t)$ at the fifth time interval defined by the grid $x_{0},\ldots,x_{N}$ obtained from the nine datasets discussed in Section \ref{sim} of the manuscript. MCMC was run for $1 \times 10^6$ iterations. The plot refers to the last $1500$ samples (thinned every $500$ samples).} 
	\label{fig:trace_kintaj}
\end{figure}

\begin{figure}
	\includegraphics[width=1.0\textwidth]{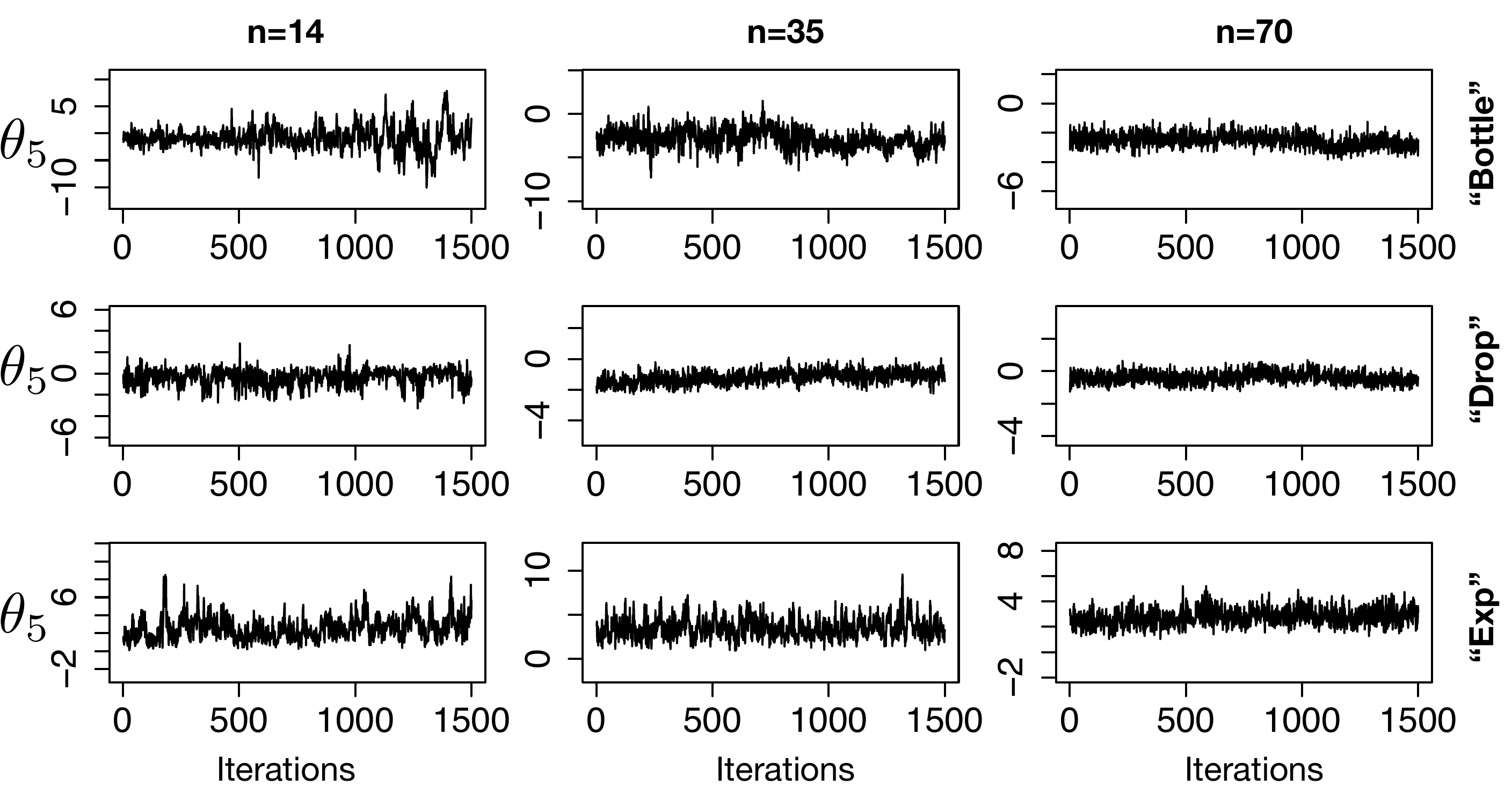}
	\caption{\small{\textbf{Trace plots of  $\theta_{5}=\log N_e(t), t \in (x_{4},x_{5})$ from the Kingman model.}  The nine panels show the trace plot of $\log N_{e}(t)$ at the fifth time interval defined by the grid $x_{0},\ldots,x_{N}$ obtained from the nine datasets discussed in Section \ref{sim} of the manuscript. MCMC was run for $1 \times 10^6$ iterations. The plot refers to the last $1500$ samples of the chain (thinned every $500$ samples). }}
	\label{fig:trace_king}
\end{figure}

\begin{figure}
	\includegraphics[width=1.0\textwidth]{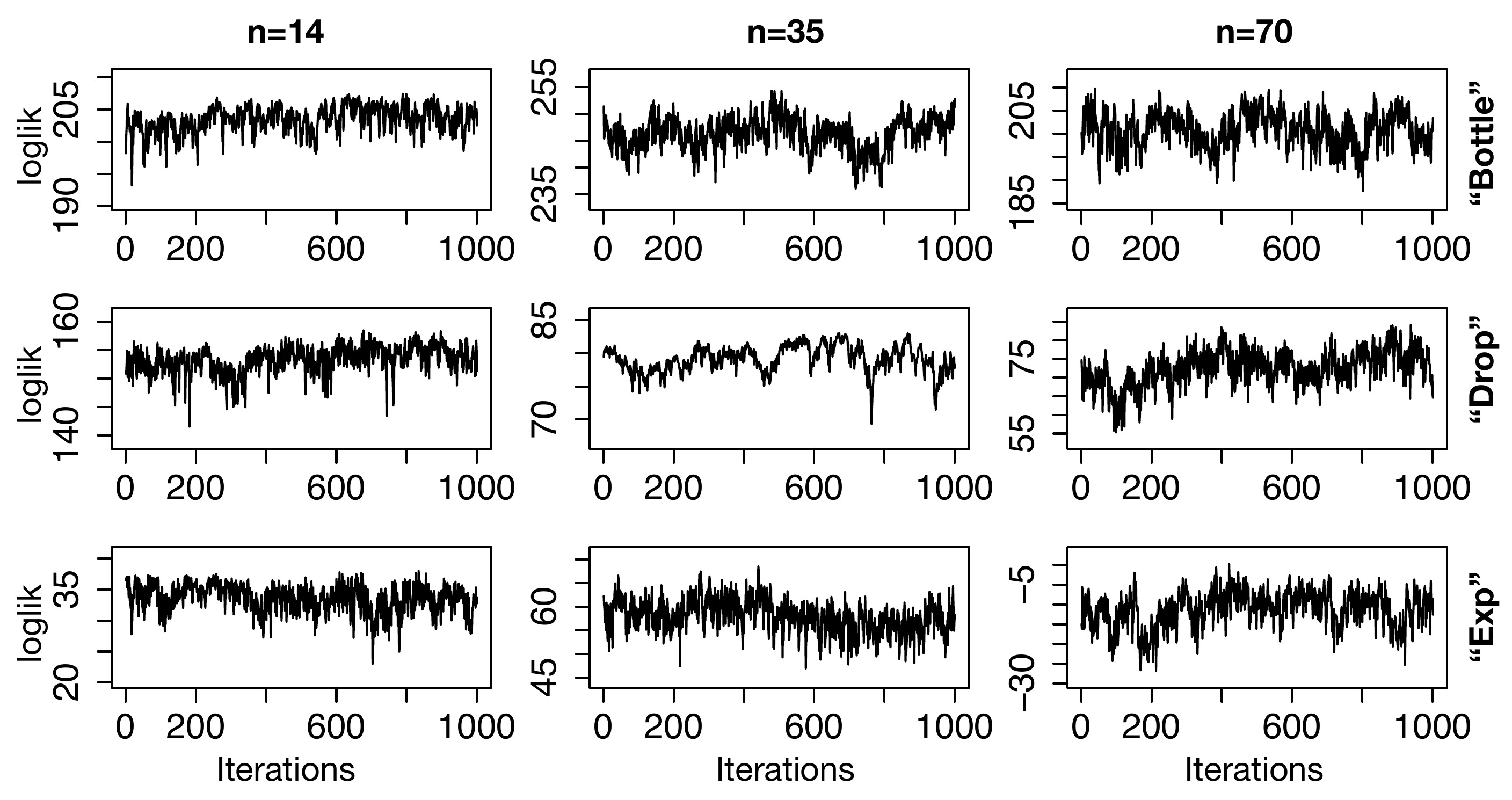}
	\caption{\small{\textbf{Trace plots of log $P (\bfY |  g^T, \bft{}, N_e, \mu) $ from the Tajima model.} The nine panels refers to the datasets discussed in Section \ref{sim} of the manuscript. MCMC was run for $1 \times 10^6$ iterations. The plot refers to the last $1000$ samples (thinned every $500$ samples).} }
	\label{fig:trace_taj_loglik}
\end{figure}

\begin{figure}
	\includegraphics[width=1.0\textwidth]{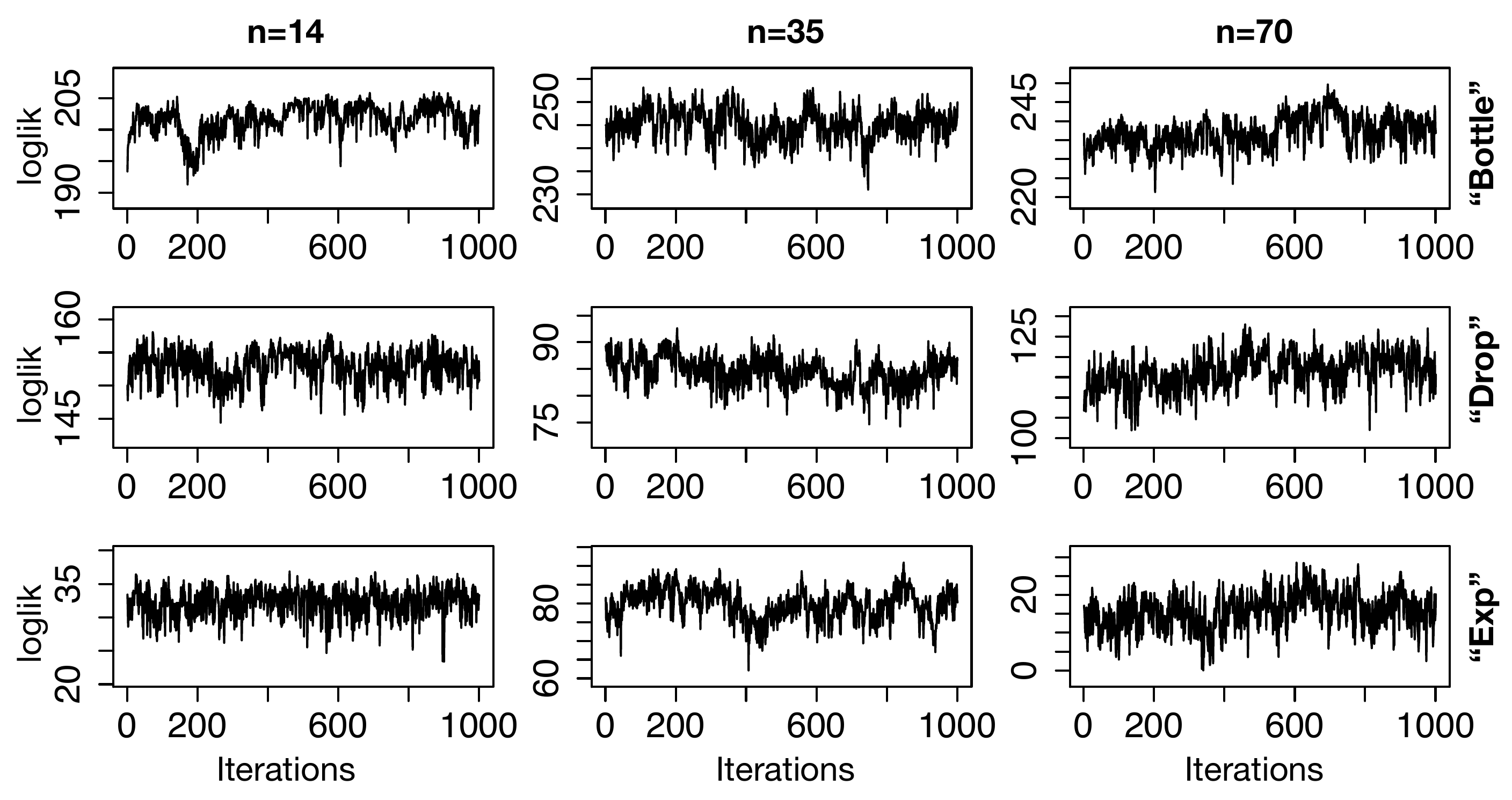}
	\caption{\small{\textbf{Trace plots of log $P (\bfY |  g^K, \bft{}, N_e, \mu) $ from the Kingman model.} The nine panels refers to the datasets discussed in Section \ref{sim} of the manuscript. MCMC was run for $1 \times 10^6$ iterations. The plot refers to the last $1000$ samples of the chain (thinned every $500$ samples).} }
	\label{fig:trace_king_loglik}
\end{figure}

We performed several checks of our model implementation in \texttt{R} and Python. Parts of our codes rely on existing work: the implementation of the coalescent density (Eq. \eqref{prior_time} in the manuscript) is publicly available in the \texttt{R} package \texttt{phylodyn} and its implementation was validated in \cite{pal13,kar17}; the splitHMC code to update $\log N_e$, along with the discretization of $\log N_e$, was proposed and validated in \cite{lan15}. The remaining parts to validate are the calculation of the data likelihood conditionally on both Tajima and Kingman topologies and the MCMC, which relies on two additional MC kernels (for \bft{} and $g$) in addition to the splitHMC. 

\textbf{Likelihood.} To validate the implementation of the likelihood, we 
fix a genealogy simulated under constant $N_{e}$ and the total number of mutations, and simulate millions of synthetic data sets, \textit{i.e.} we superimpose an a-priori-fixed number of mutations on a given genealogy. We then compute the frequency of each data set and compare these frequencies with the normalized likelihood with our code.
The two quantities should be identical.

We consider the following scenarios:  $a)$ $\bfn=(3,2)$ and $\bfs=(0,.2)$ , $b)$ $\bfn=(2,2,2,2)$ and $\bfs=(0,.15,.3)$, and $c)$ $\bfn=(2,2,2,2)$ and $\bfs=(0,.1,.2,.3,.4)$. For each scenario, we place a varying total number of mutations ($1,2,4,6$) uniformly at random along the genealogy and repeat each of the $12$ combinations four times to consider different genealogies for a given set up. We sample eight million data sets for each run. We employ the same validity check for the Tajima likelihood code and the Kingman likelihood code. In the former, we consider unlabeled data and topologies; in the latter, we have labeled data and topologies (labels assigned at random). We report the average ratio of the likelihood computed with our code to the empirical frequency. We exclude from the average those datasets that are observed less than $80$ times ($0.00001\%$ of the total number of samples). While this threshold is arbitrary, we excluded the topologies observed too few times because the empirical frequency estimators used to estimate the sampling probabilities of these rare events may not be as accurate as for large sample sizes. These resulted in excluding on average $0.008\%$ of the total observed distinct Tajima's topologies and $0.009\%$ of the total observed distinct Kingman's topologies. The mean ratio for the Tajima likelihood is $0.99$ with a mean variance of $0.0046$, the mean ratio for the Kingman likelihood is $1.016$ with a mean variance of $0.0135$. We interpret these results as reassuring about our implementation of the likelihood calculations.

\textbf{MCMC.} To validate the implementation of the MCMC, we employ two approaches: first, we check for convergence in simulations through commonly used criteria such as ESS and trace plots; second, we compare our results with state-of-the-art implementations of similar methods (see dedicated section in the Supplementary material ``Comparison of Tajima-based inference with state-of-the-art alternatives"). The trace plots and ESS used here refer to the examples considered in Section \ref{sim} of the manuscript. Figures \ref{fig:trace_kintaj} and Figure \ref{fig:trace_king} depict trace plots for 
$\log N_e(t)$ at one specific time interval $t\in (x_{4},x_{5})$. Trace plots indicate good mixing.  ESSs reported in Tables \ref{tab:sim_time_eff}  in the manuscript and Table \ref{tab:sim_ite_eff} in the supplementary material are all fairly high and confirm the evaluation of the trace plots. In few instances, the ESS is less than $100$, however they are all fairly high for this type of application. 


\newpage
\textbf{Comparison of Tajima-based inference with state-of-the-art alternatives}

\begin{figure}
	\begin{center}
		\includegraphics[scale=0.50]{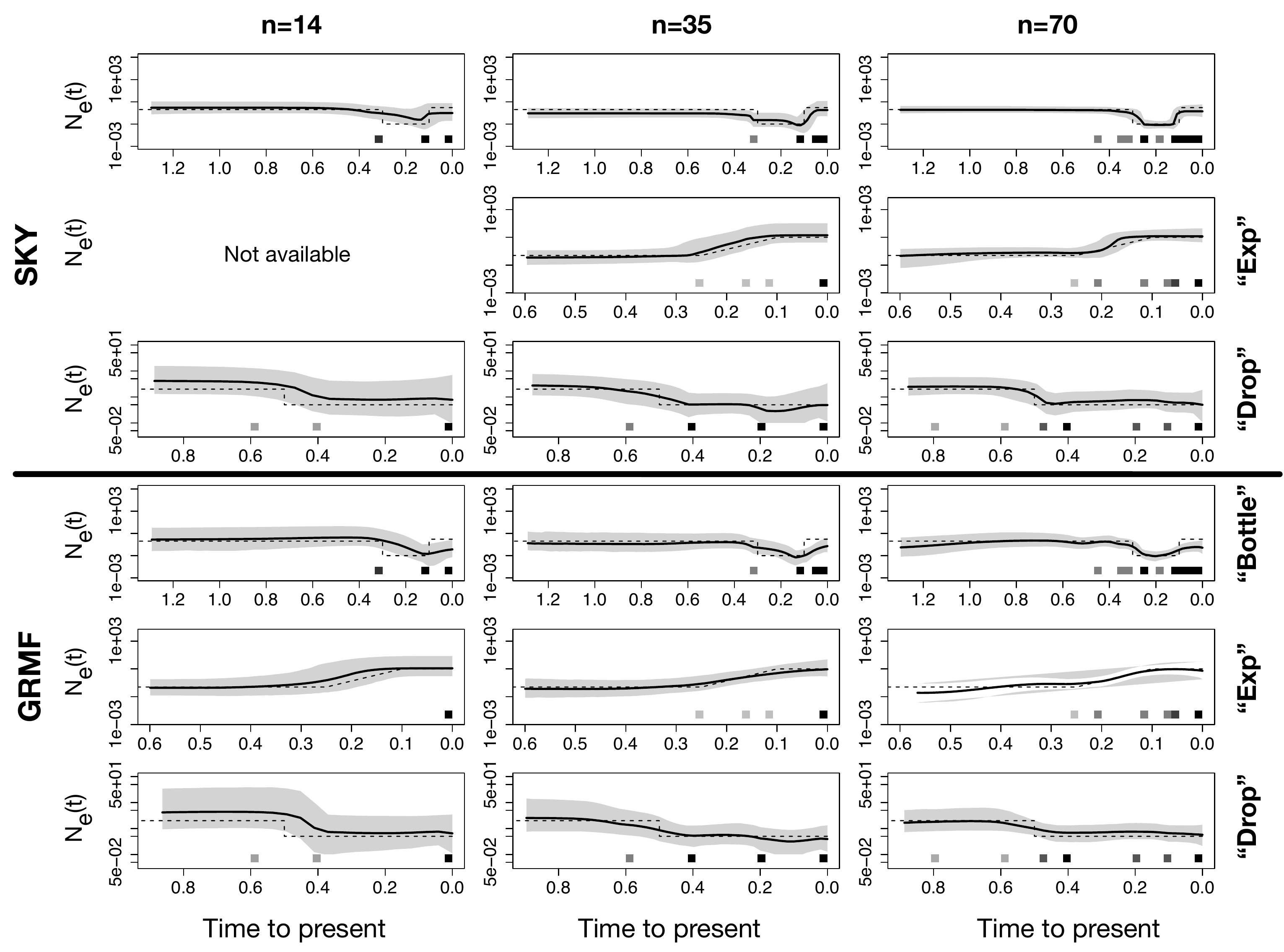}
	\end{center}
	\caption{\small{\textbf{Simulation: effective population size posterior medians from different trajectories and sample sizes.}
			$(N_e(t))_{t\geq 0}$ posterior distribution from simulated data of under three population size trajectories (rows) - bottleneck (``Bottle"), exponential growth (``Exp") and instantaneous fall (``Drop") -  different sample sizes (columns) - $n=14$, $n=35$ and $n=70$. Posterior medians are depicted as solid black lines and 95\% Bayesian credible intervals are depicted by shaded areas. \bfn{} and \bfs{} are depicted by the heat maps at the bottom of each panel: the squares along the time axis depicts the sampling time, while the intensity of the black color depicts the number of samples. Top three rows panels depict estimates obtained through SKY, bottom three rows depicts estimates obtained through GMRF. More details are given in Table 1.}}
	\label{fig:sim2}
\end{figure}

\begin{table}[h] \centering
	\renewcommand{\arraystretch}{0.7} 
	
	\caption{\small{\textbf{Simulation: performance comparison between Tajima and Oracle models.}
			We compute three statistics - envelope (ENV), sum of relative errors (SRE) and mean relative width (MRW) - for three population trajectories (Bottle, Exp, Drop) and three population sizes ($n=14,35,70$). Tajima refers to the estimation of $N_e$ through our model, SKY refers to \cite{dru05}, GMRF to \cite{min08}. Oracle refers to the method of \cite{pal12} (known \bfg). Bold depicts the method with the best performance (excluding the ``oracle"). SKY ``Exp" $n=14$ results are not included because we could not obtain convergent runs.}}
	\label{tab:sim_results_supp}
	\vspace{0.25cm}
	\scalebox{0.85}{
		\hspace{-2.5cm}
		\begin{tabular}{@{\extracolsep{5pt}} cc|c|ccc|c|ccc|c|ccc|}
			\hline \\[-1.8ex]
			& &\multicolumn{4}{c}{$\%$ENV} & \multicolumn{4}{c}{SRE} & \multicolumn{4}{c}{MRW} \\
			& n & Oracle & Tajima & SKY & GMRF & Oracle & Tajima & SKY & GMRF & Oracle & Tajima & SKY & GMRF \\
			\hline \\[-1.8ex]
			\multirow{3}{*}{\rotatebox[origin=c]{90}{Bottle}}  & 14 & 100 & \textbf{98} & 97 & 92 & 450.21 & 127.47 & \textbf{75.14} & 94.27 & 137943.26 & 6279.04 & \textbf{7.01} & 20.28 \\
			& 35 & 100 & \textbf{97} & 93 & 92 & 121.27 & \textbf{53.96} & 58.89 & 56.29 & 192.46 & 33.74 & \textbf{2.25} & 6.25 \\
			& 70 & 97 & 92 & \textbf{99} & 91 & 111.01 & 109.02 & \textbf{24.86} & 45.83 & 34.76 & 24.53 & \textbf{2.33} & 3.15 \\
			\multirow{3}{*}{\rotatebox[origin=c]{90}{Exp}} & 14 & 100 & \textbf{100} & -& \textbf{100} & 34.36 & 54.11 & - & \textbf{49.81} & 19.4 & 39.28 & - & \textbf{11.32}\\
			& 35 & 100 & 91 & \textbf{100} & \textbf{100} & 29.16 & 69.34 & 50.45 & \textbf{26.06} & 10.12 & 14.59 & 8.32 & \textbf{3.43} \\
			& 70 & 100 & \textbf{100} & \textbf{100} & \textbf{100} & 29.42 & 48.5 & 49.24 & \textbf{29.8} & 4.06 & 4.14 & 3.96 & \textbf{2.76} \\
			\multirow{3}{*}{\rotatebox[origin=c]{90}{Drop}}  & 14 & 100 & \textbf{100} & 98 & 99 & 25.65 & \textbf{32.58} & 105.93 & 85.55 & 19.38 & \textbf{5.47} & 8.06 & 14.21 \\
			& 35 & 99 & 97 & \textbf{100} & \textbf{100} & 21.53 & \textbf{17.26} & 30.9 & 21.87 & 3.22 & 3.39 & \textbf{3.32} & 4.41 \\ {}
			& 70 & 98 & 95 & \textbf{100} & 99 & 17.48 & 31.24 & \textbf{24.36} & 28.48 & 2.57 & \textbf{2.27} & 2.48 & 2.41 \\
			\hline \\[-1.8ex]
	\end{tabular} }
\end{table}

We compare our results to two popular methodologies implemented in BEAST \citep{drummond2012bayesian}: the Bayesian Skyline (SKY) \citep{dru05} and the Gaussian Markov Random Field Skyride (GMRF) \citep{min08}. For the SKY and GMRF, we assumed the Jukes-Cantor mutation model (JC) \citep{juk69} since the ISM model is not implemented in BEAST, and it is the closest mutation model to the ISM model among those implemented in BEAST. To compare our implementation to BEAST, we employ the same approach of \cite{pal19} in which we first convert the simulated incidence matrices (matrices of $0$s and $1$s assuming ISM) to sequences of nucleotides.  We first assign ancestral nucleotides uniformly at random among the four bases. This is equivalent to the uniform stationary distribution under the JC model. In the presence of a mutation, the alternative nucleotide is chosen uniformly among the alternative three bases. This is equivalent to having the same transition rate to any alternative nucleotide. Although the ISM and the JC models are different, the expected number of mutations under both models is the same. 

We note that the goal of the comparison is not to determine whether our method is superior, but rather to see if the performance of Tajima-based inference is in line with the results obtained through two popular Kingman-based methods in some challenging population scenarios. This can be interpreted as a check of the validity of our method and implementation.

We approximate the posterior distributions with $10^7$ iterations for the BEAST-implemented methods after a burn-in period of $10^6$ iterations and after thinning every $10^3$ iterations. Trace plots and ESSs all suggest convergence, except for the SKY ``Exp" $n=14$, which we exclude because we could not obtain convergent runs. 
We once more include the oracle estimator that infers $N_e$ from the true \bfg{} discussed in the main manuscript \citep{pal12}.  

Table~\ref{tab:sim_results_supp} summarizes SRE, MRW, and ENV for the $9$ simulated data sets achieved with our method (for the fixed computational budget runs), SKY, GMRF, and ``Oracle". 
First, no method unequivocally outperforms the others. The oracle methodology is the method with the best overall performance more frequently. Again, we note that the advantage of knowing \bfg{} is not as big as one would expect. Both SKY and GMRF have much narrower credible regions for the bottleneck trajectory. On the other hand, Tajima has the best overall performance in the ``drop" trajectory (low SRE and MRW). Note that $100\%$ ENV is not always an indicator of accuracy because it can be achieved with a very wide credible region.

Figure \ref{fig:sim2} plots the posterior median and credible regions of $\log N_e(t)$ for the nine scenarios considered for SKY and GMRF (our method's output is included in the manuscript). Recall that true trajectories are depicted as dashed lines, posterior medians as black lines, and $95\%$ credible regions as gray shaded areas. The figure provides a visualization of the results of Table \ref{tab:sim_results_supp}: accuracy increases with the sample size, the credible regions narrow down as $n$ increases, and the ``Drop" scenario appears to be the most challenging one.

\newpage

\textbf{Simulations: Kingman estimates of $N_e$}
\begin{figure}[!ht]
	\begin{center}
		\includegraphics[scale=0.55]{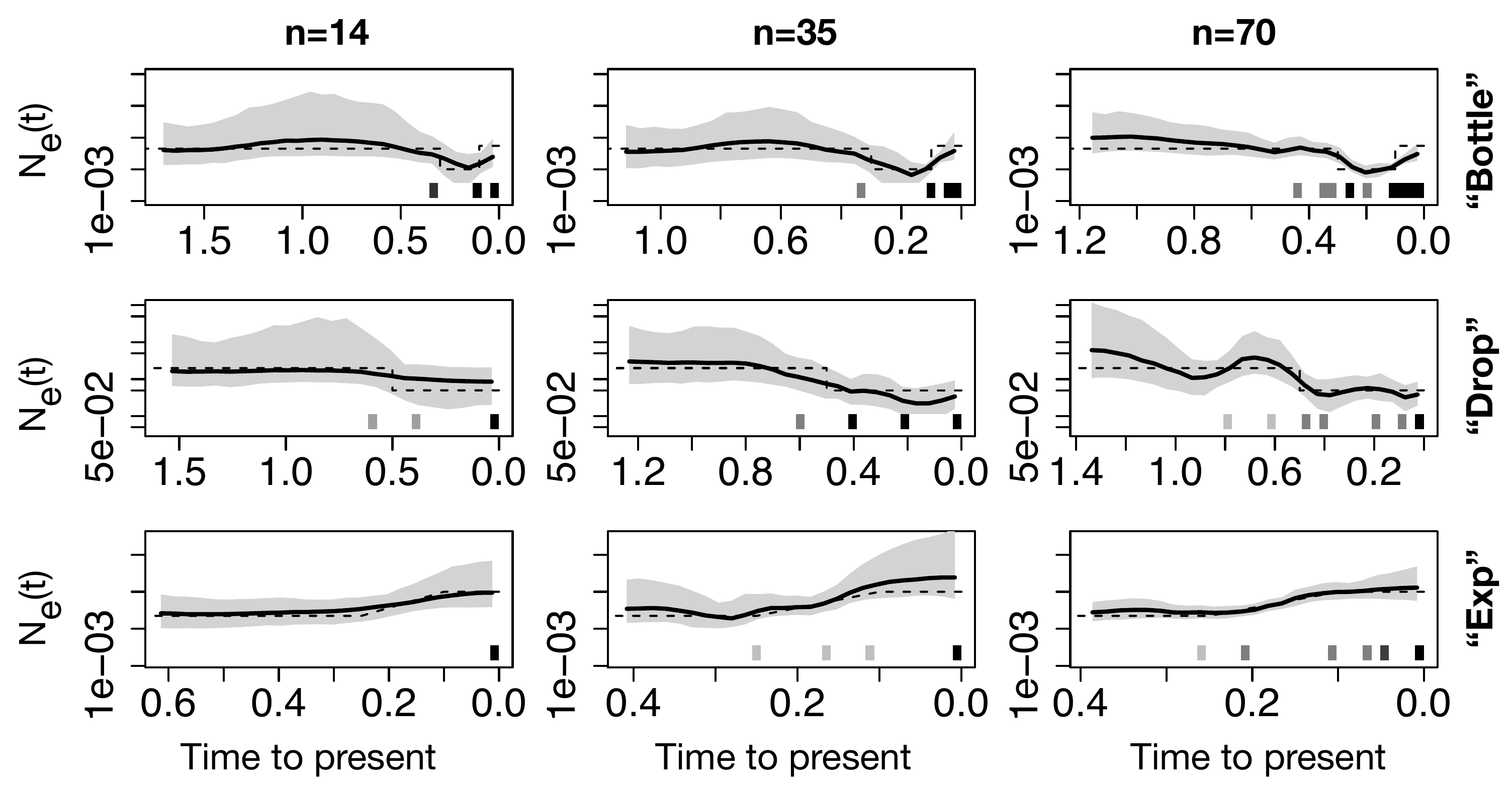}
	\end{center}
	\caption{\small{\textbf{Simulation: effective population size posterior medians from different trajectories and sample sizes for the Kingman-based model.}
			$N_e$ posterior distribution from simulated data with three population size trajectories (rows) - bottleneck (``Bottle"), exponential growth (``Exp") and instantaneous fall (``Drop") -  different sample sizes (columns) - $n=14$, $n=35$ and $n=70$. Posterior medians are depicted as solid black lines and 95\% Bayesian credible intervals are depicted by shaded areas. \bfn{} and \bfs{} are depicted by the heat maps at the bottom of each panel: the squares along the time axis depicts the sampling time, while the intensity of the black color depicts the number of samples. More details are given in Table~\ref{sim_summary} in the main manuscript.}}
	\label{fig:sim_kingman}
\end{figure}

\newpage

\textbf{Simulations: fixed number of iterations}

In Section \ref{sim} of the manuscript, we reported the results of the simulation study obtained with the Tajima and Kingman methods for a fixed computational budget ($72$ hours). Here, we report the results obtained with the two methods for a fixed number of iterations ($1\times 10^6$, with a burn-in of $4 \times 10^5$, and thinning every $500$ iterations). Table \ref{tab:sim_ite_eff} reports the mean ESS of \bft{} and of $\log N_e$, Table \ref{tab:sim_ite} reports the three criteria used to assess accuracy of the estimates (ENV, SRE, MRW). 

In terms of ESS, the two methods show more similar performance than those obtained for a fixed computational budget: with regards to the mean ESS of \bft{}, in $6$ out of $9$ data sets, the methods are tied, in $2$ Tajima is better, and in $1$ Kingman is better; with regards to the mean ESS of $\log N_e$, there are $3$ ties, Tajima shows better performance in 3, and Kingman is $3$. This analysis suggests that for a fixed number of iterations, the mixing of the two methods is comparable. 

Similarly, the empirical performance of the two methods is very similar, having ENV practically identical and a split performance in terms of SRE and MRW. The results in Table \ref{tab:sim_ite} are consistent with the findings discussed  in Section \ref{sim} of the manuscript.

\begin{SCtable}[2][h]
	
	\label{tab:sim_ite_eff}
	\caption{\small{\textbf{Simulation: mean effective sample sizes of  \bft{}  (ESS \bft{}), and $\log N_e$ (ESS $\log N_e$) of  Tajima and Kingman for a fixed computational budget.}
			Mean ESS for three population trajectories (Bottle, Exp, Drop) and three population sizes ($n=14,35,70$). Bold depicts the method with the best performance (excluding the ``oracle") or within $10\%$ of the best performance. The MCMC was run for one million iterations for both models. }}
	\scalebox{0.75}{
		\vspace{1cm}
		\renewcommand{\arraystretch}{0.7} 
		\begin{tabular}{@{\extracolsep{5pt}} cc|cc|cc}
			\hline \\[-1.8ex]
			& & \multicolumn{2}{c}{ESS \bft} & \multicolumn{2}{c}{ESS $\log N_e$}\\
			Label & n & Tajima & Kingman & Tajima & Kingman \\ 
			\hline \\[-1.8ex] 
			\multirow{3}{*}{\rotatebox[origin=c]{90}{Bottle}}& 14 & \textbf{241.15} & \textbf{232.52} & \textbf{218.3} & 79.81 \\ 
			& 35 & \textbf{1092.51} & \textbf{1201} & \textbf{191.91} & 70.01 \\ 
			& 70 & \textbf{1201} & \textbf{1201} & 161.46 & \textbf{249.95} \\ 
			\multirow{3}{*}{\rotatebox[origin=c]{90}{Drop}} & 14 &\textbf{ 58.61} & \textbf{58.81} & \textbf{334.35} & \textbf{315.83} \\ 
			& 35 & \textbf{648.35} & 440.14 & 161.6 & \textbf{315.15} \\ 
			& 70 & \textbf{1201} & \textbf{1201} & 170.36 & \textbf{268.59} \\ 
			\multirow{3}{*}{\rotatebox[origin=c]{90}{Exp}} & 14 & \textbf{172.83} & \textbf{177.05} & \textbf{226.52} & 205.88 \\ 
			& 35 & 1036.38 & \textbf{1201} & \textbf{76.65} & \textbf{76.28} \\ 
			& 70 & \textbf{1201} & 1023.59 & \textbf{60.3} & \textbf{56.56} \\ 
			\hline \\[-1.8ex]
	\end{tabular} }
	
\end{SCtable}

\begin{table} [h]\centering
	\renewcommand{\arraystretch}{0.7} 
	\caption{\small{\textbf{Simulation: performance comparison between Tajima, Kingman and Oracle models for a fixed number of MCMC iterations.}
			Envelope (ENV), sum of relative errors (SRE), and mean relative width (MRW) for three population trajectories (Bottle, Exp, Drop) and three population sizes ($n=14,35,70$). Tajima (our model), Kingman  (Kingman $n$-coalescent), Oracle \citep{pal12} (known \bfg). Bold depicts the method with the best performance (excluding the ``oracle") or within $10\%$ of the best performance. The MCMC was run for one million iterations for both models. }}
	\label{tab:sim_ite}
	\vspace{0.25cm}
	\scalebox{0.75}{
		
		\begin{tabular}{@{\extracolsep{5pt}} cc|c|cc|c|cc|c|cc}
			\hline \\[-1.8ex]
			& &\multicolumn{3}{c}{$\%$ENV} & \multicolumn{3}{c}{SRE} & \multicolumn{3}{c}{MRW} \\
			Label & n & Oracle & Tajima & Kingman & Oracle & Tajima & Kingman & Oracle & Tajima & Kingman  \\ 
			\hline \\[-1.8ex] 
			\multirow{3}{*}{\rotatebox[origin=c]{90}{Bottle}} & 14 & 100 & 98 & 100 & 408.11 & \textbf{78.67} & 207.71 & 20164.85 & \textbf{97.48} & 71982.99  \\ 
			& 35 & 99 & 96 & 92 & 155.81 & \textbf{65.16} & 88.55 & 203.52 & \textbf{189.39} & 2422.496 \\ 
			& 70 & 98 & 82 & 82 & 121.34 & 92.98 & \textbf{78.81} & 23.33 & 15.66 & \textbf{12.72}\\ 
			\multirow{3}{*}{\rotatebox[origin=c]{90}{Drop}} & 14 & 100 & 100 & 100 & 28.78 & 40.96 & \textbf{33.49} & 10.54 & \textbf{5.36} & 6.15  \\ 
			& 35 & 99 & 100 & 96 & 21.27 & 44.67 & \textbf{32.9} & 2.96 & 12.31 & \textbf{5.01}  \\ 
			& 70 & 99 & 98 & 98 & 17.1 & \textbf{11.96} & 34.38 & 2.13 & \textbf{2.11} & \textbf{4.72}  \\ 
			\multirow{3}{*}{\rotatebox[origin=c]{90}{Exp}} & 14 & 100 & 100 & 100 & 35.94 & \textbf{44.48} & \textbf{40.47} & 16.56 & 39.36 & \textbf{24.85}  \\ 
			& 35 & 100 & 95 & 100 & 35.58 & 258.32 & \textbf{135.99} & 11.41 & 777.13 & \textbf{42.76}  \\ 
			& 70 & 100 & 100 & 100 & 30.71 & 61.95 & \textbf{47.43} & 3.64 & \textbf{4.9} & 6.21  \\ 
			\hline \\[-1.8ex]
	\end{tabular} }
\end{table}

\newpage
\textbf{Simulations with multiple loci}

In Section \ref{sec:extensions}, we explain how to apply our method when the data are collected at $L$ independent loci. The assumption is that the effective population size is the same across loci. 
We also assume that all loci share the same mutation rate in our implementation, but this assumption can be easily modified.

\begin{SCfigure}[2][h]
	
	\includegraphics[scale=0.6]{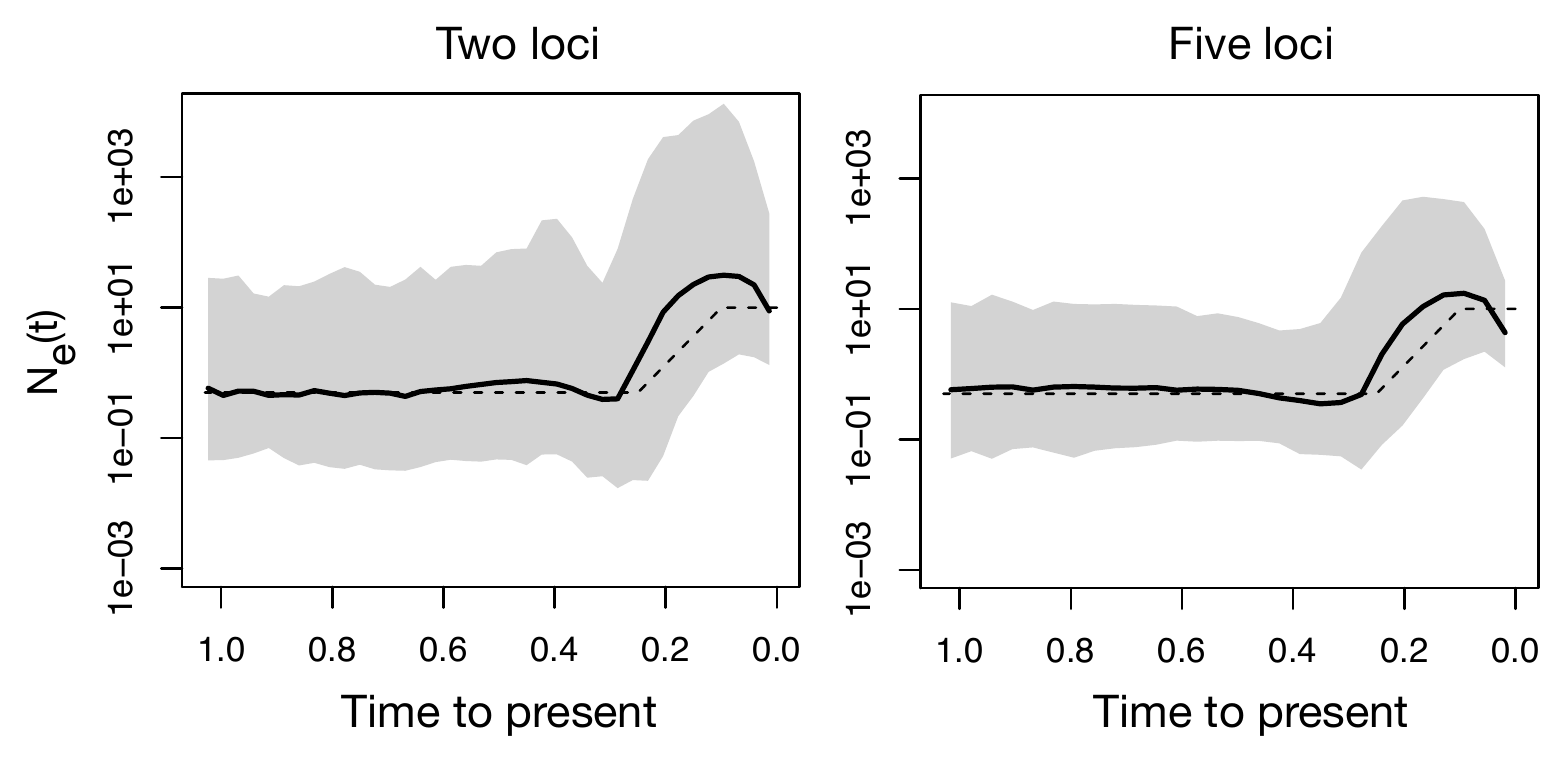}
	
	\caption{\small{\textbf{Simulation: effective population size posterior medians from different trajectories and sample sizes.}
			$N_e$ posterior distribution from simulated data of under exponential growth (``Exp") $N_e$, $n=14$, $L=2$ (first panel), and $L=5$ (second panel). Posterior medians are depicted as solid black lines and 95\% Bayesian credible intervals are depicted by shaded areas.}}
	\label{fig:multi_loci}
	
\end{SCfigure}

To test our implementation for multiple loci, we simulated $5$ datasets assuming $N_e$ has the ``Exp" trajectory \eqref{exp} ($n=14$. $s=0$). Figure \ref{fig:multi_loci} shows our results for $L=2$ and $L=5$. The case $L=5$ includes the two datasets of the $L=2$ case and an additional three. As expected, increasing the number of loci substantially reduces the width of the credible region. This is consistent with theoretical expectation.

\newpage
\textbf{Simulations with unknown mutation rate}

We replicate the analyses of Section \ref{sim} of the manuscript for the case of joint estimation of $N_e$ and $\mu$ (we previously assumed known $\mu$). The method is implemented as discussed in Section \ref{sec:extensions}. For parsimony, we consider only a subset of the datasets discussed in Section \ref{sim}. In particular, we only analyze the three datasets simulated under the ``Exp" scenario (see Supplementary material ``Simulation details"). In the case $n=14$, the dataset is isochronous, hence  $N_e$ and $\mu$ are not jointly identifiable. The case $n=35$ and $n=70$ are instead heterochronous datasets. 

\begin{figure}[h]
	\begin{center}
		\includegraphics[scale=0.92]{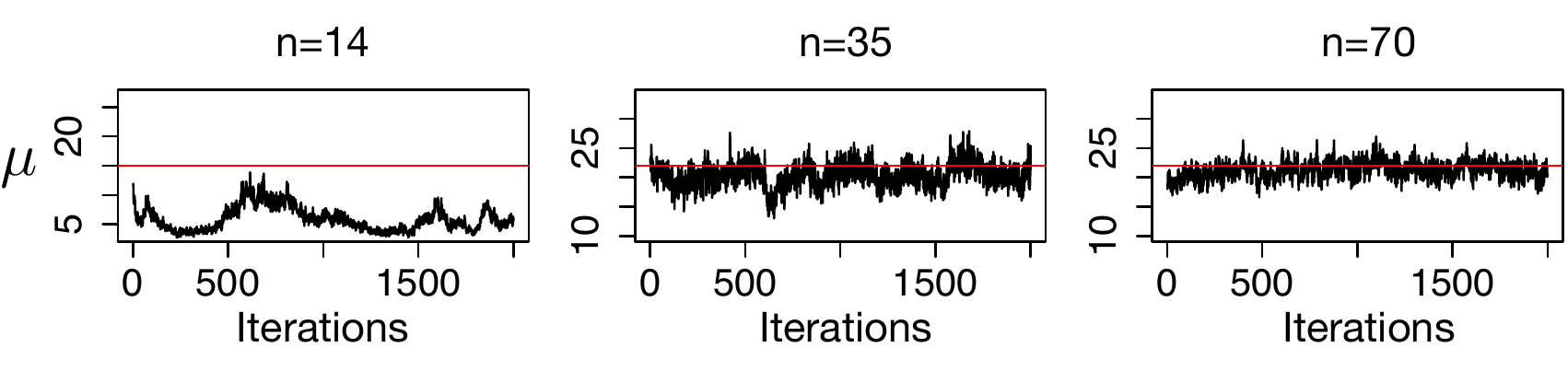}
	\end{center}
	\caption{\small{\textbf{Trace plots of $\mu$ for the three sample sizes considered.} The three panels refers to the ``exp" datasets discussed in Section \ref{sim} of the manuscript (first panel $n=14$, second panel $n=35$, and third panel $n=70$). MCMC was run for $4 \times 10^6$ iterations. The plot refers to the last $2000$ samples of the chain (thinned every $1000$ samples).}}
	\label{fig:plot_mu}
\end{figure}

\begin{figure}[h]
	\begin{center}
		\includegraphics[scale=0.52]{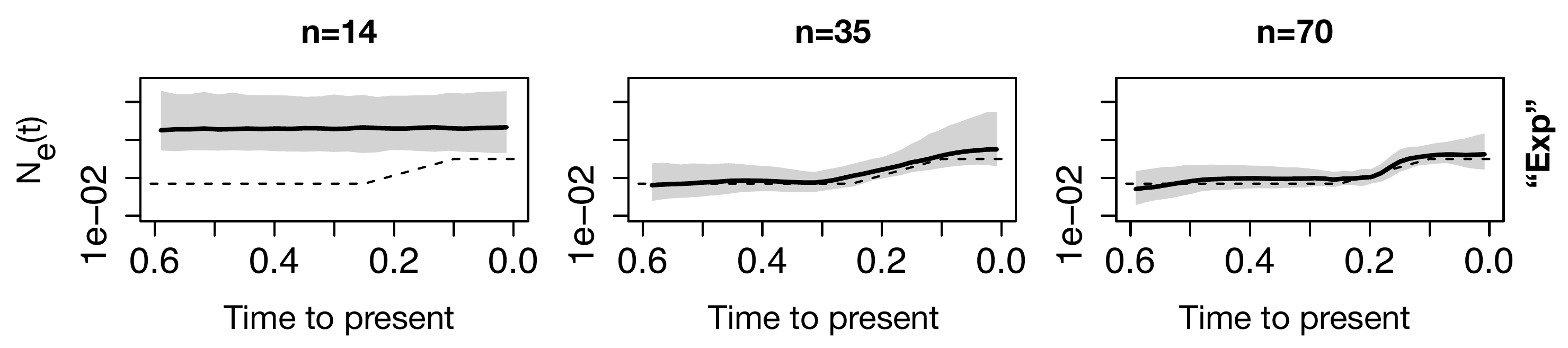}
	\end{center}
	\caption{\small{\textbf{Simulation: effective population size posterior medians from the three sample sizes in the unknown $\mu$ and  ``exp" scenario.}
			$(N_e(t))_{t\geq 0}$ posterior distribution from simulated data of under exponential growth (``Exp"), $n=14$ (first panel, isochronous samples), $n=35$ (second panel, heterochronous samples), and $n=70$ (third panel, heterochronous samples). Posterior medians are depicted as solid black lines and 95\% Bayesian credible intervals are depicted by shaded areas.} }
	\label{fig:random_mu}
\end{figure}

We run the chain for $4\times 10^6$ iterations, thinning every $1000$ iterations, and discard the first $1\times 10^6$ iterations. Figure \ref{fig:plot_mu} depicts the trace plots of $\mu$ of the last $2000$ samples from the MCMC, along with the true value (red line). Figure \ref{fig:random_mu} depicts the posterior medians and credible regions for the three datasets.

The length of chain ensured convergent runs in the $n=35$ and $n=70$ scenarios. Not surprisingly, the chain in the $n=14$ scenario does not seem to converge: this is confirmed by the trace plots (\textit{e.g} first panel of Figure \ref{fig:random_mu} -- trace plot of $\mu$). Neither $\mu$ nor $N_e$ are correctly estimated. This is consistent with the theoretical prediction, given that the $n=14$ case is not identifiable. On the other hand, we obtained convergent runs for the heterochronous data sets, as indicated by the trace plots  (\textit{e.g} second and third panels of Figure \ref{fig:random_mu} -- trace plot of $\mu$) and a mean ESS for \bft{} and $\log N_e$ above $100$. As expected, the number of MC iterations required for convergence is higher than in the fixed $\mu$ case.

The uncertainty in $\mu$ does not consistently lead to wider credible regions (as indicated by MRW). It does only in the case of $n=70$:  the MRW is $7.56$ with random $\mu$ and  $3.97$ with fixed $\mu$. In the case of $n=35$ the MRW is $57.43$ with random $\mu$ and $116.97$ with fixed $\mu$. ENV and SRE are comparable in the two settings ($n=35$ unknown $\mu$, ENV=$99$, SRE=$102.97$; $n=70$ random $\mu$: ENV=$100$, SRE=$73.71$, we refer to the main manuscript for the case of fixed $\mu$).

\newpage
\textbf{North American Bison - Details of the data set and the analysis}
\begin{SCfigure}[2][!t]
	\hspace{-0.2cm}\includegraphics[width=.5\textwidth]{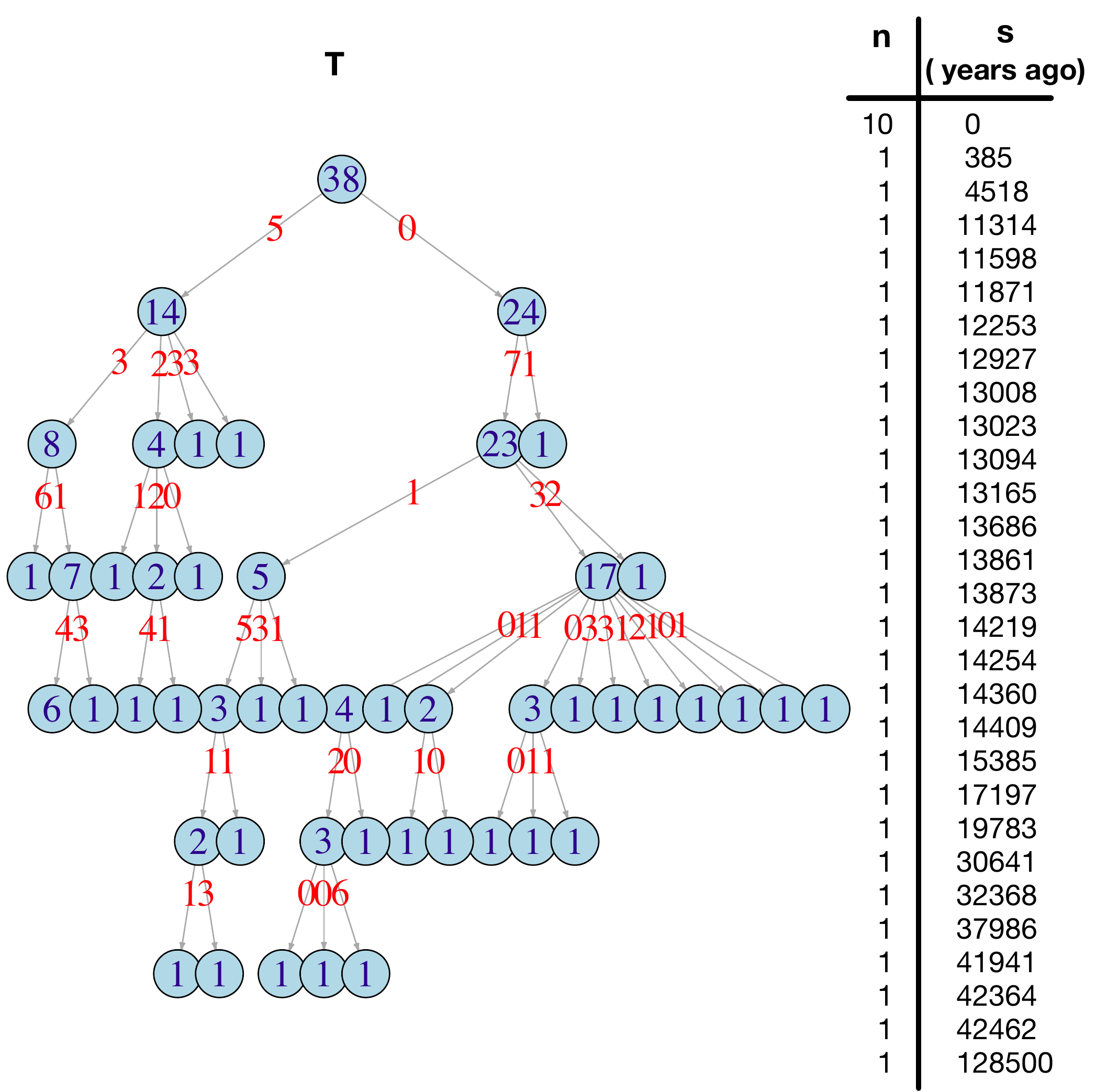}
	\caption{\small{\textbf{Bison study: data \citep{fro17}.} Perfect phylogeny \bfT{} of $38$ bison sequences selected from \cite{fro17} data-set. Node labels depict the number of sequences subtending that node. The $91$ mutations are allocated along the edges of \bfT{} (all of them are single digits). Sampling information are not written in this Figure. The two vectors \bfn{} and \bfs{} are represented by the columns to the right of \bfT{}. Sampling times are obtained by radiocarbon dating. The scale is number of years before present.
			\vspace{1.5cm}}}
	\label{fig:bison_data}
\end{SCfigure}

\cite{fro17} data comprises $50$ sequences ($14$ modern and $36$ ancient). DNA was extracted from bison specimens from Canada (28, three locations), USA (9, two locations), Siberia (7, three locations), and unknown locations (5). It includes sequences of $37$ \text{Bison priscus} (extinct ancient bison), $1$ \text{Bison latifrons} (extinct ancient bison), $11$ \text{Bison bison} (modern bison),  and $4$  \text{Bos grunniens} (control group). We selected $38$ out of $50$ sequences. We removed the control group sequences and the Siberian sequences to analyze samples from a single population (\cite{fro17} (Figure~1) suggested population structure). We removed the \text{Bison latifrons} sequence because it has $3803$ ambiguities   \textit{i.e.}, sites in a sequence that cannot be unambiguously assigned to a unique nucleotide basis at sites where all the other samples have valid entries. 
Out of the $94$ observed polymorphic sites, we retain $91$ sites compatible with the ISM assumption. To encode data in the $0-1$ incidence matrix representation $\bfY_1$, we use the root of the UPGMA tree reconstructed using \texttt{R} function \texttt{upgma} (\texttt{phanghorn}) as the ancestral state. Figure~\ref{fig:bison_data} displays the perfect phylogeny \bfT{} and the vectors \bfs{} and \bfn{}.

For our inference procedure, we set $\epsilon=0.09$, $Z_1=2$, $\sigma=0.02$, and approximated the posterior distribution with $1.5 \times 10^6$ iterations after a burn-in of $8 \times 10^5$ and after thinning every $200$ iterations. As a comparison, we ran GMRF on BEAST and approximated the posterior distribution with $1 \times 10^7$ iterations after a burn-in of $1 \times 10^6$ and after thinning every $1000$ iteration. We used the default values for all GMRF hyperparameters. We initialized both methods with the same genealogy (serial UPGMA). To compute the likelihood, we used the BEAST mutation rate estimate per site per year of $2.52 \times 10^{-8}$. Both methods are then run to estimate $N_e$ with that mutation rate fixed.

\textbf{SARS-CoV-2 - Details of the data sets and the analysis}

As we mentioned in the manuscript, we analyzed $123$ whole-genome sequences collected in France, and $32$ sequences collected in Germany and publicly available in the GISAID EpiCov database \citep{shu2017gisaid}.	We only analyzed high coverage sequences with more than 25000 base pairs and performed multiple sequence alignment with Mafft \citep{katoh2013mafft}.
To encode nucleotide data as binary sequences $\bfY_1$, we used the GenBank MN908947 \citep{wu20} sequence as the ancestral reference and eliminated sites that were not present in the ancestral sequence. The numbers of variable sites observed are $137$ and $45$ for France and Germany respectively. The observed patterns of mutations in both datasets are compatible with the ISM (no site was further removed). The Gisaid reference numbers of the sequences included in this study and data access acknowledgment are included in the supplementary material. 
We note that observed differences may be caused by sequencing errors, and these are being ignored in our study.
The heat maps included in each panel of Figure~\ref{fig:covid} show the sampling frequency information. In the French dataset, $109$ out of $123$ samples were collected in March (at least one sample every day from $03/01/20$ to $03/22/20$), $9$ in February (spread over $5$ different dates), $5$ in January (spread over $3$ days, oldest sample dated $01/23/20$). In the German dataset, $25$ out of $32$ samples were collected in March (spread over $7$ different dates and $03/16/20$ last sampling day), $6$ in February (spread over $4$ dates), $1$ in January (oldest sample $01/28/20$). We include in each dataset the reference sequence.

For our inference procedure, we set $\epsilon=0.11$, $Z_1=2$, $\sigma=0.02$, and approximate the posterior distribution with $1.4 \times 10^6$ iterations after a burn-in of $8 \times 10^5$ and after thinning every $100$ iterations. For comparison, we ran GMRF on BEAST assuming the HKY mutation model with empirically estimated frequencies \citep{hky} as proposed in previous studies \citep{scire2020phylodynamic} and approximate the posterior distribution with $5 \times 10^7$ iterations after a burn-in of $5 \times 10^6$ and after thinning every $1000$ iteration. We used the default values for all GMRF hyperparameters. We initialized both methods with the serial UPGMA genealogy \citep{dru00}. 
We first estimated the mutation rate using BEAST: BEAST estimated a mutation rate of  $5.99 \times 10^{-4}$ mutations per site per year in the French dataset, and $9.81 \times 10^{-4}$ mutations per site per year in the German dataset. We then fixed the mutation rates to the estimated values obtained with BEAST in both methods to infer the effective population size. 

\newpage 
\textbf{SARS-CoV-2 Molecular Data Description:} Data set used in the study in Section 7. We acknowledge the following sequence submitting laboratories to Gisaid.org: 
\begin{itemize}
	\item Charit\'{e} Universit\"{a}tsmedizin Berlin, Institute of Virology. Victor M Corman, Julia Schneider, Talitha Veith, Barbara M\"{u}hlemann, Markus Antwerpen, Christian Drosten, Roman W\"{o}lfel.
	\item Bundeswehr Institute of Microbiology. Mathias C Walter, Markus H Antwerpen and Roman W\"{o}lfel.
	\item Center of Medical Microbiology, Virology, and Hospital Hygiene, University of Duesseldorf. Ortwin Adams, Marcel Andree, Alexander Dilthey, Torsten Feldt, Sandra Hauka, Torsten Houwaart, Björn-Erik
	Jensen, Detlef Kindgen-Milles, Malte Kohns Vasconcelos, Klaus Pfeffer, Tina Senff, Daniel Strelow, Jörg Timm,
	Andreas Walker, Tobias Wienemann.
	\item CNR Virus des Infections Respiratoires - France SUD. Antonin Bal, Gregory Destras, Gwendolyne Burfin, Solenne Brun, Carine Moustaud, Raphaelle Lamy, Alexandre Gaymard, Maude Bouscambert-Duchamp, Florence Morfin-Sherpa, Martine Valette, Bruno Lina, Laurence Josset.
	\item National Reference Center for Viruses of Respiratory Infections, Institut Pasteur, Paris. M\'{e}lanie Albert, Marion Barbet, Sylvie Behillil, M\'{e}line Bizard, Angela Brisebarre, Flora Donati, Fabiana Gambaro, Etienne Simon-Lori\`{e}re, Vincent Enouf, Maud Vanpeene, Sylvie van der Werf, L\`{e}a Pilorge.
	\item Laboratoire Virpath, CIRI U111, UCBL1, INSERM, CNRS, ENS Lyon. Olivier Terrier, Aurélien Traversier, Julien Fouret, Yazdan Yazdanpanah, Xavier Lescure, Alexandre Gaymard, Bruno Lina, Manuel Rosa-Calatrava.
\end{itemize}

See below for a description of all sequence sampling locations and  dates. 
\begin{table}
	\centering
	\begin{tabular}{llll}
		\hline
		gisaid\_epi\_isl & date & country & division \\ 
		\hline
		EPI\_ISL\_412912 & 2020-02-25 & Germany & Baden-Wuerttemberg \\ 
		EPI\_ISL\_406862 & 2020-01-28 & Germany & Bavaria \\ 
		EPI\_ISL\_414520 & 2020-03-02 & Germany & Bavaria \\ 
		EPI\_ISL\_414521 & 2020-03-02 & Germany & Bavaria \\ 
		EPI\_ISL\_413488 & 2020-02-28 & Germany & North Rhine Westphalia \\ 
		EPI\_ISL\_414497 & 2020-02-25 & Germany & North Rhine Westphalia \\ 
		EPI\_ISL\_414499 & 2020-02-26 & Germany & North Rhine Westphalia \\ 
		EPI\_ISL\_414505 & 2020-02-27 & Germany & North Rhine Westphalia \\ 
		EPI\_ISL\_414509 & 2020-02-28 & Germany & North Rhine Westphalia \\ 
		EPI\_ISL\_417457 & 2020-03-10 & Germany & Duesseldorf \\ 
		EPI\_ISL\_417458 & 2020-03-11 & Germany & Duesseldorf \\ 
		EPI\_ISL\_417459 & 2020-03-11 & Germany & Duesseldorf \\ 
		EPI\_ISL\_417460 & 2020-03-11 & Germany & Duesseldorf \\ 
		EPI\_ISL\_417461 & 2020-03-11 & Germany & Duesseldorf \\ 
		EPI\_ISL\_417462 & 2020-03-11 & Germany & Duesseldorf \\ 
		EPI\_ISL\_417463 & 2020-03-13 & Germany & Duesseldorf \\ 
		EPI\_ISL\_417464 & 2020-03-14 & Germany & Duesseldorf \\ 
		EPI\_ISL\_417465 & 2020-03-14 & Germany & Duesseldorf \\ 
		EPI\_ISL\_417466 & 2020-03-14 & Germany & Duesseldorf \\ 
		EPI\_ISL\_417467 & 2020-03-15 & Germany & Duesseldorf \\ 
		EPI\_ISL\_417468 & 2020-03-16 & Germany & Duesseldorf \\ 
		EPI\_ISL\_419541 & 2020-03-14 & Germany & Duesseldorf \\ 
		EPI\_ISL\_419542 & 2020-03-15 & Germany & Duesseldorf \\ 
		EPI\_ISL\_419543 & 2020-03-15 & Germany & Duesseldorf \\ 
		EPI\_ISL\_419544 & 2020-03-15 & Germany & Duesseldorf \\ 
		EPI\_ISL\_419545 & 2020-03-15 & Germany & Duesseldorf \\ 
		EPI\_ISL\_419546 & 2020-03-15 & Germany & Duesseldorf \\ 
		EPI\_ISL\_419548 & 2020-03-15 & Germany & Duesseldorf \\ 
		EPI\_ISL\_419549 & 2020-03-15 & Germany & Duesseldorf \\ 
		EPI\_ISL\_419550 & 2020-03-16 & Germany & Duesseldorf \\ 
		EPI\_ISL\_419551 & 2020-03-16 & Germany & Duesseldorf \\ 
		EPI\_ISL\_419552 & 2020-03-16 & Germany & Duesseldorf \\ 
		EPI\_ISL\_402125 & 2019-12-26 & China & Hubei \\ 
		\hline
	\end{tabular}
\end{table}

\begin{table}
	\centering
	\begin{tabular}{llll}
		\hline
		gisaid\_epi\_isl & date & country & division \\ 
		\hline
		EPI\_ISL\_418412 & 2020-03-15 & France & Auvergne-Rhône-Alpes \\ 
		EPI\_ISL\_418413 & 2020-03-15 & France & Auvergne-Rhône-Alpes \\ 
		EPI\_ISL\_418414 & 2020-03-15 & France & Auvergne-Rhône-Alpes \\ 
		EPI\_ISL\_418416 & 2020-03-16 & France & Auvergne-Rhône-Alpes \\ 
		EPI\_ISL\_418417 & 2020-03-16 & France & Auvergne-Rhône-Alpes \\ 
		EPI\_ISL\_418418 & 2020-03-16 & France & Auvergne-Rhône-Alpes \\ 
		EPI\_ISL\_418419 & 2020-03-16 & France & Auvergne-Rhône-Alpes \\ 
		EPI\_ISL\_418420 & 2020-03-17 & France & Auvergne-Rhône-Alpes \\ 
		EPI\_ISL\_418422 & 2020-03-17 & France & Auvergne-Rhône-Alpes \\ 
		EPI\_ISL\_418423 & 2020-03-17 & France & Auvergne-Rhône-Alpes \\ 
		EPI\_ISL\_418424 & 2020-03-17 & France & Auvergne-Rhône-Alpes \\ 
		EPI\_ISL\_418425 & 2020-03-17 & France & Auvergne-Rhône-Alpes \\ 
		EPI\_ISL\_418426 & 2020-03-17 & France & Auvergne-Rhône-Alpes \\ 
		EPI\_ISL\_418427 & 2020-03-17 & France & Auvergne-Rhône-Alpes \\ 
		EPI\_ISL\_418428 & 2020-03-17 & France & Auvergne-Rhône-Alpes \\ 
		EPI\_ISL\_419168 & 2020-03-17 & France & Auvergne-Rhône-Alpes \\ 
		EPI\_ISL\_418429 & 2020-03-18 & France & Auvergne-Rhône-Alpes \\ 
		EPI\_ISL\_418430 & 2020-03-18 & France & Auvergne-Rhône-Alpes \\ 
		EPI\_ISL\_418431 & 2020-03-18 & France & Auvergne-Rhône-Alpes \\ 
		EPI\_ISL\_418432 & 2020-03-18 & France & Auvergne-Rhône-Alpes \\ 
		EPI\_ISL\_419169 & 2020-03-21 & France & Auvergne-Rhône-Alpes \\ 
		EPI\_ISL\_419170 & 2020-03-21 & France & Auvergne-Rhône-Alpes \\ 
		EPI\_ISL\_419171 & 2020-03-21 & France & Auvergne-Rhône-Alpes \\ 
		EPI\_ISL\_419172 & 2020-03-21 & France & Auvergne-Rhône-Alpes \\ 
		EPI\_ISL\_419173 & 2020-03-21 & France & Auvergne-Rhône-Alpes \\ 
		EPI\_ISL\_419174 & 2020-03-20 & France & Auvergne-Rhône-Alpes \\ 
		EPI\_ISL\_419175 & 2020-03-21 & France & Auvergne-Rhône-Alpes \\
		\hline
	\end{tabular}
\end{table}

\begin{table}
	\centering
	\begin{tabular}{llll}
		\hline
		gisaid\_epi\_isl & date & country & division \\ 
		\hline
		EPI\_ISL\_419176 & 2020-03-21 & France & Auvergne-Rhône-Alpes \\ 
		EPI\_ISL\_419177 & 2020-03-22 & France & Auvergne-Rhône-Alpes \\ 
		EPI\_ISL\_419178 & 2020-03-22 & France & Auvergne-Rhône-Alpes \\ 
		EPI\_ISL\_419179 & 2020-03-22 & France & Auvergne-Rhône-Alpes \\ 
		EPI\_ISL\_419180 & 2020-03-22 & France & Auvergne-Rhône-Alpes \\ 
		EPI\_ISL\_419181 & 2020-03-22 & France & Auvergne-Rhône-Alpes \\ 
		EPI\_ISL\_419182 & 2020-03-22 & France & Auvergne-Rhône-Alpes \\ 
		EPI\_ISL\_419183 & 2020-03-22 & France & Auvergne-Rhône-Alpes \\ 
		EPI\_ISL\_419184 & 2020-03-22 & France & Auvergne-Rhône-Alpes \\ 
		EPI\_ISL\_419185 & 2020-03-22 & France & Auvergne-Rhône-Alpes \\ 
		EPI\_ISL\_419186 & 2020-03-22 & France & Auvergne-Rhône-Alpes \\ 
		EPI\_ISL\_419187 & 2020-03-22 & France & Auvergne-Rhône-Alpes \\ 
		EPI\_ISL\_419188 & 2020-03-22 & France & Auvergne-Rhône-Alpes \\ 
		EPI\_ISL\_418219 & 2020-02-26 & France & Bretagne \\ 
		EPI\_ISL\_416502 & 2020-02-26 & France & Bretagne \\ 
		EPI\_ISL\_416503 & 2020-03-01 & France & Bretagne \\ 
		EPI\_ISL\_416504 & 2020-03-02 & France & Bretagne \\ 
		EPI\_ISL\_416505 & 2020-03-02 & France & Bretagne \\ 
		EPI\_ISL\_416506 & 2020-03-03 & France & Bretagne \\ 
		EPI\_ISL\_416507 & 2020-03-05 & France & Bretagne \\ 
		EPI\_ISL\_416508 & 2020-03-06 & France & Bretagne \\ 
		EPI\_ISL\_416509 & 2020-03-06 & France & Bretagne \\ 
		EPI\_ISL\_416510 & 2020-03-06 & France & Bretagne \\ 
		EPI\_ISL\_416511 & 2020-03-07 & France & Bretagne \\ 
		EPI\_ISL\_416512 & 2020-03-07 & France & Bretagne \\ 
		EPI\_ISL\_416513 & 2020-03-07 & France & Bretagne \\ 
		EPI\_ISL\_415651 & 2020-03-05 & France & Bourgogne-France-Comté \\ 
		EPI\_ISL\_415652 & 2020-03-05 & France & Bourgogne-France-Comté \\ 
		EPI\_ISL\_416757 & 2020-03-07 & France & Auvergne-Rhône-Alpes \\ 
		EPI\_ISL\_417340 & 2020-03-07 & France & Auvergne-Rhône-Alpes \\ 
		EPI\_ISL\_418222 & 2020-03-04 & France & Centre-Val de Loire \\ 
		EPI\_ISL\_416752 & 2020-03-04 & France & Auvergne-Rhône-Alpes \\ 
		EPI\_ISL\_416751 & 2020-03-05 & France & Auvergne-Rhône-Alpes \\ 
		EPI\_ISL\_414623 & 2020-02-25 & France & Grand Est \\ 
		EPI\_ISL\_414631 & 2020-03-04 & France & Grand Est \\ 
		EPI\_ISL\_414632 & 2020-03-04 & France & Grand Est \\ 
		\hline
	\end{tabular}
\end{table}

\begin{table}
	\centering
	\begin{tabular}{llll}
		\hline
		gisaid\_epi\_isl & date & country & division \\ 
		\hline
		EPI\_ISL\_418218 & 2020-02-21 & France & Hauts de France \\ 
		EPI\_ISL\_418220 & 2020-02-28 & France & Hauts de France \\ 
		EPI\_ISL\_414626 & 2020-02-29 & France & Hauts de France \\ 
		EPI\_ISL\_414627 & 2020-03-02 & France & Hauts de France \\ 
		EPI\_ISL\_414630 & 2020-03-03 & France & Hauts de France \\ 
		EPI\_ISL\_414635 & 2020-03-04 & France & Hauts de France \\ 
		EPI\_ISL\_414637 & 2020-03-04 & France & Hauts de France \\ 
		EPI\_ISL\_414638 & 2020-03-04 & France & Hauts de France \\ 
		EPI\_ISL\_415649 & 2020-03-05 & France & Hauts de France \\ 
		EPI\_ISL\_418223 & 2020-03-05 & France & Hauts de France \\ 
		EPI\_ISL\_418224 & 2020-03-08 & France & Hauts de France \\ 
		EPI\_ISL\_418225 & 2020-03-08 & France & Hauts de France \\ 
		EPI\_ISL\_415654 & 2020-03-09 & France & Hauts de France \\ 
		EPI\_ISL\_416493 & 2020-03-08 & France & Hauts de France \\ 
		EPI\_ISL\_416495 & 2020-03-10 & France & Hauts de France \\ 
		EPI\_ISL\_416496 & 2020-03-10 & France & Hauts de France \\ 
		EPI\_ISL\_416497 & 2020-03-10 & France & Hauts de France \\ 
		EPI\_ISL\_418226 & 2020-03-09 & France & Hauts de France \\ 
		EPI\_ISL\_418227 & 2020-03-12 & France & Hauts de France \\ 
		EPI\_ISL\_418228 & 2020-03-12 & France & Hauts de France \\ 
		EPI\_ISL\_418231 & 2020-03-15 & France & Hauts de France \\ 
		EPI\_ISL\_418236 & 2020-03-16 & France & Hauts de France \\ 
		EPI\_ISL\_418237 & 2020-03-16 & France & Hauts de France \\ 
		EPI\_ISL\_418238 & 2020-03-16 & France & Hauts de France \\ 
		EPI\_ISL\_418239 & 2020-03-16 & France & Hauts de France \\ 
		EPI\_ISL\_406596 & 2020-01-23 & France & Ile de France \\ 
		EPI\_ISL\_406597 & 2020-01-23 & France & Ile de France \\ 
		EPI\_ISL\_411219 & 2020-01-28 & France & Ile de France \\ 
		EPI\_ISL\_408430 & 2020-01-29 & France & Ile de France \\ 
		EPI\_ISL\_408431 & 2020-01-29 & France & Ile de France \\ 
		\hline
	\end{tabular}
\end{table}

\begin{table}
	\centering
	\begin{tabular}{llll}
		\hline
		gisaid\_epi\_isl & date & country & division \\ 
		\hline
		EPI\_ISL\_415650 & 2020-03-02 & France & Ile de France \\ 
		EPI\_ISL\_416498 & 2020-03-11 & France & Ile de France \\ 
		EPI\_ISL\_416499 & 2020-03-11 & France & Ile de France \\ 
		EPI\_ISL\_416501 & 2020-03-10 & France & Ile de France \\ 
		EPI\_ISL\_418229 & 2020-03-12 & France & Ile de France \\ 
		EPI\_ISL\_418230 & 2020-03-13 & France & Ile de France \\ 
		EPI\_ISL\_418232 & 2020-03-15 & France & Ile de France \\ 
		EPI\_ISL\_418233 & 2020-03-15 & France & Ile de France \\ 
		EPI\_ISL\_418234 & 2020-03-14 & France & Ile de France \\ 
		EPI\_ISL\_418235 & 2020-03-16 & France & Ile de France \\ 
		EPI\_ISL\_418240 & 2020-03-16 & France & Ile de France \\ 
		EPI\_ISL\_417333 & 2020-03-04 & France & Auvergne-Rhône-Alpes \\ 
		EPI\_ISL\_417334 & 2020-03-04 & France & Auvergne-Rhône-Alpes \\ 
		EPI\_ISL\_416753 & 2020-03-06 & France & Auvergne-Rhône-Alpes \\ 
		EPI\_ISL\_416754 & 2020-03-06 & France & Auvergne-Rhône-Alpes \\ 
		EPI\_ISL\_416756 & 2020-03-06 & France & Auvergne-Rhône-Alpes \\ 
		EPI\_ISL\_417337 & 2020-03-07 & France & Auvergne-Rhône-Alpes \\ 
		EPI\_ISL\_417336 & 2020-03-06 & France & Auvergne-Rhône-Alpes \\ 
		EPI\_ISL\_417339 & 2020-03-08 & France & Auvergne-Rhône-Alpes \\ 
		EPI\_ISL\_416758 & 2020-03-08 & France & Auvergne-Rhône-Alpes \\ 
		EPI\_ISL\_416747 & 2020-03-04 & France & Auvergne-Rhône-Alpes \\ 
		EPI\_ISL\_416748 & 2020-03-04 & France & Auvergne-Rhône-Alpes \\ 
		EPI\_ISL\_416750 & 2020-03-06 & France & Auvergne-Rhône-Alpes \\ 
		EPI\_ISL\_417338 & 2020-03-07 & France & Auvergne-Rhône-Alpes \\ 
		EPI\_ISL\_414624 & 2020-02-26 & France & Normandie \\ 
		EPI\_ISL\_416494 & 2020-03-04 & France & Normandie \\ 
		EPI\_ISL\_414625 & 2020-02-26 & France & Pays de la Loire \\ 
		EPI\_ISL\_416745 & 2020-03-10 & France & Auvergne-Rhône-Alpes \\ 
		EPI\_ISL\_416746 & 2020-03-03 & France & Auvergne-Rhône-Alpes \\ 
		EPI\_ISL\_416749 & 2020-03-04 & France & Auvergne-Rhône-Alpes \\ 
		\hline
	\end{tabular}
\end{table}

\end{document}